\documentclass[tightenlines,showpacs,prd,floatfix,preprintnumbers,amsmath,amssymb,nofootinbib,superscriptaddress]{revtex4}
\usepackage{amsfonts}
\usepackage[normalem]{ulem}
\usepackage{color}
\usepackage{bm}
\usepackage{amscd}
\usepackage{mathrsfs}
\def\pheq{\phantom{=}}
\newcommand{\Deqn}[1]{{Eq.~(\ref{#1})}}

\newcommand{\beq}{\begin{equation}}
\newcommand{\be}{\begin{equation}}
\newcommand{\eeq}{\end{equation}}
\newcommand{\ee}{\end{equation}}
\newcommand{\bea}{\begin{eqnarray}}
\newcommand{\eea}{\end{eqnarray}}

\newcommand{\beaa}{\begin{eqnarray*}} 
\newcommand{\eeaa}{\end{eqnarray*}} 

\newcommand{\na}{\nabla}
\newcommand{\dis}{\displaystyle} 
\newcommand{\Lie}{\mbox{\pounds}}
\newcommand{\bsube}{\begin{subequations}}
\newcommand{\esube}{\end{subequations}}

\newcommand{\hretab}{h^{\rm ret}_{\alpha\beta}}

\newcommand{\hsab}{h^{\rm s}_{\alpha\beta}}

\newcommand{\hrenab}{h^{\rm ren}_{\alpha\beta}}
\newcommand{\hsdw}{h^{\rm S}_{\alpha\beta}}

\newcommand{\pret}{\psi_4^{\rm ret}}

\newcommand{\pso}{\psi_0^{\rm s}}

\newcommand{\preto}{\psi_0^{\rm ret}}

\newcommand{\preno}{\psi_0^{\rm ren}}

\newcommand{\RET}{{\rm ret}}

\newcommand{\SING}{{\rm s}}

\newcommand{\al}{\alpha}

\begin{document}

\title{Gravitational Self-force in a Radiation Gauge}

\author{Tobias S. Keidl} 
\email{tobias.keidl@uwc.edu}
\affiliation{Department of Physics, 
University of Wisconsin--Washington County, 400 University Dr., West Bend, WI  53095, USA}

\author{Abhay G. Shah} 
\email{agshah@uwm.edu}
\email{a.g.shah@soton.ac.uk}
\affiliation{Center for Gravitation and Cosmology, Department of Physics, 
University of Wisconsin--Milwaukee, P.O. Box 413, Milwaukee, Wisconsin 53201, 
USA}

\author{John L. Friedman} 
\email{friedman@uwm.edu}
\affiliation{Center for Gravitation and Cosmology, Department of Physics, 
University of Wisconsin--Milwaukee, P.O. Box 413, Milwaukee, Wisconsin 53201, 
USA}

\author{Dong-Hoon Kim}

\email{ki1313@yahoo.com}
\affiliation{Max-Planck-Institut für Gravitationsphysik, 
Am Mühlenberg 1, D-14476 Golm,
Germany}
\affiliation{Division of Physics, Mathematics, and Astronomy,
California Institute of Technology, Pasadena, CA 91125, USA}
\affiliation{Institute for the Early Universe and Department of Physics, 
Ewha Womans University, Seoul 120-750, South Korea}

\author{Larry R. Price} 
\email{larry@gravity.phys.uwm.edu}
\affiliation{Center for Gravitation and Cosmology, Department of Physics, 
University of Wisconsin--Milwaukee, P.O. Box 413, Milwaukee, Wisconsin 53201, 
USA}

\date{\today}

\begin{abstract} 
  In this, the first of two companion papers, we present a method for finding 
the gravitational self-force in a modified radiation gauge for a particle moving on a 
geodesic in a Schwarzschild or Kerr spacetime.  An extension of 
an earlier result by Wald is used to show the spin-weight $\pm 2$ perturbed Weyl scalar
($\psi_0$ or $\psi_4$) determines the metric perturbation outside the particle 
up to a gauge transformation and an infinitesimal change in mass and angular momentum.  
A Hertz potential is used to construct the part of the retarded metric perturbation that 
involves no change in mass or angular momentum from $\psi_0$ in a radiation gauge.   
The metric perturbation is completed by adding changes in the mass and angular momentum 
of the background spacetime outside the radial coordinate $r_0$ of the particle in any 
convenient gauge. The resulting metric perturbation is singular on the trajectory of 
the particle and discontinuous across the sphere $r=r_0$. 
A mode-sum method can be used to renormalize the self-force, but the justification 
given in the published version of this paper \cite{sf2} referred to work by Sam 
Gralla \cite{gralla10} to justify the use of the renormalized self-force, and 
the radiation gauge we use does not satisfy the regularity conditions required by Gralla.   
Instead we show that the renormalized self-force, computed either from the 
retarded field for $r>r_0$ or for $r<r_0$  gives the correct equations of 
motion in a gauge smoothly related to a Lorenz gauge; and 
Pound et al.~\cite{pmb13} argue that the average of the self-force obtained 
in the way described in our paper for $r>r_0$ and for $r<r_0$ gives the 
correct equation of motion for our gauge (what Pound et al. call the 
no-string gauge).
 
In a Lorenz gauge one can renormalize the self-force by subtracting the leading and subleading terms in a mode-sum expansion of the retarded field, and the self-force 
in the two gauges just described is again obtained by subtracting only these  
leading and subleading terms.  A companion paper \cite{sf3} finds that these two 
terms coincide with the leading and subleading Lorenz-gauge terms.     

We explicitly compute the singular Weyl scalar and its mode-sum decomposition 
to subleading order in $L$  for a particle in circular orbit in a Schwarzschild 
geometry and obtain the renormalized field.  Because the singular field can be defined as 
this mode sum, the coefficients of each angular harmonic in the sum must 
agree with the large $L$ limit of the corresponding coefficients of the 
retarded field.  One may therefore compute the singular field by numerically matching 
the retarded field to a power series in $L$.  To check the accuracy of the 
numerical method, we analytically compute leading and subleading terms in the 
singular expansion of $\psi_0$ and compare the numerical and analytic values of 
the renormalization constants, finding agreement to high precision.   
Details of the numerical computation of the perturbed 
metric, the self-force, and the quantity $h_{\alpha\beta}u^\alpha u^\beta$ 
(gauge invariant under helically symmetric gauge transformations) are presented for 
this test case in the companion paper.

\end{abstract}

\maketitle

\section{\label{intro}Introduction}

Among the most important sources for LISA are extreme mass ratio 
inspirals (EMRIs) of solar mass compact objects into supermassive black holes.
LISA could potentially measure hundreds of EMRI events~\cite{Gair:2004iv} whose wide  
range of astrophysical and fundamental implications \cite{prince09,scch09} 
include determination of the Hubble constant~\cite{Schutz:1986gp,MacLeod:2007jd};  
of luminosity distance, mass and spin of galactic black holes; and 
measurements of the deviation from a Kerr geometry of the central object~\cite{Barack:2003fp}.  
Such measurements depend on accurate parameter estimation, for which it is 
essential to have accurate gravitational waveforms available; these require  
an accurate calculation of the gravitational self-force experienced by the particle.  

The gravitational self-force contains both dissipative and conservative parts.  
The dissipative part is simply the familiar contribution from the half-retarded 
minus half-advanced Green's function, smooth at the position of 
the particle.    The more difficult portion of calculating the gravitational 
self-force is the determination of the conservative part of the field.  
Estimates of its effect on the phase of the waveform are given, for example in 
\cite{burko03,2005CQGra..22S.801D,tanaka06,fh07,pp08}).  
The focus of our work is on this latter problem.  A number of authors 
have worked on this problem in recent years. Poisson's Living Review \cite{poisson04} 
is a comprehensive self-contained introduction to the subject; Barack \cite{barack09}
gives an extensive recent review; and Detweiler \cite{detweiler05} provides a 
more concise summary.  

A complicating feature of the conservative part of the self-force is a combination of 
its inherent non-locality in curved space due to scatter off 
curvature \cite{MinoSasaki,quinnwald,Detweiler:2002mi}) and the singular behavior 
of its expression near the particle.  The calculation can be made tractable through the observation 
that the field consists of both a regular part that is entirely responsible for the 
self-force and a singular part that contributes nothing to the self-force. 

The other complicating feature of self-force computations is the choice of gauge.  
The Lorenz gauge is particularly useful for sorting out formal issues but does not 
lead to separable, decoupled equations in the Kerr spacetime.  For this reason we 
choose to work in a radiation gauge that exploits the separability of the Teukolsky 
equation~\cite{1972PhRvL..29.1114T,1973ApJ...185..635T} for the perturbed Weyl scalar 
$\psi_0$ in the Kerr background.  Using the 
Hertz potential formalism developed for gravity by Kegeles and Cohen~\cite{ck74,kc79}, 
Chrzanowksi~\cite{chr75}, Stewart~\cite{stewart79} and Wald \cite{wald78} (which we will 
call the CCK formalism), 
it is possible to construct a metric perturbation from source-free solutions to the 
Teukolsky equation.  Explicit forms of the reconstruction are given for perturbations of
Schwarzschild and Kerr spacetimes, respectively, by Lousto and Whiting \cite{lw02} and Ori \cite{ori03}.
With $\psi_0$ written as a sum of angular harmonics (in Kerr, a sum of oblate spheroidal harmonics),
the reconstruction yields the part of the perturbed metric that has no change in the 
mass or angular momentum of the spacetime. 
As in the case of a static particle in flat space \cite{kfw}, there is a gauge 
in which the perturbed metric is smooth everywhere except on a
radial null ray extending inward to the horizon from each point of the particle's trajectory 
and a gauge in which the perturbed metric is smooth everywhere except on 
a radial null ray extending outward to infinity from each point of the trajectory. 
By choosing the first gauge for $r>r_0$ and the second for $r<r_0$, with $r_0$ 
the radial coordinate of the particle, we obtain a gauge in which the perturbed 
metric diverges only at the particle but which is discontinuous on the $r=r_0$ 
sphere.

There is a radiation-gauge form of a change in 
mass (an $\ell=0$ perturbation in the case of a Schwarzschild background) that arises from 
a Hertz potential in the CCK formalism, but it is singular 
on a ray through the particle \cite{bo01,kfw}.  One can find a nonsingular form for the 
metric of a mass perturbation in a radiation gauge \cite{kfw}, but we know of no advantage to using 
it.  To obtain a gauge in which the 
metric perturbation diverges
only on the particle's trajectory, we add the change in mass and angular 
momentum in a gauge for which that part of the metric perturbation is continuous.

An expression for the self-force is then found as a mode-sum in terms of the retarded metric.
To renormalize it, one must subtract off a singular part that does not contribute to the 
self-force.  We consider two alternatives, involving either an analytic or numerical 
determination of the singular part of the expression for the self-force $f^\alpha$ as 
a power series in the integer $\ell$ that labels the angular eigenvalues 
(more precisely, a series involving $L:= \ell+1/2$).  Renormalization relies on 
subtracting from the bare self-force -- from the expression for the self-force in 
terms of the retarded field -- a singular part that does not contribute to the 
self-force.  This can be checked by showing that the singular vector field $f^{\rm s\,\alpha}$ 
that one subtracts has vanishing angle-average over a small sphere about the particle. 
In the case of a circular orbit about a Schwarzschild black hole, the conservative part of 
the self force is axisymmetric about a radial ray through the particle.  We numerically compute the 
axisymmetric part of $f^{\rm s\,\alpha}$ and show that (as in a Lorenz gauge) it coincides with
the axisymmetric part of $-{\frak m}\nabla^\alpha \frac 1\rho$, with $\rho$ the geodesic 
distance to the trajectory. Because the angle-average of $\nabla^\alpha \frac 1\rho$ over a 
sphere of radius $\rho$ about the particle vanishes as $\rho\rightarrow 0$, the renormalized 
self-force gives the first-order correction to geodesic motion in our modified radiation gauge.%
\footnote{
Because of the discontinuity across $r=r_0$ of the metric perturbation, this 
claim needs justification beyond that given by Gralla \cite{gralla10}, and that has 
now been provided by Pound {\it et al.}\cite{pmb13}.}    

The paper is organized as follows:  In Sec. II, we introduce the self-force equations 
and a criterion for their use in a generic gauge; we briefly review features of mode-sum renormalization  
in a Lorenz gauge; and we review relevant parts of the Newman-Penrose~\cite{np62,np63} formalism 
and spin-weighted harmonics.    
In Sec. III, we begin with a list of the steps involved in computing the self-force in a modified 
radiation gauge.  In the subsections that follow, we obtain a simple analytic expression for the 
singular part of each of the Weyl scalars $\psi_0$ and $\psi_4$; we relate the small-distance 
behavior of the Weyl scalars to their large $\ell$ behavior, and observe that the singular behavior 
in $\ell$ of the expression for the self-force has the same behavior (involves the same powers of $\ell$) in a radiation gauge as in a Lorenz gauge; and we show that the perturbed metric obtained from $\psi_0$ 
is unique up to gauge transformations and the addition of metric perturbations corresponding to infinitesimal 
changes in mass and spin. We conclude the section by studying the parity of 
the radiation-gauge part of the perturbed metric; in particular, 
we show that spatial part of the metric (in the frame of the particle) 
is even under parity to 
leading order in the geodesic distance $\rho$ to the trajectory.
In Sec. IV, we specialize to a particle in a circular orbit around a 
Schwarzschild black hole, finding $\psi_0^{\rm ret}$ and $\psi_0^{\rm s}$, the retarded and singular 
forms of $\psi_0$ and comparing the analytic and numeric methods of renormalizing $\psi_0$.  
The substantial analytical work involved in the mode-sum expression for the leading and subleading 
terms of $\psi_0^{\rm s}$ is detailed in an appendix.  
Finally, in Sec. V, we briefly discuss the results.  

\noindent
{\em Conventions}

Greek letters early in the alphabet $\alpha, \beta, \ldots$ will be abstract spacetime 
indices; letters $\mu, \nu, \ldots$ will be concrete spacetime indices, labeling components 
in Schwarzschild or Boyer-Lindquist coordinates.  Bold-face Greek indices $\bm\mu,\bm\nu$ will label 
components along the null Newman-Penrose (NP) tetrad defined in Eq.~(\ref{eq:tetrad}) below. 
We adopt the $+---$ signature of Newman and Penrose and use the standard names 
$l^\alpha, n^\alpha, m^\alpha, \bar m^\alpha$ for the null NP tetrad and NP 
notation for the spin coefficients.    

\section{\label{genform} Review of self-force and of black-hole perturbations in an NP framework.} 

\subsection{Gravitational self-force}
We work in linear perturbation theory, for which the metric perturbation is a solution with 
point-particle source to the Einstein field equation linearized about a Kerr or Schwarzschild 
background.  With the particle's mass, trajectory and velocity denoted by $\mathfrak m$, $z(\tau)$ and $u^\alpha$, respectively, the source is given by 
\be
T^{\alpha\beta}(x) 
	= \mathfrak m u^\alpha u^\beta \int  \delta^{4}(x,z(\tau)) d\tau.
\label{db_stress_particle}\\
\ee
Here $u^\alpha = u^\alpha(x)$ when $x$ lies on the particle's trajectory, and the 
$\delta$-function is normalized by $\int \delta^4(x,z)\sqrt{|g|}d^4x = 1$.    

We denote by $\hretab$ the retarded solution to the perturbed Einstein equation with this 
source.  
As noted by Quinn and Wald and by Detweiler and Whiting \cite{DetWhiting}, in the MiSaTaQuWa prescription for finding the self-force \cite{MinoSasaki,quinnwald}, a particle follows a geodesic of 
the metric $g_{\alpha\beta}+\hrenab$, where $\hrenab$ is given by 
\be 
 \hrenab = \hretab - \hsab,
\ee
with $\hsab$ a locally defined singular field, chosen to cancel the 
singular part of $\hretab$ and to give no contribution to the self-force. 
The perturbed geodesic equation has the form
\be 
a^\alpha := u\cdot \nabla u^\alpha  
= -(g^{\alpha\delta}-u^\alpha u^\delta)
	\left(\nabla_\beta h^{\rm ren}_{\gamma\delta}
	-\frac12\nabla_\delta h^{\rm ren}_{\beta\gamma}\right)u^\beta u^\gamma.
\label{geopert}\ee 
Here $a^\alpha$ is the acceleration with respect to the background metric, 
and the self-force is, by definition, $f^\alpha = {\mathfrak m} a^\alpha$.  

We will denote by $a^{\rm ret\,\alpha}$ the expression on the the right side of Eq.~(\ref{geopert}), with 
$h^{\rm ren}_{\alpha\beta}$ replaced by $\hretab$.  Work by Gralla \cite{gralla10},
following a careful justification of the self-force equations by Gralla and Wald \cite{GrallaWald08}, gives the following characterization of $a^\alpha$, based 
on the vanishing of the angle-averaged singular part of the expression for the self-force:\\
Let $\rho$ be geodesic distance to the particle trajectory. Let $\hretab$ be the retarded metric perturbation in a gauge for which its spatial part near the 
trajectory is $O(\rho^{-1})$ and has even parity to that order. 
Then the self-force is given in 
local inertial coordinates about $P$ by
\be
 a^{\rm ren\, \mu} =  \lim_{\rho\rightarrow 0} \int_{S_\rho} a^{\rm ret\,\mu}\, d\Omega
\label{eq:aren}\ee
where $S_\rho$ is a sphere of constant $\rho$ in a geodesic surface orthogonal to the 
worldline.  That is, the first-order perturbative correction to the geodesic
equation is 
\be
u^\beta\nabla_\beta u^\alpha = a^{\rm ren\,\alpha},    
\label{eq:geopert1}\ee
with $a^{\rm ren\,\alpha}$ given by Eq.~(\ref{eq:aren}).\\
 
One can thus identify the singular part of the self-force with any vector field 
$\frak{m} a^{\rm s\,\alpha}$ near the particle trajectory for which 
$a^{\rm ret\,\alpha}-a^{\rm s\,\alpha}$ is continuous and 
\be
  \lim_{\rho\rightarrow 0} \int_{S_\rho} a^{\rm s\,\mu}\, d\Omega=0.
\label{eq:angle-av}\ee
Then 
\be
a^{\rm ren\,\alpha}(P) = \lim_{P'\rightarrow P} \{ a^{\rm ret\,\alpha}(P') - a^{\alpha} [h^s](P')\}.
\ee

A free particle in flat space has no self-force, and the form of its linearized 
field can be used to obtain $a^{\rm s\,\alpha}$ in a curved spacetime.   
Its linearized gravitational field in a Lorenz gauge is the Schwarzschild 
solution in isotropic coordinates, linearized about flat space
\beq
h_{\mu\nu} = -\frac{2\mathfrak m}{\rho}\delta_{\mu\nu}
	= \frac{2\mathfrak m}{\rho}(\eta_{\mu\nu}-2u_\mu u_\nu).
\label{eq:hs0}\eeq 
In a Lorenz gauge for a particle in geodesic motion in a curved background, 
the singular part $\hsab$ of the metric perturbation takes this form in local inertial coordinates $T,X,Y,Z$ centered at at any point $P$ along the trajectory, with ${\bm \partial}_T=\bf u$:
\beq
h_{\mu\nu}^{\rm s, Lor} = -\frac{2\mathfrak m}{\rho}\delta_{\mu\nu}
	= \frac{2\mathfrak m}{\rho}\left.(g_{\mu\nu}-2u_\mu u_\nu)\right|_P.
\label{eq:hs}\eeq 
The corresponding singular part of the self-force is given by 
\be
  f^{\rm s,Lor\,\mu} = -{\frak m}\nabla^\mu \frac1\rho.
\ee
As in flat space, the angle-average of $f^\mu$ over a sphere of constant $\rho$ vanishes 
in the small-$\rho$ limit.
In particular, although the leading correction to the flat-space coordinate expression 
$x^i/\rho^3$ can be a term of order $\rho^0$, the term has the form  
$c_{jk} x^i x^j x^k /\rho^3$; because the term is odd in $x^i$, we have
\be
     \lim_{\rho\rightarrow 0} \int_{S_\rho} \nabla^\mu \frac1\rho\ d\Omega = 0.
\ee 

Because the self-force involves only the values of $\na_\gamma\hrenab$ on the 
particle's trajectory, two tensors $\hsab$ and $\widetilde h^{\rm s}_{\alpha\beta}$ give the 
same self-force if 
$\na_\gamma (\widetilde h^{\rm s}_{\alpha\beta}-\hsab)$ vanishes on the 
particle's trajectory.  In particular, Detweiler and Whiting \cite{DetWhiting} show that 
there is a choice $\hsdw$ of the singular field that is a locally defined solution to the 
perturbed field equations with the same point-particle source.  One can choose local inertial coordinates
(THZ, coordinates, for example) for which the Detweiler-Whiting singular solution differs from the 
form (\ref{eq:hs}) by terms of order $\rho^2$ \cite{detweiler01}.  Following the notation in 
Detweiler-Whiting, we denote their form of the singular field by an upper-case $S$.

\subsection{Mode-sum renormalization}

A review of mode-sum renormalization, is given in Ref.~\cite{barack09}.  We recall some of 
its main features in a Lorenz gauge; many of these continue to hold in our  
(modified) radiation gauge.  

In mode-sum renormalization, the retarded field $\hretab$ and the corresponding expression 
$a^{\rm ret\,\alpha}$ are written as sums over angular harmonics, labeled by $\ell,m$. 
In a Schwarzschild background, these are unambiguously associated with the spherical symmetry 
of the background.  In Kerr, one can use the spherical coordinates of a Boyer-Lindquist chart 
to define the decomposition.  When the retarded field is written as a superposition of angular 
harmonics, its short-distance singular behavior (\ref{eq:hs}) is replaced by a large 
$L$-divergence of the mode sum at the position of the particle. (Appendix \ref{ap:singular} 
relates the large $L$ behavior of an function on a sphere to its singular behavior for small $\rho$.)  

For a particle at coordinate radius $r_0$, the angular harmonics have finite limits as $r\rightarrow r_0$ from $r<r_0$ or $r>r_0$.   Denoting by $h^{\rm ret}_{\mu\nu\ell}$ the sum over $m$ of all 
harmonics belonging to a given $\ell$, one has 
\be
   h^{\rm ret\pm}_{\mu\nu\ell}(P) = \widetilde A_{\mu\nu} + \widetilde B_{\mu\nu}/L + O(L^{-3}).
\label{eq:hretlor}\ee
Similarly, with $a^{\rm ret\, \alpha}_\ell$ the contribution to $a^{\rm ret\, \alpha}$ from 
$h^{\rm ret}_{\mu\nu\ell}$, the large-$L$ behavior of $a^{\rm ret\, \alpha}_\ell$ is given by 
\be
   a^{\rm ret\, \mu}_\ell(P) = A^{\pm\mu} L+ B^{\mu}  + O(L^{- 2}).
\label{eq:aretL}\ee

Explicit expressions for the {\em regularization parameters} $A_{\mu}^\pm$ and $B_{\mu}$ 
have been found for generic orbits in a Kerr background by Barack \cite{barack09}, 
who shows that the first two terms in this expansion reproduce the singular part of the acceleration, 
$\dis -\nabla^\alpha \frac1\rho$, up to terms that vanish at the particle:  
\be
   a^{\rm s\, \mu}_\ell(P) = A^{\pm\mu} L+ B^{\mu}.
\ee
The fact that the term of order $L^{-1}$ vanishes for each component 
$a^{\rm s\, \mu}$ is related to the behavior of 
a short-distance expansion in which terms of order $\rho^k$ with $k$ even occur 
with an odd number of factors of $x^i$.
The retarded acceleration, expressed as a mode-sum that diverges on the particle trajectory, 
is regularized by a cutoff $\ell_{\rm max}$ in $\ell$, 
\be
   a^{\rm reg\, \mu} = \sum_{\ell=0}^{\ell_{\rm max}}  (a^{\rm ret\, \alpha}_\ell-a^{\rm s\,\mu}_{\ell}), 
\ee
and the renormalized acceleration is given by 
\be
   a^{\rm ren\, \mu} = \lim_{\ell_{\rm max}\rightarrow\infty} 
\sum_{\ell=0}^{\ell_{\rm max}} (a^{\rm ret\, \alpha}_\ell-a^{\rm s\,\mu}_{\ell}). 
\label{eq:arenmu}\ee

\subsection{Black-hole perturbations in an NP framework}

    The present method obtains the metric perturbation from components of the Weyl tensor along 
basis vectors of an NP tetrad 
\be
e_{\bf 1}^\alpha := l^\al, \quad e_{\bf 2}^\alpha := n^\al,\quad e_{\bf 3}^\alpha := m^\al, \quad 
e_{\bf 4}^\alpha := \bar m^\al,
\ee
 whose two real null vectors $l^\al$ and $n^\al$ are along the 
principle null directions of the Kerr geometry.  In particular, the Kinnersley tetrad has 
in Boyer-Lindquist coordinates the components 
\beq
(l^\mu) = (\frac{r^2+a^2}\Delta,1,0,\frac a\Delta), \quad 
(n^\mu) = \frac{1}{2({r^2+a^2\cos^2\theta})}(r^2+a^2,-\Delta,0,a), 
\quad (m^\mu) = \frac{1}{\sqrt{2}(r+ia\cos\theta)}(ia\sin\theta,0,1,i/\sin\theta), 
\label{eq:tetrad}\eeq
where $\Delta = r^2-2Mr+a^2$.  
We denote the associated derivative operators by  
\beq
\bm D = l^\mu\partial_\mu, \qquad \bm\Delta = n^\mu\partial_\mu, \qquad \bm\delta = m^\mu\partial_\mu, \qquad\bar{\bm\delta} = \bar{m}^\mu\partial_\mu, 
\eeq
where boldface distinguishes these operators from subsequently defined scalars.  The nonzero spin coefficients are
\begin{eqnarray}
\varrho=-\frac{1}{r-ia\cos\theta},\qquad\beta=-\bar\varrho\frac{\cot\theta}{2\sqrt{2}},\qquad\pi=\frac i {\sqrt{2}} a \varrho^2\sin\theta,\qquad\tau=-\frac i {\sqrt 2} a \varrho\bar\varrho \sin\theta,\nonumber\\
\mu=\frac{1}{2} \varrho^2\bar\varrho\Delta,\qquad\gamma=\mu+\frac 1 2 \varrho\bar\varrho(r-M),\qquad\alpha=\pi-\bar\beta,\qquad\qquad \qquad
\end{eqnarray} \\
where we distinguish $\varrho$ from $\rho$ introduced before Eq.~(\ref{eq:aren}).  The spin-weight $s=\pm 2$ components, $\psi_0$ and 
$\psi_4$, of the perturbed Weyl tensor are given by
\bea
\psi_0 &=& -C_{\alpha\beta\gamma\delta}l^\alpha m^\beta l^\gamma m^\delta, \\
\psi_4 &=& -C_{\alpha\beta\gamma\delta}n^\alpha \bar{m}^\beta n^\gamma \bar{m}^\delta.
\eea
Each satisfies the decoupled Teukolsky equation appropriate to its spin weight:
\bea
{\cal T}_s\psi_s &:=&\left\{\left[\frac{(r^2+a^2)^2}{\Delta} - a^2 \sin^2 \theta \right]\frac{\partial^2}{\partial t^2}  -2s\left[\frac{M(r^2-a^2)}{\Delta}-r-ia\cos\theta\right]\frac{\partial}{\partial t}+ \frac{4M a r}{\Delta}\frac{\partial^2}{\partial t\partial\phi} -\Delta^{-s} \frac{\partial}{\partial r} \left(\Delta^{s+1}\frac{\partial}{\partial r}\right)\right.\nonumber\\
&&\left.-\frac 1 {\sin\theta} \frac{\partial}{\partial\theta}\left(\sin\theta \frac{\partial}{\partial \theta}\right)-2s\left[\frac{a(r-M)}{\Delta}+\frac{i\cos\theta}{\sin^2\theta}\right]\frac{\partial}{\partial \phi}
 + \left[\frac{a^2}{\Delta} - \frac1{\sin^2\theta}\right]\frac{\partial^2}{\partial \phi^2}+(s^2\cot^2\theta-s)\right\}\psi_s\nonumber\\
&=& 4\pi(r^2+a^2\cos^2\theta) T_s,\nonumber \\
\label{eq:teuk}\eea
where
\bsube\bea
\psi_{s=2}&=& \psi_0, \nonumber\\
 T_{s=2} &=&  2(\bm \delta +\bar\pi -\bar \alpha -3\beta-4\tau)[(\bm D-2\epsilon-2\bar\varrho)T_{13}-(\bm \delta +\bar \pi-2\bar \alpha -2\beta)T_{11}]
\nonumber \\
 &&+2(\bm D-3\epsilon+\bar\epsilon -4\varrho -\bar\varrho)[(\bm \delta+2\bar\pi-2\beta)T_{13}-(\bm D-2\epsilon+2\bar\epsilon-\bar\varrho)T_{33}],
\label{eq:psi0source}\\
\nonumber\\
\psi_{s=-2}&=&\varrho^{-4} \psi_4,\nonumber\\
T_{s=-2} &=& 2 \varrho^{-4}(\bm\Delta +3\gamma-\bar\gamma+4\mu+\bar\mu)
 		     [(\bar{\bm \delta}-2\bar\tau+2\alpha)T_{2 4}-(\bm\Delta+2\gamma-2\bar\gamma+\bar\mu)T_{44}]\nonumber \\
		&& +2 \varrho^{-4}(\bar{\bm \delta}-\bar\tau+\bar \beta+3\alpha +4\pi)[(\bm\Delta+2\gamma+2\bar\mu)T_{24}
		   -(\bar{\bm \delta}-\bar\tau+2\bar\beta + 2\alpha)T_{22}].
\eea\esube
 
The scalars $\psi_0^{\rm ret}$ and $\varrho^{-4}\pret$ can be decomposed into time and angular harmonics,
\bsube\bea
   \psi_{0\,\ell m\omega}^{\rm ret} &=& 
		{}_{2}R_{\ell m\omega}~{}_{2}S_{\ell m\omega}e^{i(m\phi-\omega t)},
\label{eq:psi0lmom}\\
     \varrho^{-4}\psi_{4\,\ell m\omega}^{\rm ret} 
  &=& {}_{-2}R_{\ell m\omega}~{}_{-2}S_{\ell m\omega}e^{i(m\phi-\omega t)},
\eea\esube
where the ${}_{s}S_{\ell m\omega}$ are oblate spheroidal harmonics and where 
 ${}_{s}R_{\ell m\omega}$ satisfies the radial equation, 
\be
  \Delta^{-s} \frac d{dr}\left(\Delta^{s+1}\frac{dR}{dr}\right) + 
\left(\frac{K^2-2is(r-M)K}\Delta+4is\omega r-\lambda\right)R = -4\pi T_{s\ell m\omega},
\label{radialt}\ee
with $K=(r^2+a^2)\omega - ma$.  The source, a distribution involving $\delta(r-r_0)$ and its first two derivatives, 
is obtained from the source term in Eq.~(\ref{eq:teuk}) using the completeness and orthogonality of the 
spin-weighted spheroidal harmonics.  
 
Each angular eigenvalue $\lambda$ is a continuous function of $a$.  For a Schwarzschild background, $\lambda$ 
has the value $(\ell-s)(\ell+s+1)$, and for Kerr it is labeled by its value of $\ell$ at $a=0$.  

For perturbations of Schwarzschild, tensor components with $s_1$ indices along $m^\alpha$ and $s_2$ indices along $\bar m^\alpha$ have angular behavior given by spin-weighted spherical harmonics 
${}_sY_{\ell m}(\theta,\phi)$ with spin-weight $s=s_1-s_2$, where
\be
{}_sY_{\ell m}= \left\{\begin{array}{ll}
              \left[ (\ell-s)!/(\ell+s)! \right]^{1/2}\eth^s Y_{\ell m}, &\ \ 0\le s\le \ell,
              \\
              (-1)^s\left[(\ell+s)!/(\ell-s)!\right]^{1/2}\bar{\eth}^{-s} Y_{\ell m},&
	 -\ell\le s\le 0,
	\end{array}\right.
 \label{green_sylm_a}\ee
with 
\begin{eqnarray}
\eth\eta &=& -\left(\partial_\theta+i\csc\theta\partial_\phi-s\cot\theta\right)\eta, 
\label{green_eth_b}
\nonumber\\
\bar{\eth}\eta&=&-\left(\partial_\theta-i\csc\theta\partial_\phi+s\cot\theta\right)\eta. 
\label{green_eth_bar_b}
\end{eqnarray}

\subsection{Reconstruction of the metric perturbation from $\psi_0$ or $\psi_4$} 
The CCK procedure for obtaining metric perturbations from perturbed Weyl scalars was developed by Chrzanowski and by Cohen and Kegeles \cite{chr75,ck74,kc79} (see also 
Stewart~\cite{stewart79}), and a simpler derivation is given in Wald \cite{wald78}.  
Discussions in the context of the self-force problem are found in \cite{ori03,lw02} 
and \cite{kfw}.

The procedure gives the metric perturbation in a radiation gauge, a gauge characterized by the conditions 
\beq
l^\beta h_{\alpha\beta} = g^{\alpha\beta}h_{\alpha\beta} = 0,
\label{eq:lrg}
\eeq
or by the corresponding conditions 
\beq
n^\beta h_{\alpha\beta} = g^{\alpha\beta}h_{\alpha\beta} = 0.
\label{eq:nrg}
\eeq
Price, Shankar and Whiting \cite{wp05,psw07} show that a radiation gauge exists locally for vacuum 
perturbations of any type D vacuum spacetime. 

The CCK construction has two parts.  Given a solution $\psi_0$ or $\psi_4$ to the source-free $s=\pm 2$ Teukolsky equation, 
one first finds a {\em Hertz potential}, a function $\Psi$ that again satisfies a source-free Teukolsky equation. 
Then, by taking derivatives of the Hertz potential, one constructs a metric perturbation for which $\psi_0$ and $\psi_4$ are 
the perturbed Weyl scalars.  

There is a striking difference between the asymptotic behavior produced by the CCK procedure and the asymptotic behavior of a metric perturbation in a Lorenz gauge that approximately satisfies Eq.~(\ref{eq:lrg}) or (\ref{eq:nrg}). 
The difference is related to 
the terminology used in the literature for the two families of radiation gauges, in which the gauge satisfying 
$l^\beta h_{\alpha\beta}=0$ is called the IRG or {\em ingoing radiation gauge} and the gauge satisfying 
$n^\beta h_{\alpha\beta}=0$ is called the ORG or {\em outgoing radiation gauge}.    
  
Outgoing solutions in a Lorenz gauge (for example modes for which the metric perturbation has asymptotic behavior 
$e^{-i\omega u}/r + O(r^2)$), however, satisfy the IRG conditions (\ref{eq:lrg}) to leading order: 
Because $l^\alpha$ is along $\nabla^\alpha u$, we have 
\be
0 = \nabla^\beta h_{\alpha\beta} = l^\beta h_{\alpha\beta} + O(r^{-2}).
\ee 
As one might expect, an asymptotically vanishing gauge vector can take one from a metric perturbation in a Lorenz gauge in which the IRG condition is approximately satisfied to a metric perturbation that exactly 
satisfies the condition:   
We exhibit in Appendix \ref{ap:gauge} an explicit, asymptotically vanishing, gauge vector from a generic outgoing
solution $h^{\rm Lor}_{\alpha\beta}$  in a Lorenz gauge to an asymptotically flat metric perturbation in an IRG.  
An analogous gauge transformation
leads for incoming radiation to an asymptotically flat metric perturbation satisfying $n^\beta h_{\alpha\beta}= 0$. 

Curiously, however, the CCK procedure yields metric perturbations in the two gauges that 
are asymptotically flat for the {\em opposite} cases:  In the IRG, with 
$l^\beta h_{\alpha\beta} =0$, the CCK procedure yields an asymptotically flat metric for {\em ingoing} radiation.  
For outgoing radiation the CCK procedure gives a metric whose dominant components are asymptotically of order $r$.  
Similarly, in the ORG gauge, with $n^\beta h_{\alpha\beta} =0$, the CCK procedure yields an asymptotically flat metric 
for {\em outgoing} radiation. For ingoing radiation the CCK procedure gives a metric whose dominant components are asymptotically of order $r$. This then is the justification for the terms ``outgoing'' and ``ingoing radiation gauge'' 
introduced by Chrzanowski and used in the subsequent literature.

For the gauge (\ref{eq:nrg}), a Hertz potential $\Psi$ is related to $\psi_0$  by four angular derivatives or four 
 radial derivatives, and both of these alternatives are listed below.  In subsequent 
sections, we will be concerned only with the specialization of these results to the Schwarzschild spacetime.  
In this case given a $\Psi$ that satisfies the Teukolsky equation for $\psi_0$, a metric perturbation 
in the radiation gauge ORG is given by 
\bea
h_{\alpha\beta}&=&\varrho^{-4}\{\, n_\alpha n_\beta (\bar{\bm\delta}-3\alpha-\bar{\beta} +5\pi)(\bar{\bm\delta}-4\alpha+\pi)+\bar{m}_\alpha\bar{m}_\beta(\bm\Delta+5\mu-3\gamma + \bar \gamma)(\bm\Delta+\mu-4\gamma) \nonumber\\
&\pheq& - n_{(\alpha}\bar{m}_{\beta)} \left[(\bar{\bm \delta} - 3\alpha +\bar \beta +5\pi + \bar \tau) (\bm \Delta + \mu - 4\gamma)+
(\bm \Delta + 5\mu -\bar \mu -3\gamma -\bar \gamma)(\bar{\bm \delta} -4\alpha + \pi) \right]\, \} \Psi+\rm{c.c.},
\label{mpschw}
\eea
which we take as the starting point for the discussion that follows.
Note that the sign in this equation is appropriate for a $+---$ signature and is opposite to that in, 
for example, Wald \cite{wald78}.  

In the ORG, $\Psi$ is related to $\psi_0$ through four angular derivatives according to
\beq
\psi_0 = \frac18[{\cal L}^4\bar{\Psi} + 12M\partial_t\Psi],
\label{eq:hertzan}\eeq
where ${\cal L} = \eth - ia\sin\theta\partial_t$.  
Equivalently, ${\cal L}^4 = {\cal L}_{1}{\cal L}_0{\cal L}_{-1}{\cal L}_{-2}$, with 
${\cal L}_s = -[\partial_\theta - s\cot\theta + i\csc\theta\partial_\phi] - ia\sin\theta\partial_t$.
There is a corresponding equation involving four radial derivatives, namely%
\footnote{$\widetilde{\bf D}$ is Chandrasekhar's $\mathscr D_0^\dagger$ and Ori's $D^\dagger$.  Lousto 
and Whiting's Eq. (28) is an incorrect version of Eq.~(\ref{eq:hertzrn}), with 
$\bm \Delta$ ($\widehat\Delta$ in their notation) instead of $\widetilde{\bf D}/2$. This is corrected 
by Whiting and Price \cite{wp05}, in which $\widehat \Delta$ is defined as the GHP prime 
\cite{ghp} of $\bm D$.  The Ori and Lousto-Whiting papers also have an incorrect factor of 
two in each of the equations for $\Psi$ that is inherited from an error in Kegeles-Cohen \cite{kc79}.
} 
\be
  \varrho^{-4}\psi_4 = \frac1{32} \Delta^2 \widetilde{\bm D}^4 \Delta^2 \bar\Psi,
\label{eq:hertzrn}\ee
where $\widetilde{\bm D}$ is proportional to $\bm \Delta$, renormalized to make it the radial derivative 
along the ingoing principal null geodesics (the $t\rightarrow-t, \phi\rightarrow -\phi$ version of $\bm D$):  
\be
\widetilde{\bm D}:=-\frac{2\Delta}{r^2+a^2\cos^2\theta}\bm\Delta = -\frac{r^2+a^2}\Delta\partial_t + \partial_r - \frac a\Delta\partial_\phi.
\ee

The corresponding equations for the (different) Hertz potential in the IRG are listed in the second line of
the summary table below, with $\widetilde {\cal L}:=\bar\eth + ia\sin\theta\partial_t$.

\begin{table}[h]
  \begin{tabular}{|c|c|c|c|}
    \hline
     \multicolumn{1}{|p{1cm}|}{\centering Gauge}
	& \multicolumn{1}{|p{3cm}|}{\centering Gauge conditions}
        & \multicolumn{1}{|p{3.5cm}|}{\centering Radial equation}
        & \multicolumn{1}{|p{4.5cm}|}{\centering Angular equation}\\
    \hline
  	ORG   &{\centering $n^\beta~h_{\alpha\beta}=0,~h=0$}
		    &	$\dis\varrho^{-4}\psi_4 = \frac1{32} \Delta^2 \widetilde{\bm D}^4 \Delta^2 \bar\Psi	$ 
	            &	$\dis\psi_0 = \frac18[{\cal L}^4\bar{\Psi} + 12M\partial_t\Psi]$ \\
     \hline
  	IRG  &{\centering $l^\beta~h_{\alpha\beta}=0,~h=0$}
		    &	$\dis\psi_{0}=\frac12 \bm D^4\bar\Psi$ 
	            &	$\dis\varrho^{-4}\psi_4 
		= \frac18[\widetilde{\cal L}^4\bar{\Psi} - 12M\partial_t\Psi]$ \\  \hline
 \end{tabular}

\label{table:psi}
\caption{Relations between the gauge-invariant Weyl scalars and the Hertz potentials in the 
two radiation gauges.}
\end{table}
The fact that $\varrho^{-4}\psi_4$ and $\psi_0$ satisfy the vacuum Teukolsky equation for spin-weights $\mp 2$ 
when $\Psi^{\rm ORG}$ or $\Psi^{\rm IRG}$ satisfy the spin-weight $\pm 2$ Teukolsky equations follows 
from the relations in the table together with the commutators
\bea
  {\cal T}_2 {\bm D}^4 &=&  {\bm D}^4 \bar {\cal T}_{-2}, \qquad\qquad {\cal T}_{-2} \tilde{\cal L}^4 = \tilde{\cal L}^4  \bar {\cal T}_{-2}, \nonumber\\
  {\cal T}_{-2} \Delta^2\tilde{\bm D}^4\Delta^2 &=&  \Delta^2\tilde{\bm D}^4\Delta^2 \bar {\cal T}_2, \qquad {\cal T}_2 {\cal L}^4 = {\cal L}^4  \bar {\cal T}_2.
\eea
These are equivalent to Eqs. (40) and (56) in Chap. 8 of Chandrasekhar \cite{Chandra} 
and their adjoint relations as defined there.       

   Explicit solutions to the equations for the Hertz potentials have been presented for a Schwarzschild background by 
Lousto and Whiting \cite{lw02} and for Kerr by Ori \cite{ori03}. Ori shows that the CCK procedure 
gives a unique Hertz potential in each gauge for each angular harmonic of $\Psi$.  That is, 
for each harmonic, there is a unique $\Psi$ that satisfies both the angular equation and the 
sourcefree Teukolsky equation; there is a unique $\Psi$ that satisfies both the radial equation
and the sourcefree Teukolsky equation; and the two solutions coincide. 
For $\psi_0$ proportional to the 
harmonic $_2S_{
\ell m\omega}$, $\Psi^{\rm ORG}$ is proportional to $_{-2} S_{\ell,-m,-\omega}$.

Note that, although Ori's metric reconstruction is done mode-by-mode, his statement of 
uniqueness of solutions does not explicitly restrict it to uniqueness of each angular harmonic.  
This is, however, a necessary restriction: 
As shown in Keidl et al. \cite{kfw}, if one requires only that $\Psi$ satisfy one of 
the radial or angular equations of Table \ref{table:psi}, together with the 
appropriate Teukolsky equation, additional freedom remains.  This freedom in the 
Hertz potential corresponds to the addition to the metric perturbation of type D solutions 
(changes of mass and spin and addition of a perturbed C-metric solution) and with gauge transformations of the perturbed metric. 

With $\psi_0$ and the ORG $\Psi$ decomposed in time and angular harmonics, Eq.~(\ref{eq:hertzan}) can be inverted algebraically as follows, at any $r$ outside the particle's orbit -- for $r_{\rm min}< r < r_{\rm max}$, 
where $r_{\rm min}$ and $r_{\rm max}$ are the perihelion and aphelion values of $r$.  
The harmonics of $\Psi$ and $\psi_0$ each have the form
\bsube\bea
\psi_{0\ell m\omega} &=& {}_{2}R_{\ell m\omega}~{}_{2}S_{\ell m\omega}e^{i(m\phi-\omega t)},
\label{eq:psi0lmo}\\
\Psi_{\ell m\omega} &=& {}_{2}\tilde{R}_{\ell m\omega}~{}_{2}S_{\ell m\omega}e^{i(m\phi-\omega t)},
\label{eq:hertzsoln1}\eea\esube
where $R$ and $S$ are solutions to the radial and angular Teukolsky equations, respectively, and $\tilde{R}$ is to be determined.  Using the identity 
${}_s\bar{S}_{\ell m\omega}=(-1)^{m+s}{}_{-s}S_{\ell\,-\! m\,-\!\omega}$, we can write the 
harmonic decomposition of $\bar\Psi$ in the form 
\bea
\bar\Psi &=& \sum_{\ell,m,\omega}\overline{{}_{2}\tilde{R}_{\ell m\omega}~{}_{2}S_{\ell m\omega}e^{i(m\phi-\omega t)}}\nonumber\\
\nonumber\\
&=& \sum_{\ell,m,\omega}(-1)^m{}_{2}\bar{\tilde{R}}_{\ell \! -m \! -\omega}~{}_{-2}S_{\ell m\omega}e^{i(m\phi-\omega t)}.
\eea 
The Teukolsky-Starobinsky identity (Eqs.~(9.59) and (9.61) of Ref.~\cite{Chandra}) has the form
\beq
{\cal L}^4~{}_2S_{\ell m\omega} = D~{}_{-2}S_{\ell m\omega},
\eeq
where $D^2=\lambda_{CH}^2(\lambda_{CH}+2)^2+8a\omega(m-a\omega)\lambda_{CH}(5\lambda_{CH}+6) + 48a^2\omega^2[2\lambda_{CH}+3(m-a\omega)^2]$ and $\lambda_{CH}$, the angular eigenvalue used by Chandrasekhar \cite{Chandra},  
is related to the separation constant $\lambda$ of Eq.~(4.9) of \cite{1973ApJ...185..635T} by $\lambda_{CH}=\lambda+s+2$.   
Because Eq.~(\ref{eq:hertzan}) mixes $\Psi$ and $\bar\Psi$, its inversion for each angular harmonic involves a linear 
combination of $\psi_{0\ell m\omega}$ and $\bar\psi_{0\ell\, -m\,-\omega}$.  We find that the 
algebraic inversion gives the ORG Hertz potential in the form
\be
  \Psi_{\ell m\omega} 
= 8\frac{(-1)^m D\bar\psi_{0\ell\,-\! m\,-\!\omega}+12iM\omega \psi_{0\ell m \omega}}{D^2+144 M^2\omega^2}.
\label{eq:hertza1}\ee
We use this inversion for circular orbits in a Schwarzschild and Kerr background.

For generic orbits, the individual harmonics $\psi_{0\ell m\omega}$ do not satisfy the sourcefree Teukolsky 
equation in the region $r_{\rm min} < r < r_{\rm max}$, where $r_{\rm min}$ and $r_{\rm max}$ are the values of $r$ at 
perihelion and aphelion; and the presence of a source invalidates the algebraic angular inversion.  
To find $\Psi$ requires one to integrate one of the radial equations of Table I. 
With the Kinnersley tetrad, the IRG radial equation for $\Psi$ has the simplest form:   
In Kerr coordinates $u,r, \theta, \tilde\phi$, 
where $u=t-r^*$, with  
\bea
\frac{dr^*}{dr}=\frac{r^2+a^2}{\Delta},  \qquad\textrm{and}\qquad
\tilde\phi = \phi + a \int_r^\infty dr' \frac{1}{\Delta(r')},
\label{eq:kerrcoords}\eea 
we have ${\bm D}f(u,r,\theta,\tilde\phi) = \partial_r f(u,r,\theta,\tilde\phi)$.  The radial equation for 
$\Psi^{\rm IRG}$ is then $\partial_r^4\bar\Psi^{\rm IRG} = 2\psi_0$, with solution 
\be
  \bar\Psi^{\rm IRG} = 2\int_r^\infty dr_4\int_{r_4}^\infty  dr_3\int_{r_3}^\infty  
		dr_2\int_{r_2}^\infty  dr_1 \psi_0(u,r,\theta,\tilde\phi),
\label{eq:hertzr1}\ee   
satisfying the vacuum Teukolsky equation for ingoing radiation when the outgoing radial null ray does 
not intersect the particle. To find $\Psi$ at points on a $t=$ constant surface, one can use 
Eq.~(\ref{eq:hertzr1}) outside the radial coordinate $r_0$ of the particle and a corresponding 
integral from the horizon to $r$ for $r<r_0$.

\subsection{Gauge transformations of the self-force}

  In the form (\ref{geopert}), the perturbed geodesic is parameterized so that its tangent 
is normalized to $1$ with respect to the background metric.  A gauge transformation of this equation 
was obtained by Barack and Ori \cite{bo01}, and Appendix \ref{ap:gauge} gives an alternative, covariant 
derivation in terms of an infinitesimal diffeo of the metric and a family of unperturbed geodesics.  With the same 
normalization, changing a background geodesic by a gauge transformation generated by $\xi^\alpha$ 
changes its tangent vector by    
\be
	\delta_{\bm \xi} u^\alpha = (\delta^\alpha_\beta -u^\alpha u_\beta) \Lie_{\bm\xi} u^\beta; 
\label{eq:gaugeu1}\ee
and $u^\alpha + \delta_{\bm \xi} u^\alpha$ satisfies a geodesic equation of the metric perturbed by 
$\Lie_{\bm\xi} g_{\alpha\beta}$.  With the acceleration defined by  
$\delta_{\bm \xi} a^\alpha:= \delta_{\bm \xi} u^\beta \nabla_\beta u^\alpha +u ^\beta\nabla_\beta \delta_{\bm \xi} u^\alpha$, we have
\be
  \delta_{\bm \xi} a^\alpha = -(\delta^\alpha_\beta -u^\alpha u_\beta )({\bf u\cdot\bm\nabla})^2\xi^\beta + R^\alpha{}_{\beta\gamma\delta}u^\beta u^\gamma \xi^\delta,
\label{eq:gaugea1}\ee
and $a^\alpha u_\alpha = 0$.  Note that the right side vanishes if $\xi^\alpha$ happens to drag a geodesic of 
the background spacetime to another geodesic of the background spacetime: This is the equation of geodesic 
deviation governing the connecting vector joining two neighboring geodesics of $g_{\alpha\beta}$. 
For general $\xi^\alpha$, the right side of Eq.~(\ref{eq:gaugea1}) then measures the failure of an 
infinitesimal diffeo generated by $\xi^\alpha$ to produce a geodesic of the background metric. 

A particle in circular orbit has 4-velocity $u^\alpha = u^t k^\alpha$, with $k^\alpha=t^\alpha+\Omega \phi^\alpha$ 
a Killing vector.  The perturbed spacetime with a particle in circular orbit is helically symmetric, symmetric 
with respect to $k^\alpha$.  We show in Appendix \ref{ap:gauge} that, for a gauge transformation that preserves 
helical symmetry (for $\Lie_k\xi^\alpha=0$), the gauge-transformed self-force is given by 

\be
  \widehat f^\alpha = f^\alpha + \frac12{\mathfrak m}(u^t)^2 \xi^\beta\na_\beta \nabla^\alpha (k^\gamma k_\gamma).    
\ee

\section{Method for computing the self-force on an orbiting mass}
\label{s:method}

\subsection{Overview}

In this section we outline the method for computing the self-force in a radiation gauge.
Subsequent sections will elaborate on the details of each step of the calculation.  The 
method is a revision of that initially suggested in Refs.~\cite{kfw} and \cite{keidl07}.
In broad terms, the steps involved in computing the self-force in a modified 
radiation gauge are:
\begin{enumerate}
\item[A.] Compute the retarded Weyl scalar, $\psi_0^{\rm ret}$ (or $\psi_4^{\rm ret}$).    
\item[B.] Use the retarded Weyl scalar to construct a Hertz potential $\Psi^{\rm ret}$ 
as a sum of angular harmonics for $r\neq r_0$.  
\item[C.] Using the CCK formalism described above, reconstruct the retarded metric perturbation in a radiation gauge, and find the perturbation in mass and angular 
momentum in an arbitrary gauge.  
\item[D.] Find the expression for the self force as a mode sum involving the retarded 
field. 
\item[E.] Find the regularization parameters for the singular part $f^{\rm s\,\alpha}$ of 
the self-force and compute  $f^{\rm ren\,\alpha}$. 
\end{enumerate}

In this approach, all fields are written as a sum of time and angular harmonics.  
The angular harmonics (\ref{eq:psi0lmom}) of $\preto$ have finite one-sided limits as 
$r\rightarrow r_0$: 
\be
     \psi_{0\ell m\omega}^{{\rm ret}\pm} := \lim_{r\rightarrow r_0^\pm} \psi^{\rm ret}_{0\ell m\omega}(t_0,r).
\ee  
For a Schwarzschild background the harmonics are 
spin-weight 2 spherical harmonics, while for Kerr they are oblate spheroidal 
harmonics whose form depends on the Kerr parameter $a$.  
For a Kerr as well as a Schwarzschild background, however,
one can write the singular field in terms of spin-weighted {\em spherical} harmonics \cite{hughes00,dolan06,wb10}, 
where each spin-weighted spheroidal harmonic is a sum of the form
\be
 _sS_{\ell m\omega}e^{im\phi} 
  = \sum_{\ell' = \ell_{\rm min}}^\infty b_{\ell'\omega}\ {}_sY_{\ell' m}, \qquad \ell_{\rm min} = \max(|s|,|m|).
\label{eq:spw2sph}\ee
The computation of $\psi^{\rm ret}_0$ is straightforward, involving an integration of the 
radial Teukolsky equation (\ref{radialt}) and the computation of spin-weighted spherical harmonics. 
In computing a Hertz potential $\Psi^{\rm ret}$ from $\preto$, 
our choice of radiation gauge is dictated by requiring that the CCK-constructed $\Psi^{\rm ret}$ vanish 
asymptotically:  We use the ORG and find $\Psi^{\rm ret}$ as the solution of Eq.~(\ref{eq:hertza1}) 
to the angular equation.  The tetrad components 
of the metric perturbation $h^{\rm ret}_{\alpha\beta}$ are then obtained from Eq.~(\ref{mpschw}) and 
are used to compute the expression for the bare self-force in terms of $h^{\rm ret}_{\alpha\beta}$, 
\be 
f^\alpha[h^{\rm ret}]/{\frak m} = a^{\rm ret\,\alpha} 
 = -(g^{\alpha\delta}-u^\alpha u^\delta)
	\left(\nabla_\beta h^{\rm ret}_{\gamma\delta}
	-\frac12\nabla_\delta h^{\rm ret}_{\beta\gamma}\right)u^\beta u^\gamma.
\label{eq:fret}\ee 

The remainder of the problem is the mode-sum renormalization of the self-force and the recovery of the 
part of the metric perturbation that does not arise from $\psi_0$.  We discuss them in turn in the 
next two subsections.  

\subsection{Mode-sum renormalization in a radiation gauge}
\label{sec:renormalization}

  One can carry out a mode sum renormalization by finding the radiation-gauge version of the 
power series (\ref{eq:aretL}) that expresses the singular behavior of  $a^{\rm\, ret}$ for 
$r>r_0$ and $r< r_0$ near the position of the particle.     
We consider two ways to proceed:\\
1. One can find an analytic expression for 
$a^{\rm s\,\alpha}$ as a sum in powers of $L$, starting from an expression
we derive below for $\psi_0^{\rm s}$, the singular part of $\psi_0^{\rm ret}$.
This analytic way follows the steps just listed in describing the path from 
$\preto$ to $a^{\rm ret\,\alpha}$, to successively obtain expressions for $\Psi^{\rm s}$, 
$h^{\rm s}_{\alpha\beta}$ and $a^{\rm s\,\alpha}$. \\ 
2.  The second method is significantly simpler: One simply numerically matches a power series in $L$ 
to $a^{\rm ret,\bm\mu}_{\ell}(P)$ for successive values of $\ell$.  
\\

There is an ambiguity in defining the singular part of the perturbed metric and 
a corresponding ambiguity in the singular part of the self-force.  Because  
$\psi_0^{\rm ret}$ is gauge invariant, its singular field 
$\psi_0^{\rm s}$ is uniquely defined in a neighborhood of the particle's 
trajectory as the perturbed Weyl scalar associated with the Detweiler-Whiting 
singular field.  In reconstructing the metric from $\psi_0^{\rm s}$, however,
there are two free functions that arise from the integrations that yield 
the corresponding Hertz potential, $\Psi^{\rm s}$. Because $\psi_0^{\rm s}$ 
is defined in a finite neighborhood, the integrals of Eq.~(\ref{eq:hertzr1}) 
are replaced by integrals in which each upper limit is a free function 
$R_n(u,\theta,\phi)$ (or, for each angular and temporal harmonic
by definite integrals whose upper limit is a free constant). 
Two of these free functions are determined by the requirement that 
$\Psi^{\rm s}$ satisfy the vacuum Teukolsky equation, and the remaining 
two constitute the ambiguity in $\Psi^{\rm s}$. (Similarly, because $\psi_0^{\rm s}$ 
is not defined on entire spheres of coordinate radius $r$, its angular 
harmonics do not determine unique angular harmonics of $\Psi^{\rm s}$.
The angular integration is again local and again gives two free 
functions.  The ambiguity in $\Psi^{\rm s}$ implies a corresponding 
ambiguity in the singular parts of the perturbed metric and the self-force.
It is present because the IRG (or ORG) gauge conditions 
do not uniquely determine a {\it local} metric perturbation, and 
Price {\rm et al.} describe the remaining gauge freedom \cite{psw}.  
\\

We begin with a discussion of the analytic method.  We find an explicit expression for $\psi_0^{\rm s}$
to subleading order in the distance to the particle's trajectory  
in terms of components of the tetrad vectors along and orthogonal to the trajectory.  
We then characterize the powers of $L$ that appear in the power series for $a^{\rm ret s\, \alpha}$.  
Finally, we turn to a description of renormalization by numerical matching.  
In Sec. IV below, we test the numerical method in a relatively simple case 
by comparing analytic and numerical renormalizations of $\psi_0$ for a particle 
in circular orbit in a Schwarzschild background.  

Note that, in a mode-sum renormalization, the singular parts of the the perturbation in the 
metric and the self-force are determined by the large-$L$ behavior of the retarded fields.  
They are, in particular, independent of the choice of gauge in which one describes 
perturbations of mass and angular momentum.

\subsubsection{Analytic method}

   The analytic method begins by finding an analytic expression for $\psi_0^{\rm s}$ or $\psi_4^{\rm s}$.
The decomposition of the metric perturbation $h_{\alpha\beta}^{\rm ret}$ in a Lorenz gauge,
\beq
h_{\alpha\beta}^{\rm ret,Lor} = h_{\alpha\beta}^{\rm ren,Lor} + h_{\alpha\beta}^{\rm s,Lor},
\label{detwhitdecomp}
\eeq
gives a corresponding gauge-invariant decomposition of the perturbed Weyl scalars. 
From the expression for the perturbed Weyl (or Riemann) tensor in terms of the perturbed metric, we have 
\bsube\bea
\psi_0^{\rm ret} &= {\cal O}_{0}^{\alpha\beta}h_{\alpha\beta}^{\rm ret}, \qquad 
\psi_0^{\rm s} &= {\cal O}_{0}^{\alpha\beta}h_{\alpha\beta}^{\rm s}, 
\label{eq:psi0}\\
\psi_4^{\rm ret} &= {\cal O}_{4}^{\alpha\beta}h_{\alpha\beta}^{\rm ret},\qquad
\psi_4^{\rm s} &= {\cal O}_{4}^{\alpha\beta}h_{\alpha\beta}^{\rm s},
\label{eq:psi4}\eea\esube
where 
\bsube
\bea
\mathcal O^{\alpha \beta}_0&=&\frac 1 2 \left( m^\alpha m^\beta l^\gamma l^\delta 
+ l^\alpha l^\beta m^\gamma m^\delta - l^\alpha m^\beta m^\gamma l^\delta 
- m^\alpha l^\beta l^\gamma m^\delta \right)\nabla_\gamma\nabla_\delta,\\
\mathcal O^{\alpha \beta}_4&=&\frac 1 2 \left( \bar m^\alpha \bar
m^\beta n^\gamma n^\delta + n^\alpha n^\beta \bar m^\gamma \bar m^\delta
- n^\alpha \bar m^\beta \bar m^\gamma n^\delta - \bar m^\alpha n^\beta n^\gamma \bar
m^\delta \right)\nabla_\gamma\nabla_\delta.
\eea\esube

Although we will not need the Detweiler-Whiting form of the singular field, 
it is worth noting that using it would yield a smooth version of $\psi^{\rm ren}_0$ 
(or $\psi^{\rm ren}_4$) that satisfies the sourcefree Teukolsky equation in a 
neighborhood of the particle's trajectory.
That is, because $h^{\rm S}_{\alpha\beta}$ satisfies the perturbed field equation
with the same source as $\hretab$, the corresponding field $\psi_0^{\rm S}$ is a 
local solution to the $s=2$ Teukolsky equation with the same source as $\preto$, 
implying that $\preto-\psi_0^{\rm S}$ satisfies the corresponding homogeneous equation. 
\\
 
Denote by 
$\epsilon :=\sqrt{T^2+X^2+Y^2+Z^2}$ the distance with respect to the Euclidean metric 
$dT^2+dX^2+dY^2+dZ^2$ of the local inertial coordinates introduced before Eq.~(\ref{eq:hs}).
We will argue below that the self-force in our modified radiation gauge, like that in a 
Lorenz gauge, has dominant singular behavior of order $\epsilon^{-2}$ and can be regularized 
by subtracting leading and subleading terms in $\epsilon$.  These arise from the leading and subleading 
terms in $\epsilon$ in $\psi_0^{\rm s}$.   Instead of directly using Eq.~(\ref{eq:psi0}) to 
compute $\pso$, it is easier simply to observe that 
${\cal O}_0^{\mu\nu}$ and $h^{\rm s}_{\mu\nu}$ have to subleading order in $\epsilon$ 
their flat space form.  It follows that the value of $\psi^{\rm s}_0$ has at subleading order in 
$\epsilon$ its form for a perturbation of flat space.  That is, in terms of $T$ and 
$\rho$, $\pso$ is the linearized Schwarzschild field of a static particle:
\beq
C_{\alpha\beta}^{{\rm s}\quad \gamma\delta} = 
   -\frac{4\mathfrak m}{\rho^3}\left( \delta_{[\alpha}^{[\gamma}\delta_{\beta]}^{\delta]} 
   + 3\delta_{[\alpha}^{[\gamma}\nabla_{\beta]}\rho \nabla^{\delta]}\rho 
   - 3\delta_{[\alpha}^{[\gamma}\nabla_{\beta]}T\,\nabla^{\delta]}T 
   - 6\nabla_{[\alpha}\rho\,\nabla_{\beta]}T\,\nabla^{[\gamma}\rho\, \nabla^{\delta]}T \right).
\eeq
From this we obtain
\bea
\psi_0^{\rm s} &=&  -\frac{6\mathfrak m}{\rho^3}\left(l^T m^\rho - l^\rho m^T \right)^2, 
\label{eq:ps0}\\
\psi_4^{\rm s} &=&  -\frac{6\mathfrak m}{\rho^3}\left(n^T \bar{m}^\rho - n^\rho \bar{m}^T \right)^2,
\label{eq:ps4}\eea
where $l^T := l^\alpha\na_\alpha T$, $l^\rho := l^\alpha \nabla_\alpha \rho$, 
$m^T := m^\alpha\na_\alpha T$, $m^\rho := m^\alpha \nabla_\alpha \rho$.
One can, of course, also obtain (\ref{eq:ps0}) and (\ref{eq:ps4}) by applying the operators 
${\cal O}^{\alpha\beta}_0$ and ${\cal O}^{\alpha\beta}_0$ of Eqs.~(\ref{eq:psi0}) and 
(\ref{eq:psi4}) to the singular metric (\ref{eq:hs}).
Eqs.~(\ref{eq:ps0}) and (\ref{eq:ps4}) are valid in general Petrov type D spacetimes.

For each of the quantities $\psi_0, \Psi, h_{\alpha\beta}$, and $a^\alpha$, we will use the subscript 
$\ell$ to denote the sum over $m$ of the contributions from angular harmonics associated with $\ell,m$,
and we suppress the index $\omega$.  The equations that relate $\psi_0$ and $\Psi$ do not mix 
spin-weighted spheroidal harmonics, and $\Psi^{\rm s}$ is most simply found as a sum of these harmonics.
The equations, (\ref{mpschw}) and (\ref{geopert}), that relate $\Psi$ to $h^{\rm s}_{\alpha\beta}$ and 
$h^{\rm s}_{\alpha\beta}$ to $a^\alpha$, however, mix spheroidal harmonics on a Kerr background. 
The large-$L$ expansions of their components can be found in terms of spin-weighted spheroidal 
harmonics, spin-weighted spherical harmonics, or spherical harmonics. 
Different choices lead to different definitions of the subleading contributions, because of the 
mixing of different values of $\ell$ in relating, for example, spin-weighted spherical harmonics to spin-weighted spheroidal harmonics.  What matters is only that the same convention is used for 
the angular harmonics of the retarded and singular fields.  

For a particle in circular orbit in a Schwarzschild background, we obtain 
a stronger result: We find that the leading, sub-leading,and sub-sub-leading contributions, $AL + B + C/L$,
to the singular part of $a^{\textrm{ret}\,r}$ are independent of whether 
$a^{\textrm{ret}\,r}$ is written as a sum over mixed spin-weighted spherical harmonics or as a sum over ordinary scalar spherical harmonics. The same applies to the leading and sub-leading contributions (usually presented as $B \, + \, C/L$ in literature) to the singular field of $h^\textrm{ret}_{ab} u^a u^b$.

Because $\pso$ involves two derivatives of $\hsab$ and can be computed from the Lorenz singular field, 
its large-$L$ behavior is greater by two powers of $L$
than the large-$L$ behavior at $r=r_0$ of $h^{\rm s, Lor}_{\alpha\beta}$ of Eq.~(\ref{eq:hretlor}):  
\be
   \psi^{\rm s\pm}_{0\ell} := \sum_m \psi^{\rm s\pm}_{0\ell m}(t_0)\, {}_2\!Y_{\ell m}(\theta_0,\phi_0) 
	= \hat A^{\pm} L^2 +\hat B^\pm L + O(L^{-1}).
\label{eq:psicoef}\ee
In Sec. IV and Appendix \ref{appsingfield}, we find the large-$L$ expansion of $\psi_0$ for 
a particle in circular orbit in a Schwarzschild background (restricting consideration to the 
part of $\psi_0$ axisymmetric about the position of the particle).

Because the Hertz potential $\Psi^{\rm ret}$ involves four integrals of $\psi_0$, 
its singular part has to subleading order in $L$ the form 
\be
	\Psi^{\rm s}_\ell = A^\pm_\Psi /L^2 + B^\pm_\Psi /L^3.
\ee
This behavior also follows from the explicit form (\ref{eq:hertza1}) of $\Psi_{\ell m\omega}$, together 
with the fact that, for large $\ell$, the spheroidal eigenvalues $\lambda$ approach their spherical values,
$\lambda/\ell^2 \rightarrow 1 $.   The leading term in the expansion is then immediate from the 
leading term in (\ref{eq:psicoef}), while subleading terms involve the expansion of $\lambda$ (found analytically or 
numerically) in terms of $L$.    

 The metric perturbation involves two derivatives of $\Psi$, implying that the singular and retarded 
fields in a radiation gauge have the same leading power of $L$ as their Lorenz counterparts, 
\be
  h^{\rm s,RG\pm}_{\ell\bm{\mu\nu}} = A_{\bm{\mu\nu}}^\pm + B_{\bm{\mu\nu}}^\pm/L + O(L^{-2}).
\ee 

Finally, the self-force $f^{\rm ren,RG\alpha}$ is computed from 
$\na_\gamma h^{\rm ren, RG}_{\alpha\beta}$, using Eq.~(\ref{geopert}).  
As in the Lorenz gauge, the additional derivative gives the 
behavior 
\be
   a^{\rm s\, RG\bm\mu}_\ell = A^{\bm\mu\pm} L + B^{\bm\mu\pm} + O(L^{-2}).
\label{eq:fsrg}\ee 
The renormalized acceleration at the position of the particle is then given by 
\be
   a^{\rm ren\, RG\bm\mu} = \lim_{\ell_{\rm max}\rightarrow\infty} 
\sum_{\ell=0}^{\ell_{\rm max}} (a^{\rm ret\,RG \bm\mu}_\ell-a^{\rm s\,RG \bm\mu}_{\ell}). 
\label{eq:arenmu1}\ee

As noted in the introduction, for a particle in circular orbit in a Schwarzschild background,  
we find that the large $L$ expansion of $a^{\rm s\, RG\alpha}$ agrees through $O(L^0)$ with 
\be
    a^{\rm s\, RG\alpha} = - \nabla^\alpha\frac1\rho,
\ee
differing only by a constant from its form in a Lorenz gauge.  This form of $ a^{\rm s\, RG\alpha}$ 
does not contribute to the self-force, because Eq.~(\ref{eq:angle-av}) is satisfied -- the 
small-$\rho$ angle average of $a^{\rm s\, RG\alpha}$ vanishes.  Thus in this case, we can 
identify the singular field with its leading and subleading terms as a power series in $L$, 
 \be
   a^{\rm s\, RG\bm\mu}_\ell = A^{\bm\mu\pm} L + B^{\bm\mu\pm}.
\label{eq:fsrg1}\ee

The published version of this paper noted that, ``for a CCK radiation gauge, there is as yet no general proof that 
$a^{\rm s\, RG\alpha}$ is given by its leading and subleading terms, '' but that 
``we expect it to hold for a radiation gauge as well [as for a Lorenz gauge], with an argument based on 
the common property that $a^{\rm s\alpha}$ can be expressed (for $r>r_0$ or $r<r_0$) 
as a power series that begins at $O(\epsilon^{-2})$ and involves positive powers of 
the coordinate differences $x^\mu - x_0^\mu$ and odd powers of $\rho$.'' 
As mentioned above, it is now known that, by using the Lorenz regularization parameters for 
$r>r_0$ and $r<r_0$, one acquires a renormalized self-force appropriate to the 
CCK gauge that is smooth for $r\neq r_0$. 

One can alternatively use the Lorenz regularization parameters on one side or the other 
of the $r=r_0$ surface to acquire a renormalized self-force that gives the 
correct perturbed equations of motion for the perturbed metric in a gauge 
that we now describe.  
Let $h^{\rm RG,ren}_{\alpha\beta}$ be a metric obtained from $\psi_0^{\rm ren}$ by the 
CCK reconstruction described in the paper. The metric $h^{\rm RG,ren}_{\alpha\beta}$ 
is defined in a finite neighborhood $N$ of the unperturbed trajectory.  
Resolve the gauge ambiguity in a radiation gauge by choosing the singular radiation-gauge 
field $h^{\rm s,RG}_{\alpha\beta}$ that agrees at $O(1)$ with the Lorenz gauge.  One can, in particular, choose 
$h^{\rm RG,ren}_{\alpha\beta}$ so that $f^{\rm RG,ren}_\alpha = f_\alpha[h^{\rm RG,ren}]$ 
is the self-force obtained from the retarded field $f^{\rm RG,ret}_\alpha$, 
using the Lorenz-gauge regularization parameters $A_\alpha$ and $B_\alpha$ and 
taking the limit $r\rightarrow r_0^+$ (or $r\rightarrow r_0^-$).  
Because $h^{\rm RG,ren}_{\alpha\beta}$ 
differs from $h^{\rm Lor,ren}_{\alpha\beta}$ by a smooth gauge transformation,
\[ 
    h^{\rm RG,ren}_{\alpha\beta} = h^{\rm Lor,ren}_{\alpha\beta} + \pounds_\xi g_{\alpha\beta},
\]  
the singular field of 
$\widetilde h^{\rm ret}_{\alpha\beta} 
	= h^{\rm Lor, ret}_{\alpha\beta}+\pounds_\xi g_{\alpha\beta}$ 
is the Lorenz-gauge singular field.  Because $\widetilde h^{\rm ret}_{\alpha\beta}$ 
is smoothly related to the Lorenz-gauge retarded field, the renormalized radiation-gauge 
self-force gives the correct perturbed equation of motion for the gauge choice 
$\widetilde h^{\rm ret}_{\alpha\beta}$.  \\

In the Schwarzschild example below, the coordinate expression for $\psi_0^{\rm s}$ to subleading order 
is a sum of more than 25 terms, and to find its large-$L$ expansion we computed the 
large-$L$ expansion of each term.  Finding the corresponding large-$L$ expansion for 
$a^{\rm s\, RG\bm\mu}$ involves finding a large-$L$ expansions of all combinations of three derivatives 
of each of these terms for each component of $a^{\rm s\, RG\bm\mu}$.  
Without significant insight, this would mean finding the large-$L$ 
expansion of about 300 terms.  Given the simple form that $a^{\rm s\alpha}$ takes for the Schwarzschild 
circular-orbit there may be a similar form in the generic case and a much simpler way to find it.  
In its absence, the numerical method is much easier to use.
   
\subsubsection{Numerical method}

The regularization parameters occurring in Eq.~(\ref{eq:fsrg}) for $a^{\rm s\, RG\mu}_\ell$ 
are coefficients in the large-$L$ expansion of $a^{\rm ret\, RG\mu}$ evaluated at $r=r_0^\pm$. 
Consequently, once one has found the numerical values of $a^{\rm ret\, RG\mu}$, 
it is not in principle necessary to carry out an additional analytic computation of 
$a^{\rm s\, RG\mu}_\ell$.  Instead, one can match to the sequence of 
values $a^{\rm ret\, RG\mu}_\ell$ a power series in $L$ of the form 
\be
 a_\ell^{\rm ret\,\bm\mu} = A^{\bm\mu\pm} L + B^{\bm\mu\pm} + \sum_{k=2}^n \frac{E^{\bm\mu}_k}{L^k},
\label{eq:fSmul}\ee  
finding the $E_k^{\bm\mu}$ that yield a best fit.
A numerical check is that subtraction of the order $L^{-k}$ term reduces the order of the 
series by one power of $L$.  And one should check that the reduction of order holds for 
values of $L$ larger than those used to obtain the coefficients in the matching.
Finally, one checks numerically that $f^{{\rm reg},\bm\mu}$ converges to a value $f^{{\rm ren},\bm\mu}$
as the cutoff $\ell_{\max}$ increases.    
The disadvantage of the numerical matching method is that, for a given desired accuracy in $a^{\rm ren\alpha}$, 
one must compute $a^{\rm ret\alpha}$ to higher values of $\ell$ than is required when one or 
more regularization parameters are known analytically.  

To identify $a^{\rm s\,\alpha}_\ell$ with $A^{\bm\mu\pm} L + B^{\bm\mu\pm}$, one must again show 
that $a^{\rm s\alpha}$ satisfies Eq.~(\ref{eq:angle-av}).  This can done if   
one can find for each component $a^{\bm\mu}$ expressions in terms of $\rho$ and the coordinate 
differences $x^\mu-x^\mu_0$ whose large-$L$ expansion is $A^{\bm\mu\pm} L + B^{\bm\mu\pm}$, and 
if this expression satisfies (\ref{eq:angle-av}).  
If the limiting angle-average has a 
finite value, that finite value must be added back, in accordance with Eq.~(\ref{eq:aren}),  
to find the self-force.     

Equivalently, if the first two terms in this $L$-expansion correspond to the terms of 
order $\epsilon^{-2}$ and $\epsilon^{-1}$ in the position-space expansion, so that the 
terms involving $E_k$ correspond to 
terms of order $\epsilon$ and higher, then these latter terms sum to zero. 
We suspect this is the case for a radiation gauge, because 
the behavior of the singular field as a power series in $L$ implies behavior 
in $\epsilon$ that is no more singular than in a Lorenz gauge and corresponds for 
$r>r_0$ and $r<r_0$ to a power series in $\epsilon$.  
This is consistent with our numerical construction of the singular part of the self-force 
for a particle in circular orbit in Schwarzschild. 

With an analytic knowledge of the first term or terms in the expansion, the method 
is still useful in finding subsequent terms, and this has been done to speed convergence in the 
self-force computations involving mode-sum renormalization in a Lorenz gauge.
In particular, once one knows that $a^{\rm s\,\bm\mu}_\ell$ can be identified with the leading 
and subleading terms $A^{\bm\mu\pm} L + B^{\bm\mu\pm}$, the computation of $a^{\rm s\,\bm\mu}_\ell$ is 
given by 
 \be
 a^{\rm ren\,\bm\mu} = \sum_{\ell=0}^{\ell_{\rm max}}\left(a^{\rm ret\,\bm\mu}_\ell -A^{\bm\mu\pm} L - B^{\bm\mu\pm} 
- \sum_{k=2}^n \frac{E^{\bm \mu}_k}{L^k}\right) 
+ \sum_{k=2}^n E^{\bm \mu}_k\sum_{\ell=0}^\infty L^{-k},
\label{eq:fretmul}\ee 
with an error of order $\ell_{\rm max}^{-(n+1)}$.  

To justify the numerical method, we present in Sec.~\ref{sec:circ} below a comparison of the analytic and numerical determination of regularization parameters for the axisymmetric part of $\psi_0$ and their contribution 
to the self-force.
The companion paper presents the full numerical computation of the self-force for a particle in circular orbit in a Schwarzschild background.  Checks of the work include a numerical computation of a quantity $h_{\alpha\beta}u^\alpha u^\beta$ that is invariant under helically symmetric gauge transformations and has previously been computed in Lorenz and Regge-Wheeler gauges.  

\subsection {
Mass, Spin and Center-of-Mass
: The Remaining Metric}
\label{sec:mj}

The CCK reconstruction of the metric perturbation from $\psi_0$ 
gives a perturbed metric for which there is no change in the mass and angular momentum.  
For a Schwarzschild background, this is immediate from the fact that $\psi^{\rm ret}_0$ 
involves only values of $\ell$ with $\ell\geq 2$, because the construction 
of the perturbed metric preserves the value of $\ell$.  For a Kerr background, both $\psi^{\rm ret}_0$
and $\Psi^{\rm ret}$ 
are a sum of spheroidal harmonics with spin-weight 2, and their expression (\ref{eq:spw2sph}) in terms of 
spin-weighted spherical harmonics involves only harmonics with $\ell \geq 2$.  A mass perturbation 
requires an $\ell=0$ part of the asymptotic perturbed metric at $O(r^{-1})$.  The operator in Eq.~(\ref{mpschw}) 
mixes different values of $\ell$, but it differs from its Schwarzschild form by terms smaller by $O(r^{-1})$ 
than its leading terms.  That is, both $\psi^{\rm ret}_0$ and $\Psi^{\rm ret}$ are $O(r^{-5})$, the spherically symmetric 
part of the operator (\ref{mpschw}) is $O(r^4)$, and terms that mix different values of $\ell$ are $O(r^{3})$, 
implying $h^{\rm ret,RG}_{\alpha\beta}$ has no $\ell=0$ part at $O(r^{-1})$.  

A perturbation of angular momentum requires an $\ell=1$ contribution at $O(r^{-2})$ in $h_{\bf 13}$. To 
see that it also vanishes, we show that the correction in a Kerr background to the Schwarzschild expression for $h^{\rm ret}_{\bf13}$ is asymptotically $O(r^{-3})$.  The operator that relates 
$h^{\rm ret}_{\bf13}$ to $\Psi$ in Eq.~(\ref{mpschw}) is 
$(\bar{\bm \delta} - 3\alpha +\bar \beta +5\pi + \bar \tau) (\bm \Delta + \mu - 4\gamma)+
(\bm \Delta + 5\mu -\bar \mu -3\gamma -\bar \gamma)(\bar{\bm \delta} -4\alpha + \pi)$.  
Now $\bm D$ is a derivative along an outgoing null ray.  In the Kerr coordinates of Eq.~(\ref{eq:kerrcoords}),
${\bm D}f(u,r,\theta,\tilde\phi) = \partial_r f(u,r,\theta,\tilde\phi)$.  Because each 
time harmonic of $\Phi$ has the form $\Phi = S(\theta,\tilde\phi) e^{-i\omega u}r^{-5} [1+O(r^{-1})]$, 
${\bm D} \Phi =  S(\theta,\tilde\phi) e^{-i\omega u}O(r^{-6})$; because the spin coefficients are 
$O(r^{-1})$ or smaller, $h^{\rm ret}_{\bf13}$ falls off like $r^{-2}$, and the correction to 
its Schwarzschild behavior is at $O(r^{-3})$.      

To complete the metric reconstruction one needs to add the contributions from a change in the 
mass and angular momentum of the spacetime outside $r=r_0$.  Satisfying the perturbed field equation 
also requires a 
contribution $h^{\rm cm}_{\alpha\beta}$ to the perturbed metric that has for $r\neq r_0$ the 
form of a gauge transformation
 associated with a change in the system's center of mass:
{
\[
h^{\rm cm}_{\alpha\beta} =  \Theta(r_0-r) \Lie_{{\bm\xi}_<} g_{\alpha\beta}+ \Theta(r-r_0) \Lie_{{\bm\xi}_>}g_{\alpha\beta}. 
\]
When the values or gradients of $\xi_<^\alpha$ and $\xi_>^\alpha$ do not coincide at $r=r_0$, 
the corresponding perturbed stress tensor is a distribution with support on the $r=r_0$ sphere.  
Note that $h^{\rm cm}_{\alpha\beta}$ is {\it not} the metric perturbation 
associated with a gauge transformation  $\Theta(r_0-r){\mathbf\xi}_< + \Theta(r-r_0) {\mathbf\xi}_>$:
A metric perturbation that is pure gauge everywhere is sourcefree.
}

  That nothing further is needed follows from an minor extension of a theorem by Wald that 
implies that a perturbed vacuum metric is determined up to gauge transformations 
and the addition of a Petrov type D perturbation of the black-hole geometry \cite{wald73}.  
There are four kinds: (i) an infinitesimal change in the black hole's mass and  
(ii) in its spin; (iii) the perturbative version of the C-metric 
and (iv) the perturbative version of the Kerr-NUT solution. 
The type D perturbations are all stationary and axisymmetric, and 
only the mass and angular momentum perturbations are smooth in the 
region exterior to the black hole: The C and Kerr-NUT perturbations 
are each singular on their axis of symmetry, coinciding for a 
Kerr background, with the axis of symmetry of the Kerr geometry.      
If we were dealing with a source-free perturbation, regularity 
at the horizon and at infinity would rule out the addition of Kerr-NUT and 
C-metric perturbations, and, after a choice of gauge, we would be left 
with changes in mass and angular momentum (and gauge transformations).  
To extend the argument to $\hretab$ in our case, we note that smoothness of each time-harmonic 
of $\psi_0^{\rm ret}$ 
for $r\neq r_0$, together with the explicit form (\ref{eq:hertza1}), implies that 
$\Psi^{\rm ret}$ is smooth for $r\neq r_0$.  That is, smoothness of $\psi_0^{\rm ret}$ implies 
that the coefficients of each angular harmonic in its decomposition fall off faster 
than any power of $L$.  Eq.~(\ref{eq:hertza1}) implies that the coefficients of the 
angular harmonics of $\Psi$ fall off still faster in $L$. Thus each time 
harmonic of $\Psi$ is smooth and has no contribution from a C or Kerr-NUT perturbation.  

 The contributions to the retarded field of mass and spin were examined by Detweiler and Poisson 
\cite{dp04} and by Price \cite{lpthesis}.  With the Hertz potential restricted 
to $\ell\geq 2$ (for a Schwarzschild or Kerr background), there is no local contribution to 
mass and angular momentum.  For $r<r_0$, this is the appropriate solution for the retarded 
field.  For $r>r_0$, one has to determine in any gauge four parameters corresponding to 
changes in mass and spin (two correspond to a change in the spin direction and are gauge transformations); 
and three parameters associated with gauge transformations for $r>r_0$ that eliminate an asymptotic dipole.
The change in mass and in the magnitude of angular momentum along its direction in the 
background Kerr spacetime can be found from the integrals 
\bsube
\bea
  \delta M = \int_S (2T^\alpha_{\ \beta} -\delta^\alpha_\beta T)t^\beta dS_\alpha, \\
  \delta J = -\int_S T^\alpha_{\ \beta}\phi^\beta dS_\alpha
\eea\esube
over any hypersurface intersecting the trajectory.  (For a Schwarzschild background, all components 
of $\delta \bf J$ are determined in this way.)  The remaining   
parameters are determined by jump conditions in the field equations across $r=r_0$.

Although the metric perturbations corresponding to changes in mass and spin can be written in a 
radiation gauge, we do not see a good reason to do so.  In particular, the radiation gauge form of 
a mass perturbation that arises from a Hertz potential has a singularity on the radial ray through 
the particle.  (There is an alternative radiation-gauge form of a mass perturbation that is nonsingular 
on the axis of symmetry \cite{keidl07}, but it has no obvious advantage over a mass perturbation 
in another gauge.)   Instead we use for $\hretab$ a radiation gauge for $\ell\geq 2$, together with 
an arbitrary convenient gauge for $\ell < 2$.  

\subsection{Leading order parity of $\psi_0$, $\Psi$ and $h_{\alpha\beta}$} 

We now consider the parity of the perturbed metric in a radiation gauge. Gralla's criterion 
is that the projection of the perturbed metric to a surface orthogonal to the particle's 
4-velocity is even under parity to leading order in $\rho$.  Parity here means the locally 
defined diffeo that exchanges points at the same geodesic distance on opposite sides of 
each geodesic orthogonal to the particle trajectory; and the associated hypersurface spanned 
by geodesics orthogonal to a point $P$ of the trajectory is the surface onto which 
$h_{\alpha\beta}$ is projected.   Because the invariance 
under parity is only to leading order, we can define parity in terms of local 
inertial coordinates $T,X,Y,Z$ as the diffeo 
${\cal P}: (T,X, Y, Z)\mapsto (T,-X,-Y,-Z)$.  
The Weyl tensor has to leading order in $\rho$ its flat-space form, implying that 
it is even (invariant) under both parity and time-reversal, where time reversal, 
$\cal T$, exchanges points with coordinates $\pm T, X,Y,Z$. (Its behavior 
under both parity and time-reversal will be needed in the argument below.)     

We first show that these symmetries are retained by the singular form of 
$\psi_0$, given by Eq.~(\ref{eq:ps0}).  Because the tetrad vector fields 
$l^\alpha$ and $m^\alpha$ (and $n^\alpha$) are smooth, they are constant 
near a point $P$ on the trajectory to leading order in $\rho$ (as long 
as the particle is not on the $\theta=0,\pi$ axis, where $m^\alpha$ is not defined). 
The corresponding scalars $l^\alpha \nabla_\alpha T$ and $m^\alpha\nabla_\alpha T$ 
are then even to leading order under parity.  Because the Cartesian components 
of $\nabla_\alpha\rho$ have opposite signs at diametrically opposite points $(T,X,Y,Z)$ and 
$(T,-X,-Y,-Z)$, the scalars $l^\alpha \nabla_\alpha\rho$ and $m^\alpha\nabla_\alpha \rho$  
are odd under parity.  Then $(l^T m^\rho - l^\rho m^T)^2$ is even and hence $\psi_0$ 
is even under $\cal P$ to leading order in $\rho$.    

Similarly, the scalars $l^\alpha \nabla_\alpha T$ and $m^\alpha\nabla_\alpha T$ are 
odd to leading order under time-reversal; and $l^\alpha \nabla_\alpha \rho$ and 
$m^\alpha\nabla_\alpha \rho$ are even, implying that
$(l^T m^\rho - l^\rho m^T)^2$ and $\psi_0$ are even to leading order 
under time reversal.    

To show that $\Psi$ is even under parity to leading order in $\rho$ requires 
additional steps.  First, because $\Psi$ is obtained from $\psi_0$ as a 
sum of angular harmonics, we use the fact that the leading order parts of 
$\psi_0$ and $\Psi$ in $\rho$ for small $\rho$ are associated by the transform 
to angular harmonics with the leading order in $L$ part of its angular harmonics, 
for large $L$, as described in Appendix \ref{ap:singular}.
In particular the leading terms in $\rho$ of $\psi_0$ and $\Psi$ are, respectively, 
$O(\rho^{-3})$ and $O(\rho)$, and they correspond to the large-$L$ terms of 
order $O(L^2)$ and $O(L^{-2})$ in the angular harmonics.  Second,  
Eq.~(\ref{eq:hertza1}), expressing $\Psi$ in terms of $\psi_0$, involves angular harmonics 
of $\psi_0$ on a surface of constant $t$, a surface that is not perpendicular to the particle trajectory;  because of this, the plane tangent to the surface is not invariant under parity.  
It is, however, invariant under $\cal PT$, 
implying that the restriction of $\psi_0$ to a constant $t$ surface is invariant to leading 
order under $\cal PT$ (Restricted to the constant $t$ surface, $\cal PT$ is a parity 
transformation about $P$.)   
It follows that $\Psi^s$ and $\hsab$ are also invariant to leading order under ${\cal PT}$. 
We give the argument in terms of the the angular equation (\ref{eq:hertzan}) for $\Psi$.
The right side of this equation is dominated for large $L$ by its first term, 
having to leading order the form 
\be
   \psi_0 = \frac18 {\cal L}^4 \bar\Psi; 
\ee
that is, because ${\cal L}^4$ is, for large $L$, quartic in $\lambda$ and $\omega$, it dominates 
the second term. (One cannot look at the large $L$ limit with $\omega$ fixed, because $\omega$ 
is not independent of $L$: For a circular orbit, for example, $\omega=m\Omega$.)  
For the particle at $\theta_0\neq 0,\pi$, $\cal L$ has to leading order in $\rho$ the form 
${\cal L} =  \partial_\theta + i\csc\theta_0\,\partial_\phi  -ia\sin\theta_0\,\partial_t$.  
In Boyer-Lindquist coordinates, $\cal PT$ is given to leading order in $\rho$ by 
$(t,r,\theta,\phi)\rightarrow (2t_0-t, 2r_0-r, 2\theta_0-\theta, 2\phi_0-\phi)$, 
implying that each term in this leading form of $\cal L$ is odd under $\cal PT$. Then 
${\cal L}^4$ is even to leading order:  
\be 
{\cal L}^4{\cal PT}={\cal PT}{\cal L}^4.
\ee
The parts of $\Psi$ that are even and odd under $\cal PT$ then have sources that differ by one 
order in $L$:
\bsube\bea
  \frac12(1+{\cal PT})[{\cal L}^4 (\bar\Psi^{\rm odd}+\bar\Psi^{\rm even})+12M\partial_t\Psi ]&=&  {\cal L}^4\bar\Psi^{\rm even}[1+O(L^{-1})] = 8\psi_0^{\rm even},\\
  \frac12(1-{\cal PT})[{\cal L}^4( \bar\Psi^{\rm odd}+\bar\Psi^{\rm even})+12M\partial_t\Psi] 
	&=&  {\cal L}^4 \bar\Psi^{\rm odd} + O(\bar\Psi^{\rm even}\times L^3)
	 = 8\psi_0^{\rm odd}.
\eea\esube 
With a source smaller by one order in $L$, the algebraic inversion (\ref{eq:hertza1}) then gives $\Psi^{\rm odd}$ 
smaller than $\Psi^{\rm even}$ by one power of $L$.  

Next note that, because $\rho$ is independent of $T$ and the tetrad vectors are smooth (and hence 
constant to lowest order in $\rho$), $\psi_0$ is independent of $T$ to lowest order in $\rho$.  
That is, translating $\psi_0$ by $\Delta T$ changes it only by a term of order $\psi_0 \Delta T$,  
from the $O(\Delta T)$ change in the tetrad vectors.  Now 
time-translating $\Psi$ from a point on the $t=0$ surface to a point on the relatively 
boosted $T=0$ surface through the same point $P$ of the trajectory involves a translation by 
a time $\Delta T$ proportional to $\rho$ and hence changes $\Psi$ only to subleading order
in $\rho$. Thus $\Psi$ is invariant under $\cal P$ to leading order in $\rho$.  

Finally, a tensor $T_{\alpha\beta}$ is invariant under $\cal P$ if 
$T_{\alpha\beta} = {\cal P}^* T_{\alpha\beta}$, where the pullback ${\cal P}^*T_{\alpha\beta}$
has components in coordinates $\{x^\mu\}$ given by $
 ({\cal P}^* T)_{\mu\nu}(Q)
	:= \partial_\mu {\cal P}^\sigma \partial_\nu {\cal P}^\tau T_{\sigma\tau} [{\cal P}(Q)] 
$.
(We have used ${\cal P}^{-1}=\cal P$.) In terms of the coordinates $(T,X,Y,Z)$,  
the requirement that the spatial projection of $h_{\alpha\beta}$ is invariant under parity 
is equivalent to the condition that each spatial component $h_{ij}$ is even under parity 
\be
h_{ij}(T,X,Y,Z)=({\cal P} h)_{ij}(T,X,Y,Z) = h_{ij}(T,-X,-Y,-Z).
\label{eq:hparity}\ee
That this condition is satisfied follows from Eq.~(\ref{mpschw}) for $h_{\alpha\beta}$ in terms of $\Psi$ and 
the fact that the leading, $O(\rho^{-1})$, part of $h_{\alpha\beta}$ comes entirely from terms quadratic 
in the derivative operators, in $\bm\Delta$, $\bm\delta$ and $\bar{\bm\delta}$. Because $\Psi$ 
is independent of $T$ to leading order in $\rho$, each derivative operator 
along a tetrad vector involves only the spatial ($X^i$) components of the vector, 
implying that the quadratic derivatives $\bm\Delta^2\Psi, \ldots, \bm{\bar\delta}^2\Psi$ all 
have even parity to leading order.  Finally, the lowest-order constancy of the tetrad vectors implies 
that the products of their components $n_i n_j, \ldots, \bar m_{(i}\bar m_{j)}$ in local inertial coordinates are even to leading order in $\rho$ under 
parity.  We conclude that the projection of $h_{ij}^{\rm ret,RG}$ orthogonal 
to the 4-velocity is even under parity at leading order in 
$\rho$: to $O(\rho^{-1})$.%
\footnote{In fact, the argument shows that each component $h_{\mu\nu}$, 
regarded as a scalar, has even parity at leading order in $\rho$.}

\section{Particle in circular orbit in a Schwarzschild geometry}
\label{sec:circ}
  
    As a simplest explicit example of the method, we consider a particle of mass $\mathfrak m$ in circular orbit at radial coordinate $r_0$ about a Schwarzschild black hole of mass $M$.  In this 
section we compare the numerical and analytic renormalization methods by looking at the 
mode-sum renormalization of the axisymmetric part of $\psi_0$.  We first compute the retarded 
field; we then find an analytic expression for the singular field to subleading order; finally,
we obtain the regularization parameters of the singular field numerically, finding agreement to 
high accuracy with their analytic values. The numerical computation of the self-force is described in the companion paper.  

We work in Schwarzschild coordinates and adopt the notation    
\beq
ds^2 = f dt^2 -f^{-1}dr^2 - r^2(d\theta^2 + \sin^2\theta d\phi^2),
\eeq
where 
\be
  f(r):=1-\frac{2M}r.
\ee
The Kinnersley tetrad vectors have components 
\beq
(l^\mu) = (1/f(r),1,0,0), \qquad (n^\mu) = \frac{1}{2}(1,-f(r),0,0), \qquad (m^\mu) = \frac{1}{\sqrt{2r}}(0,0,1,i/\sin\theta).
\label{eq:ktet}\eeq
With this choice of tetrad the nonzero spin coefficients are
\beq
\varrho=-\frac{1}{r}, \qquad \beta=-\alpha=\frac{\cot\theta}{2\sqrt{2}r}, \qquad 
\gamma=\frac{M}{2r^2}, \qquad \mu=-\frac{1}{2r} \left(1-\frac{2M}{r}\right),
\label{spincoefficients}
\eeq
with corresponding Christoffel symbols 
\bsube\bea
\Gamma^{\bf 1}_{\bf 12} &=&-\Gamma^{\bf 2}_{\bf 22} = 2\gamma \\ 
\Gamma^{\bf 1}_{\bf 43} &=& \Gamma^{\bf 1}_{\bf 34} = \Gamma^{\bf 4}_{\bf 24} = \Gamma^{\bf 3}_{\bf 23} = \mu \\ 
\Gamma^{\bf 3}_{\bf 13} &=& \Gamma^{\bf 2}_{\bf 43} = \Gamma^{\bf 2}_{\bf 34} = \Gamma^4_{\bf 14} = -\varrho \\ 
\Gamma^{\bf 3}_{\bf 33} &=& \Gamma^{\bf 4}_{\bf 44} = -\Gamma^{\bf 4}_{\bf 43} = -\Gamma^{\bf 3}_{\bf 34} = 2\beta.
\eea\label{eq:gamma}\esube
The only nonzero component of the background Weyl tensor is
\beq
\Psi_2 = -\frac{M}{r^3}.
\eeq

The particle's 4-velocity is 
\beq
u^\alpha = u^t(t^\alpha+\Omega\phi^\alpha), 
\eeq
with $t^\alpha$ and $\phi^\alpha$ timelike and rotational Killing vectors and with
$ u^t = \sqrt{1-3M/r_0}$.  Its energy and angular momentum per unit mass, 
$E:= -u_\alpha t^\alpha$ and $J:= u_\alpha \phi^\alpha$, 
are given by 
\beq
E = \frac{r_0-2M}{\sqrt{r_0^2-3Mr_0}}, \quad J^2 = \frac{r_0^2M}{r_0-3M}.
\eeq

From Eq.~(\ref{mpschw}), the nonzero components of the metric perturbation are
\bea
h_{\bf11} &=& -\frac{r^2}{2}(\bar{\eth}^2\Psi + \eth^2\bar \Psi), \\ 
h_{\bf33} 
&=& -r^4\left[\frac14\left(\partial_t^2-2f\partial_t\partial_r+f^2\partial_r^2\right)
	- \frac{3(r-M)}{2r^2}\partial_t + f\frac{3r-2M}{2r^2}\partial_r+\frac{r^2-2M^2}{r^4}\right]\Psi,\\
h_{\bf13} &=&  \frac{r^3}{2\sqrt{2}}\left(\partial_t-f\partial_r-\frac{2}{r}\right)\bar{\eth}\Psi.
\eea

To compute the self-acceleration (\ref{geopert}) in terms of these tetrad components, we use the relations 
\be
	t^\alpha = \frac12f l^\alpha + n^\alpha, \qquad 
	\phi^\alpha = \frac{-i r}{\sqrt{2}}(m^\alpha - \bar{m}^{\alpha}), \qquad
\nabla^\alpha r = -\frac12 f l^\alpha + n^\alpha,
\ee
to write
\be
a^r = (u^t)^2 \left( \frac12f_0 l^\alpha - n^\alpha \right) 
	\left( \frac12 f_0 l^\beta + n^\beta 
	-i \frac{\Omega r_0}{\sqrt{2}}( m^\beta - \bar m^\beta) \right) 	
	      \left( \frac12f_0 l^\gamma + n^\gamma 
	-i \frac{\Omega r_0}{\sqrt{2}}( m^\gamma - \bar m^\gamma) \right)
	(\nabla_\beta h_{\alpha \beta} - \frac{1}{2}\nabla_\alpha h_{\beta \gamma}).\qquad
\ee
Then, expanding the covariant derivatives and using Eqs.~(\ref{spincoefficients}) and (\ref{eq:gamma}), 
we find
\bea
a^r&=& (u^t)^2\left\{f_0^2\left[ \frac1{16}f_0 {\bm D} +\frac38\,\bm\Delta  
	+\frac i8\Omega(\eth-\bar\eth)-\frac12\frac{M }{r_0^2} \right] h_{\bf11}\right.
 \nonumber\\
&&\left.\phantom{xxxxx}f_0\left[\left(\frac18\frac{M}{r_0} {\bm D} - \frac14\frac{M}{r_0f_0}\bm\Delta 
	+ \frac12\frac{M}{r_0^2} \right)h_{\bf33} 
 + \left(-\frac i{\sqrt{2}}\Omega r_0 \bm\Delta -\frac1{4\sqrt2}\frac{M}{r_0^2}(\eth-\bar{\eth})
	+\frac i{2\sqrt2}\Omega \right)h_{\bf13}+c.c.\right]\right\}.
\eea

\subsection{\label{retfield}The retarded field}

The retarded fields $\pret$ and $\preto$ are simplest to compute in coordinates for which 
the unperturbed orbit lies in the $\theta = \pi/2$ plane.  
The particle's trajectory is then given by $\phi=\Omega t$, where $\Omega=\sqrt{M/r_0^3}$. 
From Eq.~(\ref{db_stress_particle}), its stress-energy tensor is 
\bea
T^{\alpha\beta} &=& \frac{{\mathfrak m}}{u^tr_0^2} u^\alpha u^\beta \delta(r-r_0)\delta(\cos(\theta)) \delta(\phi-\Omega t) \nonumber\\
&=& \sum_{\ell,m}\frac{{\mathfrak m}}{u^tr_0^2} u^\alpha u^\beta \delta(r-r_0){}_sY_{\ell m}(\theta,\phi){}_s\bar{Y}_{\ell m}\left(\frac{\pi}{2},\Omega t\right) .
\label{eq:tab}\eea
Because the source and the background metric are both helically symmetric, Lie derived by $t^\alpha + \Omega \phi^\alpha$, the retarded fields -- metric and Weyl tensor -- will also be 
helically symmetric.  Because the tetrad vectors (\ref{eq:ktet}) are 
also helically symmetric, the symmetry is shared by the scalars $\preto$ and $\pret$, 
which therefore only involve $\phi$ and $t$ in the combination $\phi-\Omega t$.
In the harmonic decomposition, $\phi$ and $t$ then occur only in the combination 
$e^{im(\phi-\Omega t)}$, and the frequency associated with each value of $m$ is $\omega = m\Omega$.  

The scalars $\pret$ and $\preto$ satisfy the Bardeen-Press equation \cite{bp71}, 
the $a=0$ form of the Teukolsky equation, namely
\beq
{\cal T}_s\psi:= \left[\frac{r^4}\Delta\partial_t^2 -2s\left(\frac {Mr^2}\Delta -r\right)\partial_t 
-\Delta^{-s} \frac{\partial}{\partial r} 
\left( \Delta^{s+1} \frac{\partial} {\partial r} \right) 
- \bar{\eth}\eth \right]\psi = 4\pi r^2T_s. 
\label{bp_circ}
\ee
We will work with $\preto$, whose source, from Eq.~(\ref{eq:psi0source}), has the form
\bea
 T_{s=2}  &=& -2(\bm\delta-2\beta)\bm\delta T_{\bf 11} 
	+ 4(\bm D-4\varrho)(\bm\delta-2\beta)T_{\bf 13} 
	-2(\bm D-5\varrho)(\bm D-\varrho)T_{\bf 33}\nonumber\\
&=:& T^{(0)} + T^{(1)} + T^{(2)}, 
\label{teuksource}
\eea
where the superscripts indicate the maximum number of radial derivatives in each term.
From Eq.~(\ref{eq:tab}), these terms have the explicit forms
\bea
T^{(0)} & =& -\sum_{\ell, m}\frac{{\mathfrak m} u^t}{r_0^4}\delta(r-r_0) [(\ell-1)\ell(\ell+1)(\ell+2)]^{1/2}{}_2Y_{\ell m}(\theta,\phi)\bar{Y}_{\ell m}\left(\frac{\pi}{2},\Omega t\right),\\
T^{(1)} &=& 2\sum_{\ell, m} \frac{{\mathfrak m}\Omega  u^t}{r_0^2}\left[i\delta'(r-r_0)  
	+\left(\frac{m\Omega}{f_0}+\frac{4 i}{r_0}\right)\delta(r-r_0)\right][(\ell-1)(\ell+2)]^{1/2}
	{}_2Y_{\ell m}(\theta,\phi){}_1\bar{Y}_{\ell m}\left(\frac{\pi}{2},\Omega t\right),\\
T^{(2)} &=& \sum_{\ell,m}{\mathfrak m} \Omega^2 u^t\Biggl[\delta''(r-r_0) + 
\left(\frac{6}{r} 
	- \frac{2im\Omega}{f}\right)\delta'(r-r_0)\nonumber\\
&\pheq& - \left(\frac{m^2\Omega^2}{f_0^2} +\frac{6im\Omega}{r_0 f_0}-\frac{2imM\Omega}{r_0^2f_0^2}-\frac4 {r_0^2}\right)\delta(r-r_0)\Biggr]\,{}_2Y_{\ell m}(\theta,\phi) {}_2\bar{Y}_{\ell m}\left(\frac{\pi}{2},\Omega t\right).
\eea

Each mode of $\psi_0$ or $\psi_4$,
\beq
\psi = e^{-i\omega t} R(r)\, {}_s\!Y_{\ell m}(\theta,\phi),
\eeq
has radial eigenfunction $R_0$ or $R_4$ satisfying the radial equation corresponding to 
its spin-weight:
\bea
\Delta R_0^{\prime\prime} + 6(r-M) R_0^\prime + \left[ \frac{\omega^2 r^4}{\Delta} + \frac{4i\omega r^2(r-3M)}{\Delta} - (\ell-2)(\ell+3) \right]R_0 &=& 0,\label{teuk0rad}\\
\Delta R_4^{\prime\prime} - 2(r-M) R_4^\prime + \left[ \frac{\omega^2 r^4}{\Delta} - \frac{4i\omega r^2(r-3M)}{\Delta} - (\ell-1)(\ell+2) \right]R_4 &=& 0,
\eea
where the prime denotes a derivative with respect to the radial coordinate $r$.  Solutions to these 
equations are related by
\beq
R_0 = \frac{\bar{R}_4}{r^4f^2}.
\eeq

  To compute $\preto$, it is helpful to define a Green's function $\hat G(r,r')$ as the solution to 
\beq
\Delta\hat{G}^{\prime\prime} + 6(r-M)\hat{G}^\prime + \left( \frac{\omega^2 r^4}{\Delta} + \frac{4i\omega r^2(r-3M)}{\Delta} - (\ell-2)(\ell+3) \right)\hat{G} = \frac{\delta(r-r^\prime)\Delta^{1/2}}{r^3}.
\eeq
namely
\beq
\hat G(r,r^\prime) 
	= - \sum_{\ell,m}A_{\ell m}\frac{[\Delta(r^\prime)]^{5/2}}{r^{\prime 3}}R_H(r_<)R_\infty(r_>),
\eeq
where $R_{\rm H}$ and $R_{\rm \infty}$ are solutions to the radial equation for $\psi_0$ that are regular at 
the horizon and at infinity, respectively, and the quantity 
\beq
A_{\ell m} := \frac{1}{\Delta^3 (R_{\rm H} R_{\rm \infty}^\prime - R_{\rm \infty} R_{\rm H}^\prime)}
\eeq
is a constant, independent of $r$.
The full spatial Green's function $G(x,x')\equiv G(r,\theta,\phi;r^\prime,\theta^\prime,\phi^\prime)$ is then given by
\beq
G(x,x')
	 = - \sum_{\ell,m}A_{\ell m}\frac{[\Delta(r^\prime)]^{5/2}}{r^{\prime 3}}R_H(r_<)R_\infty(r_>){\,}_2Y_{\ell m}(\theta,\phi){\,}_2\bar{Y}_{\ell m}(\theta^\prime, \phi^\prime).
\eeq

The Weyl scalar  $\psi_0$ is defined and smooth everywhere except on the trajectory of 
the particle.  It is given in terms of the source and the Green's function $G(x,x')$ by 
\bea
\psi_0 &=& 4\pi\int T(t,x')G(x,x')r'^2dV'\nonumber\\
&=& 4\pi\int \left(T_0^{(0)}+T_0^{(1)}+T_0^{(2)}\right)G(x,x')r'^2dV'\nonumber\\
&=:& \psi_0^{(0)}+\psi_0^{(1)}+\psi_0^{(2)},
\label{eq:psiGT}\eea
where the superscripts on $\psi_0$ correspond to the three terms in Teukolsky source function defined 
in \Deqn{teuksource}.  The three terms in \Deqn{eq:psiGT} have outside the particle trajectory%
\footnote{Note that the formal integral of the Green's function also gives a $\delta$-function contribution 
with support on the trajectory, namely 
$-4\pi{\frak m}\Omega^2 u^t f_0^{-1} \delta(r-r_0)\delta(\cos\theta)\delta(\phi-\Omega t)$.}
the form
\bea
\psi_0^{(0)} &=& 4\pi {\mathfrak m} u^t \frac{\Delta_0^2}{r_0^2}\sum_{\ell m}A_{\ell m}[(\ell-1)l(\ell+1)(\ell+2)]^{1/2}R_{\rm H}(r_<)R_{\rm \infty}(r_>){}_2Y_{\ell m}(\theta,\phi)\bar{Y}_{\ell m}\left(\frac{\pi}{2},\Omega t\right), \\
\psi_0^{(1)} &=& 8\pi i{\mathfrak m} \Omega u^t \Delta_0 \sum_{\ell m} A_{\ell m}[(\ell-1)(\ell+2)]^{1/2}
	{}_2Y_{\ell m}(\theta,\phi){}_1\bar{Y}_{\ell m}\left(\frac{\pi}{2},\Omega t\right) \times \\ \nonumber
& & \quad \Bigl\{[im\Omega r_0^2 + 2 r_0]R_{\rm H}(r_<)R_{\rm \infty}(r_>) 
	+ \Delta_0[R_{\rm H}'(r_0)R_{\rm \infty}(r)\theta(r-r_0) \\ \nonumber
& & \qquad + R_{\rm H}(r)R_{\rm \infty}'(r_0)\theta(r_0-r)]\Bigr\},\\
\psi_0^{(2)} &=& -4\pi {\mathfrak m}\Omega^2 u^t \sum_{\ell m} A_{\ell m}
	{}_2Y_{\ell m}(\theta,\phi){}_2\bar{Y}_{\ell m}\left(\frac{\pi}{2},\Omega t\right) \times \\ \nonumber
& & \biggl\{[30r_0^4 - 80Mr_0^3 + 48M^2r_0^2 - m^2\Omega^2 r_0^6 -2\Delta_0^2 - 24\Delta_0 r_0(r_0-M)+ 6im\Omega r_0^4(r_0-M)]
	R_{\rm H}(r_<)R_{\rm \infty}(r_>)
 \\ \nonumber
& & \qquad  + 2(6r_0^5 - 20Mr_0^4 + 16M^2r_0^3 - 3r_0\Delta_0^2 + im\Omega \Delta_0 r_0^4)[R_{\rm H}'(r_0)R_{\rm \infty}(r)\theta(r-r_0) 
 + R_{\rm \infty}'(r_0)R_{\rm H}(r)\theta(r_0-r)] \\ \nonumber
& & \qquad + r_0^2\Delta_0^2[R_{\rm H}''(r_0)R_{\rm \infty}(r)\theta(r-r_0) + R_{\rm \infty}''(r_0)R_{\rm H}(r)\theta(r_0-r)]\Biggr\}.
\eea

\subsection{\label{singfield}The singular field}

Because the conservative part of the self-force 
is radial, it is axisymmetric about a radial ray through the particle.  
We will compare the analytic to the numerical determination of the singular part of a Weyl scalar 
by looking at the axisymmetric part of $\psi_0$ and (in Appendix~\ref{ap:sf}) its contribution to the self-force.  
We outline the calculation of the leading and subleading terms in the axisymmetric part of the 
singular field $\psi_0^\SING$ as a sum of angular harmonics ${}_2Y_{\ell 0}(\Theta)$ whose coefficients are polynomials 
in $\ell$, with angular coordinates $\Theta$ and $\Phi$ chosen so that the 
$\Theta=0$ line (at fixed $t$) is the radial line through the particle.
Details of the conversion from a small distance expansion to a 
large $L$ expansion are left to Appendix~\ref{appsingfield}.  
 
The analytic expression for the resulting regularization parameters is then compared to a 
numerical determination by matching the retarded field to a power series in $L$.  Remarkably, although the 
subleading part of $\psi_0^\SING$ is a lengthy expression -- Eq.~(\ref{finalpsi0sing}), 
we will see that its axisymmetric part, written as a sum over angular harmonics, vanishes.  
Because the angular harmonics ${}_2Y_{\ell m}$ are complete in $L_2(S^2)$, this means that, as a 
distribution, $\psi_0^\SING$ has support at $\Theta=0$, where ${}_2Y_{\ell 0}(\Theta)=0$.  

The expression for the retarded field is simplest for 
coordinates in which the orbit is in the $\theta=\pi/2$ plane.   Expressing the singular 
field $\pso$ of Eq.~(\ref{eq:ps0}) as a sum of angular harmonics is simplest if angular
coordinates $\Theta$ and $\Phi$ are chosen with the particle at $\Theta=0$, as we have just 
described.  To compute the difference $\preno =\preto - \pso$, one must rotate $\preto$  
to the coordinate position of or $\pso$ or vice-versa. 
Following the conventions of Detweiler et al. \cite{dmw03} (henceforth DMW),  
$\Theta$ and $\Phi$ are related to $\theta$ and $\phi$ by a rotation of the form 
\bea
\sin \theta \cos \phi = \cos \Theta, \nonumber \\
\sin \theta \sin \phi = \sin \Theta \cos \Phi, \nonumber \\
\cos \theta = \sin \Theta \sin \Phi.
\eea
With the usual association of Cartesian coordinates $x,y,z$ to $r,\theta,\phi$ and of 
$\hat x, \hat y, \hat z$ to $r,\Theta,\Phi$, the map is $x=\hat z, y=\hat x, z=\hat y$.   

Eq.~(\ref{eq:ps0}) for $\pso$ involves the components $l^T=l^\alpha\na_\alpha T$ 
and $l^\rho = l^\alpha\nabla_\alpha \rho$.  We obtain these to subleading order 
in terms of Schwarzschild coordinates: That is, with $\epsilon$ the distance from the particle's position $P$ with respect to the positive-definite metric 
$g_{\alpha\beta}+2u_\alpha u_\beta$, the leading and subleading terms in $T$ and $\rho$ 
are $O(\epsilon)$ and $O(\epsilon^2)$, respectively.  The corresponding leading 
and subleading terms of $l^T$ and $l^\rho$ are then $O(1)$ and $O(\epsilon)$.     

Expansions of $\rho$ and of the local inertial coordinates $T,X,Y,Z$ about a point $P$ 
in terms of Schwarzschild coordinates are given, for example, in Ref.~\cite{dmw03}.
To subleading order, $T$ has the form
\beq \label{T eqn}
T = \left( E (t-t_0) - J \sin\Theta \cos\Phi \right) + \left( \frac{E M}{r_0^2 f_0}(t-t_0)(r-r_0)- \frac{J}{r_0}(r-r_0)\sin\Theta\cos\Phi \right).
\eeq
It is convenient to work with $\rho^2$ instead of $\rho$; to subleading order, we 
have $\rho^2 = \rho^{(2)} + \rho^{(3)}$, where the order $\epsilon^2$ and $\epsilon^3$ 
contributions are, respectively,
\beq \label{rho eqn1}
\rho^{(2)} = \frac{(r-r_0)^2}{f_0} + (r_0^2 + J^2 \cos^2\Phi)\sin^2\Theta -2EJ \sin\Theta\cos\Phi(t-t_0) + \frac{J^2f_0}{r_0^2}(t-t_0)^2,
\eeq
\bea \label{rho eqn2}
\rho^{(3)} &=& \frac{M}{r_0^2}\left(1+\frac{2J^2}{r_0^2}\right)(t-t_0)^2(r-r_0) + \frac{2JE(M-r_0)}{f_0 r_0^2}(t-t_0)(r-r_0)\sin\Theta\cos\Phi - \frac{M}{r_0^2 f_0^2}(r-r_0)^3\nonumber\\
&\pheq&+ r_0 \sin^2\Theta(r-r_0) + 2 r_0 \sin^2\Theta\sin^2\Phi(r-r_0) + \frac{2E^2r_0}{f_0}\sin^2\Theta\cos^2\Phi(r-r_0).
\eea

We use Eq.~(\ref{eq:ps0}).  This expression omits $\delta$-functions with support at the position of the 
particle.  These do not contribute to the renormalization of the retarded field if the  
renormalization is done by subtracting the singular field from the 
retarded field in a neighborhood of the particle, averaging over a sphere surrounding the 
particle, and then taking the limit as the radius of the sphere shrinks to zero (the Quinn-Wald 
prescription \cite{quinnwald}).  In a mode-sum regularization, we discard $\delta$-functions with 
support at the particle in both the singular and the retarded field.
  
The background Kinnersley tetrad written in terms of $\Theta$ and $\Phi$ is 
\beq \label{Ktetrad}
l^\alpha = \left(\frac{1}{f},1,0,0\right),\quad m^\alpha = \frac 1 {\sqrt 2 r}\left(0,0,1,\frac i{\sin\Theta}\right), \quad \mbox{where again }f = \left(1-\frac{2M}{r}\right)
\eeq

We now expand the needed tetrad components to subleading (quadratic) order in Schwarzschild coordinates, 
about their values at $P$, using superscripts $(0)$ and $(1)$ as above to denote orders in $\epsilon$ 
and writing 
$l^\alpha = l^{(0)\alpha} + l^{(1)\alpha}$, $m^\alpha = m^{(0)\alpha} + m^{(1)\alpha}$. 
Using Eqs.~(\ref{T eqn}), (\ref{rho eqn1}), (\ref{rho eqn2}), and (\ref{Ktetrad}),  we have
\bea \label{allstuff}
l^{(0)T} &=& \frac{E}{f_0}, \\ \nonumber
l^{(1)T} &=& - \frac{E M}{r_0^2 f_0^2}(r-r_0) + \frac{E M}{r_0^2 f_0}(t-t_0) - \frac{J}{r_0}\sin\Theta\cos\Phi, \\  \nonumber
m^{(0)T} &=& -\frac{J}{\sqrt{2}r_0}e^{-i\Phi},\\ \nonumber
m^{(1)T} &=& 0, \nonumber \\ \nonumber
l^{(1)\rho} &=& \frac{J^2}{ r_0^2\rho}(t-t_0) - \frac{JE}{ f_0\rho}\sin\Theta\cos\Phi + \frac{1}{f_0\rho}(r-r_0), \\ \nonumber
l^{(2)\rho} &=& \frac{M}{r_0^2f_0\rho}\left(1+\frac{2J^2}{r_0^2}\right)(t-t_0)(r-r_0) + \frac{JE(M-r_0)}{f_0^2r_0^2\rho}(r-r_0) \sin \Theta \cos \Phi + \frac{M(r_0^2+2J^2)}{2r_0^4\rho}(t-t_0)^2 \\ \nonumber
&+& \frac{JE(M-r_0)}{f_0r_0^2\rho}(t-t_0)\sin\Theta\cos\Phi - \frac{3M}{2r_0^2f_0^2\rho}(r-r_0)^2 - \frac{r_0}{2\rho}\sin^2\Theta + \frac{r_0}{\rho}\sin^2\Theta\sin^2\Phi \\ \nonumber
&+& \frac{E^2r_0}{f_0\rho}\sin^2\Theta\cos^2\Phi -\frac{2MJ^2}{r_0^4f_0\rho}(r-r_0)(t-t_0) - \frac{MJE}{f_0^2r_0^2\rho}(r-r_0)\sin\Theta\cos\Phi, \\ \nonumber
m^{(1)\rho} &=& \frac{\sqrt{2}}{2 r_0 \rho }\left(r_0^2 + J^2\cos\Phi e^{-i\Phi}\right)\sin\Theta - \frac{\sqrt{2}JE}{2 r_0 \rho}(t-t_0)e^{-i\Phi}, \\ \nonumber
m^{(2)\rho} &=& \frac{\sqrt{2}J^2e^{-i\Phi}\cos\Phi}{2 r_0^2\rho}(r-r_0)\sin\Theta - \frac{\sqrt{2}JEMe^{-i\Phi}}{2 r_0^3f_0\rho}(t-t_0)(r-r_0).
\eea
The terms in parentheses in Eq.~(\ref{eq:ps0}) are then given to subleading order by
\bea
\left(l^Tm^{\rho}-m^Tl^{\rho}\right)^2 &=&  \left( l^{(0)T}m^{(1)\rho} - m^{(0)T}l^{(1)\rho} \right)^2 \nonumber\\
										   &&+ 2 \left( l^{(0)T}m^{(1)\rho} - m^{(0)T}l^{(1)\rho} \right) \left( l^{(0)T}m^{(2)\rho} - m^{(0)T}l^{(2)\rho} + l^{(1)T}m^{(1)\rho} - m^{(1)T}l^{(1)\rho} \right);
\qquad\eea
and $\rho^{-5}$ is given by 
\beq
\frac{1}{\rho^5} = \frac{1}{\rho^{(2)\frac{5}{2}}} - \frac{5}{2}\frac{\rho^{(3)}}{\rho^{(2)\frac{7}{2}}},
\eeq
implying
\bea
\psi_0^\SING &=& -\frac{3}{2} \frac{\mathfrak m}{\rho^{(2)\frac{3}{2}}}
\left( l^{(0)T}m^{(1)\rho} - m^{(0)T}l^{(1)\rho} \right)^2  + \frac{15}{4}\frac{\mathfrak m}{\rho^{(2)\frac{5}{2}}}\rho^{(3)}\left( l^{(0)T}m^{(1)\rho} - m_T^{T(0)}l^{(1)\rho} \right)^2  \nonumber \\
&&- 3\frac{\mathfrak m}{\rho^{(2)\frac{3}{2}}}\left(l^{(0)T}m^{(1)\rho} - m^{(0)T}l^{(1)\rho} \right) \left( l^{(0)T}m^{(2)\rho} - m^{(0)T}l^{(2)\rho} + l^{(1)T}m^{(1)\rho} - m^{(1)T}l^{(1)\rho} \right)+O(\epsilon^{-1}).\qquad
\eea

The expression for $\psi_0^{\rm s}$ as a mode sum is obtained from the value of this expression at 
$t=t_0$.  The full expression, including terms involving $t-t_0$ is given in Appendix (\ref{appsingfield}).   
We denote by $\psi_0^{\rm s-L}$ and $\psi_0^{\rm s-SL}$ the leading and subleading parts of the singular 
field, respectively.  Using Eqs. (\ref{rho eqn1}), (\ref{rho eqn2}) and (\ref{allstuff}), we obtain for 
$\psi_0^{\rm{s}}$ to subleading order the explicit form  
\bea
\psi_0^{\textrm{s-L}}&=& -\frac{3{\mathfrak m} E^2 r_0^2}{f_0^2 }\frac{\sin^2\Theta}{\tilde{\rho}^{5}} + - \frac{3{\mathfrak m} J^2 e^{-2i\Phi}}{f_0^2 r_0^2}\frac{(r-r_0)^2}{\tilde{\rho}^{5}} - \frac{3{\mathfrak m} E J e^{-i\Phi}}{f_0^2} \frac{(r-r_0) \sin \Theta}{\tilde{\rho}^{5}}, 
\label{psi0s-L0}\\
\psi_0^{\textrm{s-SL}} &=& -\frac{15{\mathfrak m}J^2Me^{-2i\Phi}}{2f_0^4r_0^4}\frac{(r-r_0)^5}{\tilde{\rho}^7}  - \frac{15{\mathfrak m}JMEe^{-i\Phi}}{f_0^4r_0^2}\frac{\sin\Theta(r-r_0)^4}{\tilde{\rho}^7} \nonumber\\ 
&& + \frac{15{\mathfrak m}e^{-2i\Phi} J^2\left((-i\sin\Phi e^{i\Phi}) + \frac{J^2}{r_0^2}\cos^2\Phi\right)}{r_0f_0^2}\frac{(r-r_0)^3\sin^2\Theta}{\tilde{\rho}^7}  \nonumber\\ 
&& + \frac{15{\mathfrak m}JEe^{-i\Phi}(r_0^2+2J^2\cos^2\Phi)}{f_0^2r_0}\frac{(r-r_0)^2\sin^3\Theta}{\tilde{\rho}^7} + \frac{15 {\mathfrak m} r_0 E^2 \left(J^2+r_0^2+J^2 \cos 2\Phi \right)}{2 f_0^2}\frac{(r-r_0)\sin^4\Theta}{\tilde{\rho}^7} \nonumber\\ 
&& + \frac{9{\mathfrak m} \left(e^{-2 i \Phi } J^2 M\right) }{f_0^3 r_0^4}\frac{(r-r_0)^3}{\tilde{\rho}^5} + \frac{3{\mathfrak m} e^{i \Phi } J r_0 E}{f_0}\frac{\sin^3\Theta}{\tilde{\rho}^5} + \frac{15{\mathfrak m} e^{-i \Phi } M J E}{r_0^3 f_0^3}\frac{(r-r_0)^2\sin\Theta}{\tilde{\rho}^5} \nonumber \\ 
&& + \frac{9 \mathfrak m J^2}{r_0 f_0}\frac{(r-r_0)\sin^2\Theta}{\tilde{\rho}^5}
\label{psi0s-SL0}\eea

The axisymmetric part of each of these terms is proportional to an expression of the form
\be
    \frac{\delta^{k_1}\sin^{k_2}\Theta}{\delta^2+1-\cos\Theta)^{k+1/2}}, 
\ee 
where $k_1$, $k_2$ and $k$ are positive integers, and $\delta$, given by Eq.~(\ref{eq:rhodelta}), is proportional to $r-r_0$.
In Appendix~{\ref{appsingfield}, following DMW, we use the generating function for Legendre polynomials, 
\begin{equation} \label{DMW0}
\frac{1}{(e^T+e^{-T}-2u)^{1/2}} = \sum_\ell e^{-(\ell+1/2)|T|}P_\ell (u),\qquad T\neq 0,
\end{equation}
and its derivatives to express each term as a sum of Legendre polynomials and their derivatives.
We then use a relation between the spin-weighted harmonics $_sY_{\ell m}$ and Legendre polynomials 
to write the series in terms of the harmonics $_2Y_{\ell 0}$.   
The leading order part of $\psi_0^{\rm s}$ then has the form 
\beq
\langle\psi_0^\SING\rangle_{r_0}(\Theta) = \frac{-{\mathfrak m}(r_0-3M)^{3/2}}{r_0^2(r_0-2M)^{5/2}}\left\langle \frac{1}{\chi^{5/2}} \right\rangle\sum_{\ell=2}^{\infty}\sqrt{\frac{4\pi(\ell+2)!}{(\ell-2)!(2\ell+1)}}{\,}_2Y_{\ell0}(\Theta,0),
\label{eq:psi_sing_final0}\eeq
where $\langle\psi_0^\SING\rangle$ is the axisymmetric part of $\psi_0^\SING$.  

Finally, each subleading term is proportional as a distribution to the sum 
$\dis \sum_{l=0}^\infty (l+1/2)P_l(\cos\theta)$.  That sum is a $\delta$-function with support at 
$\Theta=0$ \cite{arfken}, and its projection along $_2Y_{\ell 0}$ therefore vanishes for all $\ell$.
Eq.~(\ref{eq:psi_sing_final0}) thus gives $\langle\psi_0^\SING\rangle$ to subleading order in $L$.

\subsection{\label{numersingfield}Comparison with numerical determination of $\psi_0^\SING$}
We complete this section with a comparison of the analytic form of $\langle\psi_0^\SING\rangle$ with its 
numerical value obtained by matching $ \psi_0^{\rm ret}$ to a power series in $L$.  A comparison of 
the numerically and analytically determined contributions to the self-force from the axisymmetric part 
of $\psi_0$ is given in Appendix \ref{ap:sf}.  

The retarded Weyl scalar $\psi_{0,l}^\RET$ is computed by integrating the radial Teukolsky equation for each 
value of $\ell$.  The coefficients of $_2Y_{\ell 0}$ are matched to a power-series in $L$ of the form
\beq
\psi_{0,l}^\RET = \sqrt{\frac{4\pi(\ell+2)!}{(2\ell+1)(\ell-2)!}}\left( A + \frac{B}{L} + \frac{C}{L^2} + \cdots \right).
\eeq
Shown below is a table of the fractional error in $A$ and $B$ when found numerically and compared to the analytic 
form given by Eq.~(\ref{eq:psi_sing_final0}).  Details of the numerical methods and checks of numerical accuracy 
are given in the companion paper.

\begin{table}[h]
  \begin{tabular}{|c|c|c|c|c|}
    \hline

                \multicolumn{1}{|p{0.75cm}|}{\centering $r_0/M$}
              & \multicolumn{1}{|p{2.5cm}|}{\centering $A_\textrm{analytic}$}
              & \multicolumn{1}{|p{2.5cm}|}{\centering $A_\textrm{numerical}$}
              & \multicolumn{1}{|p{2.25cm}|}{\centering $|\Delta A/A|$}
              & \multicolumn{1}{|p{2.25cm}|}{\centering $|\Delta B|$}\\
    \hline
     8   &	-0.002507548110466834	 &	-0.002507548108782573	 &	 $6.717\times10^{-10}$  &  $8.496\times10^{-10}$\\
    10   &	-0.001214016915072354	 &  -0.001214016915092580	 &   $1.666\times10^{-11}$  &  $1.245\times10^{-11}$\\
    15   &  -0.0003356104323965837	 &	-0.0003356104323973954	 &   $2.419\times10^{-12}$  &  $6.003\times10^{-13}$\\
    20   &  -0.0001370231924969076	 &	-0.0001370231924969057   &   $1.365\times10^{-14}$  &  $3.853\times10^{-14}$\\
    25   &  -0.00006882828667366571	 &	-0.00006882828667362643  &   $5.706\times10^{-13}$  &  $7.462\times10^{-15}$\\
    30   &  -0.00003933557520091981	 &	-0.00003933557520088896	 &   $7.843\times10^{-13}$  &  $1.067\times10^{-14}$\\
    35   &  -0.00002455304484596332	 &	-0.00002455304484594233	 &   $8.549\times10^{-13}$  &  $8.341\times10^{-15}$\\
    40   &  -0.00001634080095822354	 &	-0.00001634080095821053  &   $7.960\times10^{-13}$  &  $5.325\times10^{-15}$\\
    45   &  -0.00001141847437793787	 &	-0.00001141847437792919	 &   $7.605\times10^{-13}$  &  $3.673\times10^{-15}$\\
    50   &  -0.000008290448479296679 &	-0.000008290448479290278 &   $7.722\times10^{-13}$  &  $2.807\times10^{-15}$\\
    55   &  -0.000006208226467936966 &	-0.000006208226467932644 &   $6.961\times10^{-13}$  &  $1.912\times10^{-15}$\\
    60   &  -0.000004768831841073202 &	-0.000004768831841069988 &   $6.739\times10^{-13}$  &  $1.433\times10^{-15}$\\
    70   &  -0.000002990259098529844 &	-0.000002990259098527988 &   $6.209\times10^{-13}$  &  $8.488\times10^{-16}$\\
    80   &  -0.000001996831417921701 &	-0.000001996831417920439 &   $6.316\times10^{-13}$  &  $5.893\times10^{-16}$\\
    \hline
 
 \end{tabular}
\caption{The fractional error in the renormalization coefficient $A$ and the error in $B$ for $\langle\psi_0\rangle$ 
 is given here for a particle in circular orbit in a Schwarzschild background at radius $r_0$. 
$\Delta A$ and $\Delta B$ are the differences between the coefficients obtained numerically and by using the analytic expression (\ref{eq:psi_sing_final}). The analytic value of $B$ is zero.} 
\end{table}

\section{Brief discussion}

The methods discussed in this paper for finding the self-force in a radiation gauge have been used 
to find the self-force on a particle in circular orbit in a Schwarzschild spacetime, and work on 
orbits in a Kerr background is now underway. The advantage of a radiation gauge is the ease with which 
the retarded field can be computed.  A disadvantage is the difficulty in analytically computing the singular field 
$\hsab$ from $\psi_0^{\rm s}$.  We have avoided this difficulty by using a numerical matching procedure 
to find the singular field, and the companion paper shows that the numerical matching reproduces the 
regularization parameters for gauge-invariant quantities to machine accuracy, for the Schwarzschild 
example. 

As described in Sect.~\ref{s:method}, defining a radiation gauge 
singular field by using using either a gauge transformation from the Lorenz gauge 
or using the CCK procedure to construct a singular metric from $\psi_0^{\rm s}$ 
entails an ambiguity associated with a lack of uniqueness in choosing local radiation 
gauge.  Using an angle average to define a renormalized field is equivalent
to setting the self-force equal to the average of the renormalized 
self-force computed for $r>r_0$ and for $r<r_0$ using the Lorenz-gauge 
regularization parameters, and Pound {\it et al.}\cite{pmb13} provide a 
justification for using this average to find the trajectory in the gauge 
for the retarded field we have adopted here -- constructed using the CCK 
procedure to produce a smooth retarded metric perturbation for $r>r_0$ and 
for $r<r_0$. This average value has now been computed by Merlin and Shah \cite{ms14}.            

We have shown that the perturbed metric in a radiation gauge generically has even parity to leading 
order in geodesic distance $\rho$ to the particle trajectory.
Using the renormalized field to compute the perturbed geodesic then relies on showing that the  
singular field gives no contribution to the self-force. 
The companion paper checks this for a particle in circular orbit in a Schwarzschild spacetime, 
in which the self-force is symmetric about a radial line through the particle.  We numerically 
compute the axisymmetric part of the singular field and find that to subleading order 
it coincides with the axisymmetric part of $-{\frak m}\nabla \rho^{-1}$.

This regular behavior of the singular part of the self force may seem remarkable, given the 
line singularity in a radiation gauge that arises when one includes the perturbed mass 
in a radiation-gauge metric obtained from a Hertz potential.   
It is less surprising, however, if one considers a particle at rest in flat space.
When principal null directions are chosen to have an origin not on the particle trajectory,    
$\psi_0$ is nonzero, and the $\ell\ge 2$ part of the metric can be reconstructed 
in closed form in a radiation gauge using the CCK procedure.   The self-force of course 
vanishes, with the contribution from the singular field coinciding with that from the 
retarded field; but the contribution from each is nonzero and gauge invariant under time-independent
gauge transformations.  This implies that the singular part of the expression for the self-force in a 
radiation gauge has its Lorenz form $-{\frak m}\nabla^\alpha \rho^{-1}$.  
For a circular orbit, the result is implied by the fact that the gauge transformation of 
the self force can be written in a form, Eq.~(\ref{eq:fgauge}), that involves no derivatives 
of the gauge vector $\xi^\alpha$, 
together with the fact that, 
although $\xi^\alpha$ is logarithmically divergent, its 
contribution to a mode sum is finite \cite{pmb13}.

\begin{acknowledgments}
For a number of helpful discussions, we thank Leor Barack, Steven Detweiler, Samuel Gralla, Scott Hughes, 
Eric Poisson, Robert Wald, Bernard Whiting, and Alan Wiseman. Both Barack and an anonymous referee 
provided corrections and suggested improvements to an earlier version of this paper.     
This work was supported in part by NSF Grant PHY 0503366.  D.H.K's work was supported by the Alexander
von Humboldt Foundation's Sofja Kovalevskaja Programme funded by the
German Federal Ministry of Education and Research and by WCU (World
Class University) program of NRF/MEST (R32-2009-000-10130-0).
\end{acknowledgments}

\appendix
\section{Gauge transformations}
\label{ap:gauge}
\subsection{Gauge transformation from Lorenz to radiation gauge}

We consider here the perturbed radiation-gauge metrics that describe 
asymptotically flat vacuum metric perturbations involving no linear 
change in mass or angular momentum.  For harmonic time dependence, 
outgoing perturbations of this kind have, in a Lorenz (transverse-tracefree) 
gauge, the asymptotic behavior
$h_{\alpha\beta} = \hat h_{\alpha\beta}e^{-i\omega u}$, with 
\be
	\hat h_{\bf 11} = O(r^{-3}),\quad \hat h_{\bf 12}=O(r^{-2}), \quad  \hat h_{\bf 13}=O(r^{-2}), 
\label{eq:lor_asymp} \ee
and they therefore satisfy the IRG condition to $O(r^{-2})$:
\be
  l^\beta \hat h_{\alpha\beta} = O(r^{-2}), \quad h=0.
\ee  

We first find a corresponding asymptotically flat radiation-gauge metric perturbation 
by exhibiting a gauge transformation from the given Lorenz-gauge metric 
to an asymptotically flat metric satisfying the exact IRG conditions. 

We then show that there is an asymptotically vanishing gauge transformation to 
an asymptotically flat ORG metric perturbation.  Uniqueness of the gauge transformation
implies that this coincides with the asymptotically flat ORG metric 
obtained from the ORG Hertz potential of Eq.~(\ref{eq:hertza1}).  Here, for asymptotic 
flatness, we are requiring only that the tetrad components $h_{\bm{\mu\nu}}$ 
have, for harmonic time dependence the form $e^{-i\omega u} O(r^{-1})$; and 
for a time-independent perturbation involving no change in mass, angular 
momentum, or asymptotic dipole moment, $h_{\bm{\mu\nu}} = o(r^{-1})$.            

\noindent{\em IRG metric for outgoing radiation}.  \\
We obtain as follows the gauge transformation from a Lorenz gauge to an IRG
metric perturbation,  
\be
h^{\RET,RG}_{\alpha\beta} = \hretab + \na_\alpha \xi_\beta + \nabla_\beta \xi_\alpha.
\ee
We begin with the transformation for a Schwarzschild background and then generalize 
it to Kerr. 
The transformation is described most simply in coordinates $u,r,\theta,\phi$, with $u$ 
the outgoing null coordinate. The harmonics of $\xi^\alpha$ and 
$\hretab$ are then given by 
\be
\xi_{\bm 1} = \tilde\xi_{\bm{1}} Y_{\ell m} e^{-i \omega u}, \qquad 
\xi_{\bm 2} = \tilde\xi_{\bm{2}} Y_{\ell m} e^{-i \omega u}, \qquad
\xi_{\bm 3} = \tilde\xi_{\bm{3}}\, {}_{1}\!Y_{\ell m} e^{-i \omega u}, 
\label{eq:ximu}\ee
with the corresponding spin-weighted harmonics for the tetrad components of the metric perturbation,  
\be
	h^\RET_{\bm{\mu\nu}} = \tilde h_{\bm{\mu\nu}}\, {}_s\!Y_{\ell m}e^{-i\omega u}.
\label{eq:hmunu}\ee
The radiation gauge condition $l^\beta h_{\alpha\beta}=0$ has components
\bsube\bea
          \nabla_{\bm 1} \xi_{\bm 1} &=& -\frac12 h^\RET_{\bm{11}} \\
	\nabla_{\bm 1} \xi_{\bm 2}+ \nabla_{\bm 2} \xi_{\bm 1}&=& -h^\RET_{\bm{12}}\\
\nabla_{\bm 1} \xi_{\bm 3}+ \nabla_{\bm 3} \xi_{\bm 1}&=& -h^\RET_{\bm{13}}.
\eea\esube
These equations have the explicit form,
\bsube\bea
\partial_{r}\tilde\xi_{\bm1} & = & -\frac{1}{2} \tilde h_{\bm1\bm1}, \\
\partial_r \tilde\xi_{\bm2} + \left(i\omega -\frac12 f\partial_r -\frac M{r^2}\right)\tilde\xi_{\bm 1} 
	& = & - \tilde h_{\bf 12}, 	\\
(\partial_{r} - \frac1r)\tilde\xi_{\bm 3} -\frac1r[\ell(\ell+1)/2]^{1/2} \tilde \xi_{\bm 1} &=& -\tilde h_{\bm{13}},
\eea\label{eq:xischw0}\esube
with solution 
\bsube\bea
\tilde\xi_{\bf 1} &=& \frac12\int_r^\infty dr' \tilde h_{\bf 11}(r')  \\
\tilde\xi_{\bf 2} &=& \int_r^\infty dr' \left[\tilde h_{\bf 12}(r')
	+\frac14 f(r')\tilde h_{\bf 11}(r')
	-\left(i\omega+\frac M{r'^2}\right)\xi_{\bf 1}(r')\right] \\
\tilde\xi_{\bf 3} &=& r\int_r^\infty dr' \left[\frac1{r'}\tilde h_{\bf 13}
	-\frac1{r'^2} [\ell(\ell+1)/2]^{1/2} \, \tilde\xi_{\bf 1}\right]. 
\eea\label{eq:xischw}\esube

A result of Price et al. \cite{psw07} shows, for a vacuum perturbation satisfying the 
condition $l^\beta h_{\alpha\beta}=0$, that the remaining gauge condition, $h=0$, is 
also satisfied. That is, when $h_{\bf 1\bm\mu}=0$, we have 
$h = -2h_{\bf34}$; and from Eq.~(16) of that paper, the perturbed 
Einstein equation $\delta (G_{\bf 11} - 8\pi T_{\bf 11})=0$ implies 
\be
h_{\bf 34} 
  = a^0\left(\frac\varrho{\bar\varrho} + \frac{\bar\varrho}\varrho\right) 
	+ b^0(\varrho+\bar\varrho),
\label{eq:h34}\ee
where $a^0$ and $b^0$ are functions of $u,\theta,\phi$.  Now $\tilde h_{\bf 34} = \nabla_{\bf 3} \xi_{\bf 4} + \nabla_{\bf 4}\xi_{\bf 3}$, and Eq.~(\ref{eq:xischw}) implies $\xi_{\bf 3} = O(r^{-1})$, whence 
$\tilde h_{\bf 34}= O(r^{-2})$.  Since the right side of Eq.~(\ref{eq:h34}) is $O(r^{-2})$ only if 
$a^0=b^0=0$, we have $h_{\bf 34}=0$.

For each harmonic, it is not difficult to show that Eqs.~(\ref{eq:xischw}) give the unique solution 
to Eqs.~(\ref{eq:xischw0}) for which the components 
$h^{IRG}_{\bm{\mu\nu}}$ vanish asymptotically for $\omega\neq 0$ and 
vanish faster than $r^{-1}$ for $\omega=0$: Any other gauge 
transformation differs from the solution (\ref{eq:xischw}) by a solution to the homogeneous 
equations, to Eqs.~(\ref{eq:xischw0}) with $h_{\bm{\mu\nu}}=0$.  
Their general solution is
\bsube\bea
\tilde\xi_{\bf 1} &=& k_{\bf 1}  \\
\tilde\xi_{\bf 2} &=& -(i\omega+M/r)k_{\bf 1}+k_{\bf 2}, \\
\tilde\xi_{\bf 3} &=& -[\ell(\ell+1)/2]^{1/2}k_{\bf 1}+k_{\bf 3}r.
\eea\esube 
For $\omega$ nonzero, $h_{\bm{22}}$ vanishes asymptotically only if $k_{\bf 1}=0$ and 
$k_{\bf 2}=0$; and $h_{\bm{33}}$ vanishes asymptotically only if $k_{\bf 3}=0$.
Similarly, for $\omega=0$, 
$h_{\bm{23}}=o(r^{-1})$
 only if $k_{\bf 1}=k_{\bf 2}=0$; and $h_{\bm{33}}=o(r^{-1})$ only if $k_{\bf 3} =0$.

For a Kerr background, we similarly find the harmonics of the gauge vector in Kerr 
coordinates $u,r,\theta,\tilde \phi$ of Eq.~(\ref{eq:kerrcoords}).
Harmonics of the gauge vector have the form 
\be
\xi_{\bm 1} = \tilde\xi_{\bm{1}}(r,\theta) e^{i(m \tilde\phi-\omega u)}, \quad 
\xi_{\bm 2} = \tilde\xi_{\bm{2}}(r,\theta) e^{i(m \tilde\phi-\omega u)}, \quad
\xi_{\bm 3} = \tilde\xi_{\bm{3}}(r,\theta) e^{i(m \tilde\phi-\omega u)}. 
\label{eq:xikerr0}\ee
The corresponding harmonics for the metric perturbation are 
\be
	h^{\rm ret}_{\bm{\mu\nu}} = \tilde h_{\bm{\mu\nu}}(r,\theta) e^{i(m \tilde\phi-\omega u)}.
\ee
The gauge transformation for a Kerr background is governed by the equations
\bsube\bea
{\bm D}\xi_{\bm 1} &=& - \frac 1 2  h^{\rm ret}_{\bm {11}},\label{ker_gt1}\\
{\bm D}\xi_{\bm 2} + ({\bf \Delta}-\gamma -\bar\gamma)\xi_{\bm 1} + (\bar \tau - \pi)\xi_{\bm 3} + (\tau -\bar\pi)\xi_{\bm 4}&=& - h^{\rm ret}_{\bm {12}},\label{ker_gt2}\\
({\bm \delta}-2\bar\pi)\xi_{\bm 1} + ({\bm D}+\bar \varrho)\xi_{\bm 3} &=& -h^{\rm ret}_{\bm {13}},\label{ker_gt3}
\eea\label{eq:dxikerr}\esube
The components $\xi_{\bm \mu}$ are given successively by
\bsube
\bea
\tilde\xi_{\bm 1}&=& \frac 1 2 \int^\infty_r dr'\tilde{h}_{\bm {11}},\\
\tilde\xi_{\bm 3}&=& -\frac1{\bar\varrho} \int_r^\infty dr'\bar\varrho \left[\tilde h_{\bm {13}}
	- (\bm\delta-2 \bar\pi)\tilde\xi_{\bf 1}\right],\\
\tilde\xi_{\bm 2}&=& \int_r^\infty dr'\left[ \tilde h_{\bm {12}} 
	+ (\bm\Delta -\gamma-\bar\gamma)\tilde\xi_{\bm 1} + (\bar\tau-\pi)\tilde\xi_{\bm3} + (\tau-\bar\pi)\tilde\xi_{\bm4}\right].
\eea\label{eq:xikerr}\esube

Asymptotic regularity follows from the asymptotic behavior (\ref{eq:lor_asymp}) 
of components along $l^\alpha$ of outgoing waves in a Lorenz gauge. And the Price 
et al. result again implies $h=0$. \\ 

\noindent{\em IRG perturbed metric for ingoing radiation} \\

The Hertz potential construction yields an asymptotically flat 
IRG form for each ingoing asymptotically flat metric perturbation.  
We find the gauge transformation from a Lorenz gauge to 
this asymptotically flat IRG.  For simplicity, we restrict consideration 
to a Schwarzschild background.  

In Eqs.~(\ref{eq:ximu}) and (\ref{eq:hmunu}) the outgoing null coordinate $u$ is replaced 
by the ingoing null coordinate $v=t+r^*$.  In Eqs.~(\ref{eq:xischw0}) $\partial_r$ is 
the replaced by $e^{2im\omega r^*}\partial_r e^{-2i\omega r^*}$, and the solution 
has the form 
\bsube\bea
\tilde\xi_{\bf 1} &=& \frac12e^{2i\omega r^*}\int_r^\infty dr' e^{-2i\omega r'^*}\tilde h_{\bf 11}(r')  \\
\tilde\xi_{\bf 2} &=& e^{2i\omega r^*}\int_r^\infty dr' e^{-2i\omega r^*}\left[\tilde h_{\bf 12}(r')+\frac14 f(r')\tilde h_{\bf 11}(r')
	-\left(i\omega+\frac M{r'^2}\right)\tilde \xi_{\bf 1} \right]\\
\tilde\xi_{\bf 3} &=& re^{2i\omega r^*}\int_r^\infty dr' e^{-2i\omega r^*} \left[\frac1{r'}\tilde h_{\bf 13}
	-\frac1{r'^2} [\ell(\ell+1)/2]^{1/2} \, \tilde\xi_{\bf 1}\right]. 
\eea\label{eq:xischw2}\esube
Now, however, because the radiation is ingoing, $\tilde h_{\bf 11}\sim e^{i\omega r^*}/r$, when 
$\omega\neq 0$.   
Asymptotic flatness follows from the relation 
\[
   \int_r^\infty dr \frac {e^{ikr^*}}{r^n} = \frac {e^{ikr^*}}{r^n}\left[\frac1{ik}+O(r^{-1})\right].
\]
When $\omega=0$, asymptotic flatness follows from the asymptotic conditions (\ref{eq:lor_asymp}).  
Again, because $\xi_{\bf 3} = O(r^{-1})$, we have $\nabla_{\bf 3}\xi_{\bf 4}=O(r^{-2})$, and the 
Price et al. relation then implies $h=-2h_{\bf 34}=0$.    \\

\subsection{Gauge transformations of the self force}  
A gauge transformation of the self-force was obtained by Barack and Ori \cite{bo01}. We 
give an alternate, covariant derivation, mention a second kind of gauge freedom, 
 and obtain a simpler form of a gauge transformation of the self-force for a particle in circular orbit.  

A gauge transformation is an infinitesimal diffeomorphism that drags an unperturbed 
geodesic of the background metric to a neighboring curve that is a geodesic of the 
dragged-along metric. This can be stated precisely in terms of a congruence of 
timelike geodesics through a neighborhood of a point $P$.  The dragged metric at $P$ 
differs from the original metric 
by $h_{\alpha\beta} = \Lie_{\bm\xi} g_{\alpha\beta}$, and the perturbed geodesic through
$P$ has 4-velocity altered by $\tilde\delta u^\alpha = \Lie_{\bm\xi} u^\alpha$.  The 
perturbed geodesic equation associated with a perturbation that is pure gauge has the form  
\be
\tilde\delta (u^\beta\nabla_\beta u^\alpha) = \Lie_{\bm\xi} (u^\beta\nabla_\beta u^\alpha) = 0.
\ee
Writing 
\be
  \Lie_{\bm\xi} (u^\beta\nabla_\beta u^\alpha) = (\Lie_{\bm\xi} u^\beta)\nabla_\beta u^\alpha
+  u^\beta\nabla_\beta \Lie_{\bm\xi} u^\alpha+ u^\beta[\Lie_{\bm\xi},\nabla_\beta] u^\alpha 
\ee
and  
\be
   [\Lie_{\bm\xi},\nabla_\beta] u^\alpha 
	= u^\gamma\nabla_\beta\nabla_\gamma\xi^\alpha - R^\alpha{}_{\gamma\beta\delta}u^\gamma \xi^\delta, 
\ee 
and with the perturbed acceleration defined by
$\dis \tilde\delta a^\alpha := \tilde\delta u^\beta \nabla_\beta u^\alpha + u^\beta\nabla_\beta \tilde\delta u^\alpha$, we have 
\be  
   \tilde \delta a^\alpha 
      = - ({\bf u\cdot\bm\nabla})^2\xi^\alpha + R^\alpha{}_{\beta\gamma\delta}u^\beta u^\gamma \xi^\delta.
\label{eq:tildea}\ee

In this form, $u^\alpha + \tilde\delta u^\alpha$ is normalized to $1$ 
 with respect to the perturbed 
metric $g_{\alpha\beta}+h_{\alpha\beta}$, and the geodesic equation is affinely parameterized with 
respect to the perturbed metric.  If the geodesic is parameterized so that its tangent 
is normalized to $1$ with respect to the background metric, and we denote by $\delta_{\bm \xi} u^\alpha$  the
change in $u^\alpha$ with that normalization, then 
\be
	\delta_{\bm \xi} u^\alpha = (\delta^\alpha_\beta -u^\alpha u_\beta)\tilde \delta u^\beta = (\delta^\alpha_\beta -u^\alpha u_\beta) \Lie_{\bm\xi} u^\beta. 
\label{eq:gaugeu}\ee
With $\delta_{\bm \xi} a^\alpha:= \delta_{\bm \xi} u^\beta \nabla_\beta u^\alpha+u ^\beta\nabla_\beta \delta_{\bm \xi} u^\alpha$,  Eq.~(\ref{eq:tildea}) implies 
\be
  \delta_{\bm \xi} a^\alpha = -(\delta^\alpha_\beta -u^\alpha u_\beta )({\bf u\cdot\bm\nabla})^2\xi^\beta + R^\alpha{}_{\beta\gamma\delta}u^\beta u^\gamma \xi^\delta,
\label{eq:gaugea}\ee
and $a^\alpha u_\alpha = 0$.  Note that the right side vanishes if $\xi^\alpha$ happens to drag a geodesic of 
the background spacetime to another geodesic of the background spacetime: This is the equation of geodesic 
deviation governing the connecting vector joining two neighboring geodesics of $g_{\alpha\beta}$. 
For general $\xi^\alpha$, the right side of Eq.~(\ref{eq:gaugea}) then measures the failure of $\xi^\alpha$ to 
produce a geodesic of the background metric. 

The effect of a gauge transformation on the perturbed geodesic 
equation (\ref{geopert}) is to replace $\delta u^\alpha$ by $\delta u^\alpha+\delta_{\bm \xi} u^\alpha$ and 
$a^\alpha$ by $a^\alpha + \delta_{\bm \xi} a^\alpha$. 

There is a second kind of gauge freedom, an infinitesimal change in the background geodesic to which 
one compares a geodesic in the perturbed spacetime.  The perturbed geodesic at an initial point 
of the trajectory can then be changed from $u^\alpha$ to $u^\alpha + \delta u^\alpha$, with 
$a^\alpha = 0$.  This allows one to regard the right side of Eq.~(\ref{eq:gaugea}) as the change 
in the perturbed geodesic equation for a geodesic through an initial point $P$ with the 
same initial tangent vector as that in the original gauge.

For a particle in circular orbit, the gauge transformation of the self-force takes a simpler form 
for a gauge vector that is helically symmetric.    
Writing $u^\alpha = u^t k^\alpha$, with $k^\alpha$ the Killing vector $t^\alpha +\Omega\phi^\alpha$,
and using the relations $\Lie_{\bf k} \xi^\alpha = 0$, $\Lie_{\bf k} (k^\beta\nabla_\beta\xi^\alpha) = 0$, 
and $\nabla_\beta\nabla_\gamma k^\alpha = R^\alpha{}_{\gamma\beta\delta}k^\delta$, 
we obtain 
\be
   -({\bf k}\cdot\nabla)^2 \xi^\alpha 
	 = \frac12\xi^\beta\nabla_\beta \nabla^\alpha(k^\gamma k_\gamma) 
		- R^\alpha{}_{\beta\gamma\delta} k^\beta k^\gamma \xi^\delta.
\ee
The last term cancels the last term in Eq.~(\ref{eq:gaugea}) to give 
\be
    \delta_{\bm \xi} a^\alpha = \frac12 (u^t)^2 \xi^\beta \nabla_\beta \nabla^\alpha (k^\gamma k_\gamma), 
\label{eq:dxia}\ee
with corresponding gauge-transformed self-force 
\be
  \widehat f^\alpha = f^\alpha + \frac12{\mathfrak m}(u^t)^2 \xi^\beta\na_\beta \nabla^\alpha (k^\gamma k_\gamma).    
\label{eq:fgauge}\ee
For a particle in circular orbit in a Schwarzschild background, Eq.~(\ref{eq:dxia}) takes the form \cite{2007PhRvD..76l4036B}
\be
 \delta_{\bm \xi} a_r = 3\Omega^2/(1-3M/r) \xi^r.
\ee

\section{
 singularity and large $\ell$ behavior}
\label{ap:singular}

Mode-sum renormalization involves relations, like the formal harmonic decomposition 
\be
  (1-\cos\theta)^{-1/2} = \sum_{\ell=0}^\infty \sqrt{\frac{4\pi}{\ell+1/2}}Y_{\ell0}(\theta),
\label{eq:1mct}\ee  
between the short-distance (ultraviolet) singular behavior of a function 
and the large-$\ell$ behavior of its harmonics.  The angle $\theta$ is geodesic 
distance on the unit 2-sphere $S$ 
to the origin $\theta=0$, and in this example, a function that behaves like 
$\theta^{-1}$ has angular harmonics that behave like $L^{-1/2}$, where, as in the 
body of the paper, $L = \ell+1/2$.  In this appendix, we review how one characterizes 
the large $\ell$ behavior of functions whose explicit angular harmonics 
are not known; and we give a precise meaning, in terms of distributions on the 
2-sphere, to formal expressions like the divergent 
right side of Eq.~(\ref{eq:1mct}) and to the angular harmonics of functions like 
$f=\theta^{-n}$ for which the integral $\int d\Omega f \bar Y_{\ell m}$ diverges. 
We initially restrict the discussion to ordinary spherical harmonics and then 
generalize it to spin-weighted harmonics.  

The angular harmonics of the retarded fields are limits as $r\rightarrow r_0$ 
of angular harmonics of expressions that are nonsingular on spheres of radius 
$r\neq r_0$. One therefore defines the angular harmonics of the singular 
field in the same way; and we end by showing that our formalism reproduces the 
harmonic decomposition defined in this way. 

The general relation between small-angle and large-$\ell$ behavior is essentially 
identical to the relation between a short-distance singularity in a function $f({\bf x})$ and 
the behavior of its Fourier transform $\hat f({\bf k})$ at large $k$.  
In one-dimension, for example, the $n^{\rm th}$ derivative 
of a function $f({\bf x})$ is square integrable if and only if 
$k^n \hat f(k)$ is square integrable, because $\widehat{f^{(n)}}=(ik)^n\hat f$:  
\be
  \int dx  \left|\frac {d^n}{dx^n} f(x)\right|^2 < \infty\quad \Longleftrightarrow \quad
  \int dk  \left| k^n \hat f(k)\right|^2 < \infty.
\ee
Similarly, the $n^{\rm th}$ derivative of a function 
$f(\theta,\phi) = \sum_{\ell m}f_{\ell m}Y_{\ell m}$ on $S$ is square integrable 
if and only if $\ell^n f_{\ell m}$ is square summable:  With $D_a$ the 
covariant derivative operator of the metric $ds^2 = d\theta^2 + \sin^2\theta d\phi^2$ on $S$, 
the angular Laplacian is $D^2 := D_a D^a $, and we have 
\bea
&& \int d\Omega D_{a_1}\cdots D_{a_n} \bar f D^{a_1}\cdots D^{a_n}f 
    = \int d\Omega \bar f (-D^2)^n f < \infty \\
&& \phantom{xxxxxxx} \Longleftrightarrow \qquad \sum_{\ell>0, m} [\ell(\ell+1)]^n|f_{\ell m}|^2 < \infty.	 
\label{eq:df}\eea

   One extends this relation to functions like $(1-\cos\theta)^{-3/2}$ that are not square 
integrable by regarding them as distributions obtained by taking derivatives of functions 
like $(1-\cos\theta)^{1/2}$ that are square integrable. 
If $f$ is any distribution on $S$, $f_{\ell m} = \int d\Omega f\bar Y_{\ell m}$ exists, 
because $Y_{\ell m}$ is smooth.  Thus, for example, writing\\ 
\be
 (\frac2{\alpha^2}D_a D^a +\frac{\alpha+2}{2\alpha})(1-\cos\theta)^{\frac\alpha2}
	= (1-\cos\theta)^{\frac\alpha2 -1} 
\label{eq:fexample}\ee
gives $f=(1-\cos\theta)^{-3/2}$ as a distribution with  
\[
   f_{\ell 0} = \int d\Omega  (1-\cos\theta)^{-1/2} (-2D_a D^a +1/2) Y_{\ell 0}\ \mbox{ (already well defined)},
\] 
or 
\be
 f_{\ell 0} 
= \int d\Omega  (1-\cos\theta)^{1/2} (2D_a D^a+3/2) (-2D_a D^a +1/2) Y_{\ell 0}.
\ee      
For any distribution $f$, the right side of 
Eq.~(\ref{eq:df}) is finite for some negative $n$, with increasingly singular 
distributions corresponding to increasingly negative values of $n$.
To obtain a theorem that relates in this way the short-distance and 
large-$\ell$ behavior of distributions, we define the standard 
spaces of functions whose first $n$ derivatives are square integrable, the 
Sobolev spaces $H_n$, and then extend the definition to spaces $H_r$ of distributions, 
where $r$ can be negative (and need not be an integer). 

Recall that the Hilbert space $L_2(S)$ is the completion of smooth ($C^\infty$) 
functions in the norm $\| f\|$ defined by $\dis \| f\|^2 = \int d\Omega |f|^2$.  
The operator $-D^2+\frac14$ is positive definite with eigenvalues $L^2$. \\         
\noindent
{\bf Definition}.  For positive integers $n$, the Sobolev space $H_n(S)$ is the 
completion of smooth functions in the norm $\|\cdot\|_n$ defined by 
\be
   \| f \|_n^2 = \int d\Omega \left| f \left(-D^2+\frac14\right)^n f \right| .
\ee 

Because the right-hand side is a sum of terms of the form 
\[
C \dis \int d\Omega \bar f D^{2k}f,\quad k=0,\ldots, n,
\]
a function has finite norm if and only if the function and its first $n$ derivatives are square integrable. In particular, $H_0(S) = L_2(S)$.  

Because the operator $-D^2+\frac14$ is positive definite, it has a well-defined square root, 
and we can write the definition of the norm in the more concise form
\be
 \| f\|_n = \|{\cal D}^n f\|, \mbox{ with } {\cal D}:=\left(-D^2+\frac14\right)^{1/2}.
\ee
The action of the operator $\cal D$ on a distribution
$f$ is given by 
\be
  {\cal D} f = \sum_{\ell m} L f_{\ell m}Y_{\ell m}. 
\ee 
Equivalently, ${\cal D} f$ is defined by its action on smooth functions $g$, 
\be
  {\cal D} f(g) = f({\cal D} g), \ \ \mbox{or } 
\int d\Omega\, \bar g{\cal D} f = \int d\Omega\, ({\cal D} \bar g) f
= \sum_{\ell m} L \bar g_{\ell m}f_{\ell m}.
\ee

We can now define $H_r(S)$ for all real $r$:\\
{\bf Definition}.  The Sobolev space $H_r (S)$ is the 
completion of smooth functions in the norm $\|\cdot\|_r$ defined by 
\be
   \| f\|_r =  \left|{\cal D}^r f\right|.
\ee 
With inner product defined by 
\be
\langle g|f\rangle_r = \int d\Omega {\cal D}^r \bar g {\cal D}^r f, 
\ee
the space $H_r$ is a Hilbert space. 

The relation 
\be 
  \| f\|_r^2 = \sum_{\ell m} L^{2r} |f_{\ell m}|^2, 
\ee 
implies that a distribution $f$ is in $H_r$ if and only if the sequence $(f_{\ell m})$ of 
its angular harmonics has finite norm 
\be
  \| (f_{\ell m})\tilde\|_r^2 := \sum_{\ell m} L^{2r} |f_{\ell m}|^2. 
\ee 
Formally, for each $r$, one can turn the set of sequences $(f_{\ell m})$ of 
complex numbers into a Hilbert space%
\footnote{The inner product is 
$\dis \sum_j L^{2r} \bar g_{\ell m} f_{\ell m}$, and the Hilbert space is 
the completion in the norm $\| \cdot \tilde\|_r$ of sequences for which 
$\lim_{\ell\rightarrow\infty} |f_{\ell m}| \ell^n =0$ (these are the sequences 
corresponding to smooth functions).}    
with norm $\| (f_{\ell m})\tilde\|_r$.

A function $f$ is smooth if and only if it is in $H_r$ for all $r$, implying that 
the elements of a sequence $(f_{\ell m})$ are the angular harmonics of a smooth function 
if and only $f_{\ell m}$ falls off faster than any power of $\ell$: 
$\lim_{\ell\rightarrow\infty}\ell^n|f_{\ell m}| = 0$.  A second immediate 
consequence of the correspondence is the fact that ${\cal D}$ maps 
$H_r$ to $H_{r+1}$ and that the Laplacian maps $H_r$ to $H_{r+2}$. 
(In fact, these maps are isomorphisms.)      

The function $\theta^{-1}$ is nearly square-integrable, with the integral 
$\int d\Omega\, \theta^{-2}$ diverging only logarithmically and 
$\int d\Omega\, \theta^{-2+\epsilon}$ finite for all $\epsilon>0$.
This suggests that $\theta^{-1}\in H_{-\epsilon}$ for all $\epsilon>0$, 
and that is in fact the case:  From Eq.~(\ref{eq:1mct}), the function
$f=(1-\cos\theta)^{-1/2}$ satisfies
\be
  \| (f_{\ell m})\tilde\|_{-\epsilon} 
	= \sum_{\ell m} |f_{\ell m}|^2 L^{-2\epsilon} 
	= 4\pi \sum_{\ell} L^{-1-2\epsilon} <\infty, \mbox{ all } \epsilon>0.
\ee 
(Again, for $\epsilon=0$, the sum diverges logarithmically.)
Thus $(1-\cos\theta)^{-1/2}\in H_{-\epsilon}$.  Because 
$[2(1-\cos\theta)]^{-1/2}$ differs from $\theta^{-1}$ by $O(\theta^2)$, 
the function $\theta^{-1}$ and any other function with the same singular behavior belong to $H_{-\epsilon}$.  
From Eq.~(\ref{eq:fexample}), successive applications of $D^2$ imply  
$\theta^{-n}\in H_{-n-\epsilon}$.  If a function has singular behavior 
$\theta^{-n}$ for integer $n$ and if $f_{\ell m}$ has singular behavior 
$L^s$, for some $s$, it follows that $\dis s=n-\frac12$.   
   
Finally, a formal sum $f = \sum_{\ell m} f_{\ell m}Y_{\ell m}(\theta,\phi)$, 
like the right side of Eq.~(\ref{eq:1mct}), has the meaning 
$f(g) = \sum \bar g_{\ell m} f_{\ell m}$, for smooth $g$. 

Finally we must relate this formalism to angular harmonics of functions 
$f(r,\theta,\phi)$ that are singular at $r=r_0$ and smooth for $r\neq r_0$, 
when those harmonics are found as 

\be 
  \lim_{r\rightarrow r_0} \int d\Omega\, f(r,\theta,\phi)\, {}_s\!Y_{\ell m} .
\ee

We suppose that $f$ can be written in the form $D^r F$, where $F(r,\theta,\phi)$ is 
continuous everywhere and smooth for $r\neq r_0$ and where $D$ is 
an operator for which $D$ and $D^\dagger$ have domains that include $C^\infty(S)$.  
Then, for $g$ smooth, 
\beaa
\int d\Omega D^n F g &=& \int d\Omega  F D^{\dagger n} g\quad \Longrightarrow \\
\lim_{r\rightarrow r_0}\int d\Omega D^n F g 
	&=& \lim_{r\rightarrow r_0}\int d\Omega F D^{\dagger n}   g 
	 =  \int d\Omega \lim_{r\rightarrow r_0} F D^{\dagger n}   g\\
	&=& \int d\Omega F(r_0) D^{\dagger n}  g(r_0).
\eeaa
Because this last expression is, by definition, the action of the distribution 
$f(r_0) = D^n F(r_0)$ on $g$, we have the claimed equivalence
\be 
  \lim_{r\rightarrow r_0}\int d\Omega  f g  = 
	\int d\Omega f(r_0) g(r_0). 
\ee 


\section{\label{appsingfield} Mode-sum representation of $\langle\psi_0^\SING\rangle$}

We present here details of the computation $\psi_0^\SING$ as a mode sum, outlined in Sect.~(\ref{singfield}).
We first give the expression for $\psi_0^\SING$ to subleading order in Schwarzschild coordinates, 
including terms involving $t-t_0$.  
As in Sect.~(\ref{singfield}) we denote by $\psi_0^{\rm{s}}$ by $\psi_0^{\textrm{s-L}}$ and $\psi_0^{\textrm{s-SL}}$, 
respectively.

Eqs. (\ref{rho eqn1}), (\ref{rho eqn2}) and (\ref{allstuff}) yield for $\psi_0^{\rm{s}}$ to subleading order the explicit form 
\bea
\psi_0^{\textrm{s-L}} &=& -\frac{3{\mathfrak m} e^{-2i\Phi}}{f_0^2r_0^2 \tilde{\rho}^5}\left(E r_0^2 \sin\Theta e^{i\Phi} + J(r-r_0) - Jf_0(t-t_0)e^{-i\Phi} \right)^2 \nonumber \\
\psi_0^{\textrm{s-SL}} &=& \frac{15{\mathfrak m}}{4\tilde{\rho}^7}\Bigg[-\frac{2J^2Me^{-2i\Phi}}{f_0^4r_0^4}(r-r_0)^5 - \frac{4JMEe^{-i\Phi}}{f_0^4r_0^2}\sin\Theta(r-r_0)^4 + \frac{4J^2Me^{-2i\Phi}}{f_0^3r_0^4}(t-t_0)(r-r_0)^4 \nonumber\\ \nonumber
&& - \frac{2ME^2}{f_0^4}(r-r_0)^3\sin^2\Theta + \frac{2J^2(2J^2\cos^2\Phi + r_0^2)e^{-2i\Phi}}{f_0^2r_0^3}(r-r_0)^3\sin^2\Theta + \frac{4e^{-2i\Phi}J^4M}{f_0^2r_0^6}(r-r_0)^3(t-t_0)^2  \\ \nonumber
&& + \frac{4JEe^{-2i\Phi}(e^{i\Phi}M r_0^2+J^2(M-r_0)\cos\Phi)}{f_0^3r_0^4}(r-r_0)^3(t-t_0)\sin\Theta + \frac{4JEe^{-i\Phi}(r_0^2+2J^2\cos^2\Phi)}{f_0^2r_0}(r-r_0)^2\sin^3\Theta \\ \nonumber 
&& + \frac{4J^2e^{-2i\Phi}[r_0\{r_0-M+f_0r_0+e^{2i\Phi(r_0-M)}\}E^2+J^2f_0^2\cos2\Phi]}{f_0^3r_0^3}(t-t_0)(r-r_0)^2\sin^2\Theta \\ \nonumber 
&& + \frac{4 e^{-3 i \Phi } \left(1+2 e^{2 i \Phi }\right) E(M-r_0) J^3}{r_0^4 f_0^2}(t-t_0)^2(r-r_0)^2\sin\Theta - \frac{4 e^{-2 i \Phi } J^2 M \left(2 J^2+r_0^2\right)}{f_0 r_0^6}(r-r_0)^2(t-t_0)^3 \\ \nonumber 
&& + \frac{2 r_0 E^2 \left(J^2+r_0^2+J^2 \cos 2\Phi \right)}{f_0^2}(r-r_0)\sin^4\Theta + \frac{2 e^{-i \Phi } J^3 E\left(\left(5-e^{-2 i \Phi }\right)-\left(3+e^{2 i \Phi }\right) \frac{r_0}{M}\right) }{r_0 f_0}(t-t_0)(r-r_0)\sin^3\Theta \\ \nonumber 
&& + \frac{1}{f_0^2 r_0^3}2 e^{-2 i \Phi }\left(r_0 \left(J^2 (-2 M+(2+f_0) r_0)+e^{2 i \Phi } r_0 \left(2 J^2+M r_0\right)\right) E^2+f_0^2 J^4 \cos 2 \Phi \right)(r-r_0)(t-t_0)^2\sin^2\Theta \\ \nonumber
&& - \frac{4 e^{-2 i \Phi } (r_0-M) J^3 E\left(3e^{i \Phi }+e^{-i \Phi }\right) }{2r_0^4 f_0}(t-t_0)^3(r-r_0)\sin\Theta + \frac{2 e^{-2 i \Phi }J^4(r_0-M) }{r_0^6}(t-t_0)^4(r-r_0) \Bigg] \\ \nonumber
&& - \frac{3{\mathfrak m}}{\tilde{\rho}^5} \Bigg[- \frac{3 \left(e^{-2 i \Phi } J^2 M\right) }{f_0^3 r_0^4}(r-r_0)^3 - \frac{e^{i \Phi } J r_0 E}{f_0}\sin^3\Theta + \frac{3 e^{-2 i \Phi } J^2 M }{r_0^4}(t-t_0)^3 - \frac{5 e^{-i \Phi } M J E}{r_0^3 f_0^3}(r-r_0)^2\sin\Theta \\ \nonumber 
&& - \frac{5 e^{-2 i \Phi } J^2 M }{r_0^4f_0^2}(r-r_0)^2(t-t_0) - \frac{3 J^2}{r_0 f_0}(r-r_0)\sin^2\Theta + \frac{6 e^{-i \Phi } M J  E }{r_0^2f_0^2}(r-r_0)(t-t_0)\sin\Theta \\ 
&& - \frac{3 e^{-2 i \Phi } M J^2 }{r_0^4f_0}(r-r_0)(t-t_0)^2 + \frac{3 J^2 }{r_0}(t-t_0)\sin^2\Theta \Bigg] + O(\epsilon^{-1}).
\label{finalpsi0sing}\eea

We repeat the form at $t=t_0$ given in the text, in order to label each term with a subscript for later reference.    
\bea
\psi_0^{\textrm{s-L}}&=& \left[-\frac{3{\mathfrak m} E^2 r_0^2}{f_0^2 }\frac{\sin^2\Theta}{\tilde{\rho}^{5}}\right]_1 + \left[- \frac{3{\mathfrak m} J^2 e^{-2i\Phi}}{f_0^2 r_0^2}\frac{(r-r_0)^2}{\tilde{\rho}^{5}}\right]_2 + \left[- \frac{3{\mathfrak m} E J e^{-i\Phi}}{f_0^2} \frac{(r-r_0) \sin \Theta}{\tilde{\rho}^{5}}\right]_3, 
\label{psi0s-L}\\
\psi_0^{\textrm{s-SL}} &=& \left[-\frac{15{\mathfrak m}J^2Me^{-2i\Phi}}{2f_0^4r_0^4}\frac{(r-r_0)^5}{\tilde{\rho}^7}\right]_1 + \left[- \frac{15{\mathfrak m}JMEe^{-i\Phi}}{f_0^4r_0^2}\frac{\sin\Theta(r-r_0)^4}{\tilde{\rho}^7}\right]_2 \nonumber\\ 
&& + \left[\frac{15{\mathfrak m}e^{-2i\Phi} J^2\left((-i\sin\Phi e^{i\Phi}) + \frac{J^2}{r_0^2}\cos^2\Phi\right)}{r_0f_0^2}\frac{(r-r_0)^3\sin^2\Theta}{\tilde{\rho}^7} \right]_3 \nonumber\\ 
&& + \left[\frac{15{\mathfrak m}JEe^{-i\Phi}(r_0^2+2J^2\cos^2\Phi)}{f_0^2r_0}\frac{(r-r_0)^2\sin^3\Theta}{\tilde{\rho}^7}\right]_4 + \left[\frac{15 {\mathfrak m} r_0 E^2 \left(J^2+r_0^2+J^2 \cos 2\Phi \right)}{2 f_0^2}\frac{(r-r_0)\sin^4\Theta}{\tilde{\rho}^7}\right]_5 \nonumber\\ 
&& + \left[\frac{9{\mathfrak m} \left(e^{-2 i \Phi } J^2 M\right) }{f_0^3 r_0^4}\frac{(r-r_0)^3}{\tilde{\rho}^5}\right]_6 + \left[\frac{3{\mathfrak m} e^{i \Phi } J r_0 E}{f_0}\frac{\sin^3\Theta}{\tilde{\rho}^5}\right]_7 + \left[\frac{15{\mathfrak m} e^{-i \Phi } M J E}{r_0^3 f_0^3}\frac{(r-r_0)^2\sin\Theta}{\tilde{\rho}^5}\right]_8 \nonumber \\ 
&& + \left[\frac{9 \mathfrak m J^2}{r_0 f_0}\frac{(r-r_0)\sin^2\Theta}{\tilde{\rho}^5}\right]_9
\label{psi0s-SL}\eea
 
The axisymmetric part of each of these terms is to be written as a sum over 
${\,}_2Y_{\ell0}(\Theta,0)$ at r = $r_0$. Each term in the singular field involves the leading part of $\rho^2$, 
namely 
\be
 \tilde{\rho}^2  :=   A(r-r_0)^2+ B(1-\cos\Theta) = \rho^{(2)}|_{t=t_0} +O(\Theta^4),  
\ee
where
\be  A = \frac{r_0}{r_0-2M}, \qquad B = 2r_0^2\frac{r_0-2M}{r_0-3M}\chi(\Phi), \qquad 
	\chi(\Phi) = 1- \frac{M\sin^2\Phi}{r_0-2M}.
\ee

We follow the notation of DMW, writing 
\be
\tilde\rho^2 := B(\delta^2+1-\cos\Theta), \qquad \delta^2 := A(r-r_0)^2/B.
\label{eq:rhodelta}\ee
Then the leading term in $\pso$ is given by 
\beq
\psi_0^{\SING\textrm{\mbox{-}L}} = -\frac{3{\mathfrak m} E^2 r_0^2}{f_0^2 B^{5/2}}\frac{\sin^2\Theta}{(\delta^2 + 1-\cos\Theta)^{5/2}} - \frac{3{\mathfrak m} J^2 e^{-2i\Phi}}{f_0^2 r_0^2 A B^{3/2}}\frac{\delta^2}{(\delta^2 + 1-\cos\Theta)^{5/2}} - \frac{3{\mathfrak m} E J e^{-i\Phi}}{f_0^2 B^2 \sqrt{A}}\frac{\delta \sin\Theta}{(\delta^2 + 1-\cos\Theta)^{5/2}}.
\label{eq:psing-leading}\eeq
 The axisymmetric part of the above expression is achieved by angle averaging over $\Phi$. 
Substituting the values of A, B, E and J in the above expression, we obtain
\bea
\langle \psi_0^{\SING\textrm{-L}}\rangle_{r_0}(\Theta) &=& \frac{-3{\mathfrak m}(r_0-3M)^{3/2}}{2^{5/2}r_0^2(r_0-2M)^{5/2}} \left\langle \frac{1}{\chi^{5/2}}\right\rangle \lim_{\delta\rightarrow 0}\frac{\sin^2\Theta}{(\delta^2 + 1-\cos\Theta)^{5/2}} \nonumber\\ 
&&+ \frac{-3{\mathfrak m} M (r_0-3M)^{1/2}}{2^{3/2}r_0^2(r_0-2M)^{5/2}}\left\langle \frac{e^{-2i\Phi}}{\chi^{3/2}}\right\rangle \lim_{\delta\rightarrow 0}\frac{\delta^2}{(\delta^2 + 1-\cos\Theta)^{5/2}} \nonumber\\ 
&&+ \frac{-3{\mathfrak m} M^{1/2}(r_0-3M)}{4r_0^2(r_0-2M)^{5/2}} \left\langle \frac{e^{-i\Phi}}{\chi^2}\right\rangle  
	\lim_{\delta\rightarrow 0} \frac{\delta \sin\Theta}{(\delta^2 + 1-\cos\Theta)^{5/2}},
\eea
where $\langle f(\Phi)\rangle = (2\pi)^{-1} \int_0^{2\pi} f(\Phi)d\Phi$.

We start with a form of generating function of the Legendre polynomials given by
\begin{equation} \label{DMW}
\frac{1}{(e^T+e^{-T}-2u)^{1/2}} = \sum_\ell e^{-(\ell+1/2)|T|}P_\ell (u),\qquad T\neq 0,
\end{equation}
and set $u=\cos\Theta$, $|T| =\sqrt{2}\delta$.  In the limit $T\rightarrow 0^\pm$, the sum does not converge, 
but it is well-defined as a distribution: 
\begin{equation} 
\lim_{T\rightarrow 0}\frac{1}{(e^T+e^{-T}-2u)^{1/2}} \doteq \sum_\ell P_\ell (u),
\label{eq:dist0}
\end{equation}
where the symbol $\doteq$ means equality of both sides as distributions on the sphere.  
In particular, the regularization proceeds by imposing a cutoff $\ell_{\max}$ on the singular field and on 
the retarded field; the projection ${\cal P}$ of the distribution (\ref{DMW}) onto the subspace $\ell\leq\ell_{\max}$ is the smooth function 
\begin{equation} 
\lim_{T\rightarrow 0}{\cal P}\frac{1}{(e^T+e^{-T}-2u)^{1/2}} = \sum_{l=0}^{\ell_{\max}} P_\ell (u), 
\label{eq:cutoff}
\end{equation}
and the subsequent functions obtained by taking derivatives can be regarded as projections of the 
corresponding derivatives of the distribution (\ref{eq:dist0}).  In particular, taking successive derivatives with 
respect to $T$ gives the relation (Eq. (D11) of DMW) 
\be
  \lim_{T\rightarrow 0} \frac{1}{(e^T+e^{-T}-2u)^{k+1/2}} \doteq \sum_\ell \frac{(2\ell+1)}{2(2k-1)T^{2k-1}} P_\ell (u).
\label{eq:dmwd11}\ee
   
Consider now the expression (\ref{eq:psing-leading}) for the leading term in $\psi_0^\SING$. The first term is proportional 
to $\frac{\sin^2\Theta}{(\delta^2+1-\cos\Theta)^{5/2}}$, where $\delta$ is proportional to $T$.  
We express this term as a sum of Legendre polynomials by  
differentiating (\ref{DMW}) twice with respect to $u$ to obtain
\begin{equation}
\frac{1}{(e^T+e^{-T}-2u)^{5/2}} = \frac{1}{3}\sum_{l=0}^{\infty}e^{-(\ell+1/2)T}P_\ell ''(u).
\end{equation}

The second term is proportional to $\frac{\delta^2}{(\delta^2+1-\cos\Theta)^{5/2}}$.  The fact that it is  
$O(\epsilon^{-3})$ suggests that it can be written as a linear combination of derivatives of 
$\frac{1}{(e^T+e^{-T}-2u)^{1/2}}$,  
\be
\frac{T^2}{(e^T+e^{-T}-2u)^{5/2}} 
		= \alpha \partial_u \frac{1}{(e^T+e^{-T}-2u)^{1/2}} 
		 + \beta\partial_T^2 \frac{1}{(e^T+e^{-T}-2u)^{1/2}} ,
\ee
and solving for $\alpha$ and $\beta$ gives $\alpha = \beta = \frac{1}{2+u}$.  Then
\be
\frac{T^2}{(e^T+e^{-T}-2u)^{5/2}} 
	= \frac{1}{2+u}\sum_\ell e^{-(\ell+1/2)T}\left[P_\ell '(u)+\left(\ell+\frac{1}{2}\right)^2P_\ell (u)\right],
\ee
or
\begin{equation}
\lim_{\delta\rightarrow 0}\frac{\delta^2}{(\delta^2 +1 -\cos\Theta)^{5/2}} 
	\doteq \frac{2^{3/2}}{3}\sum_\ell \left[P_\ell '+\left(\ell+\frac{1}{2}\right)^2P_\ell \right].
\label{eq:leadingterm2}\end{equation}
The axisymmetric part of the third term (its angle average over $\Phi$) vanishes.\\

We next use the same techniques to express the {\em subleading} terms in $\psi_0^\SING$ as power series in $P_\ell$ with 
coefficients polynomial in $\ell$.  The axisymmetric parts of the second, fourth, seventh and eight terms vanish.
In each of the remaining terms, Eq.~(\ref{eq:dmwd11}) is used to expand an expression involving a power 
of  $\delta^2+1-\cos\Theta$, and in each case we find that the term is proportional as a distribution to the 
sum $\dis \sum_{l=0}^\infty (l+1/2)P_l(\cos\theta)$; that sum is a $\delta$-function with support at 
$\Theta=0$ \cite{arfken}: 
\be 
\sum_{l=0}^\infty (l+1/2)P_l(\cos\theta) = \delta(1-\cos\theta). 
\ee
Because ${\,}_2Y_{l,0}$ vanishes at $\Theta=0$, the expansion of the singular field in terms of $_2Y_{\ell 0}$ has 
no subleading contribution.  

We verify the claimed form of each of the subleading terms in Eq.~(\ref{psi0s-SL}) as follows:\\
The first term is proportional to $\delta^5/(\delta^2+1-\cos\Theta)^{7/2}$, and Eq.~(\ref{eq:dmwd11}) gives
\begin{equation}
\lim_{\delta\rightarrow 0+}\frac{\delta^5}{(\delta^2+1-\cos\Theta)^{7/2}} 
	\doteq \frac{2}{5}\sum_\ell \left(\ell+\frac{1}{2}\right)P_\ell (\cos\Theta).
\end{equation}
The third term is proportional to  
\begin{equation}
\frac{\delta^3\sin^2\Theta}{(\delta^2+1-\cos\Theta)^{7/2}} 
	= \frac{2\delta^3}{(\delta^2+1-\cos\Theta)^{7/2}} - \frac{2\delta^5}{(\delta^2+1-\cos\Theta)^{7/2}} +O(\epsilon^{-1}), 
\end{equation}
and Eq.~(\ref{eq:dmwd11}) gives
\begin{equation}
\lim_{\delta\rightarrow 0+}\frac{\delta^3\sin^2\Theta}{(\delta^2+1-\cos\Theta)^{7/2}} 
	\doteq \frac{8}{15}\sum_\ell \left(\ell+\frac{1}{2}\right)P_\ell (\cos\Theta).
\end{equation}
The fifth term is proportional to 
\begin{equation}
\frac{\delta\sin^4\Theta}{(\delta^2+1-\cos\Theta)^{7/2}} 
	= \frac{4\delta}{(\delta^2+1-\cos\Theta)^{3/2}} - \frac{4\delta^5}{(\delta^2+1-\cos\Theta)^{7/2}} 
		- \frac{4\delta^3\sin^2\Theta}{(\delta^2+1-\cos\Theta)^{7/2}},
\end{equation}
and Eq.~(\ref{eq:dmwd11}) gives
\begin{equation}
\lim_{\delta\rightarrow 0+}\frac{\delta\sin^4\Theta}{(\delta^2+1-\cos\Theta)^{7/2}} 
	\doteq \frac{64}{15}\sum_\ell \left(\ell+\frac{1}{2}\right)P_\ell (\cos\Theta).
\end{equation}
The sixth term is proportional to 
\begin{equation}
\lim_{\delta\rightarrow 0+}\frac{\delta^3}{(\delta^2+1-\cos\Theta)^{5/2}} \doteq \frac{2}{3}\sum_\ell\left(\ell+\frac{1}{2}\right)P_\ell (\cos\Theta).
\end{equation}
The ninth term is proportional to 
\begin{equation}
\frac{\delta\sin^2\Theta}{(\delta^2+1-\cos\Theta)^{5/2}}   \frac{2\delta\,[1-\cos\Theta + O(\Theta^4)]}{(\delta^2+1-\cos\Theta)^{5/2}} \\
  = \frac{2\delta}{(\delta^2+1-\cos\Theta)^{3/2}} - \frac{2\delta^3}{(\delta^2+1-\cos\Theta)^{5/2}} + O(\epsilon^{-1});
\end{equation}
again using Eq.~(\ref{eq:dmwd11}) we have
\begin{equation}
\lim_{\delta\rightarrow 0+}\frac{\delta\sin^2\Theta}{(\delta^2+1-\cos\Theta)^{5/2}} 
\doteq \frac{8}{3}\sum_\ell \left(\ell+\frac{1}{2}\right)P_\ell (\cos\Theta).
\end{equation}
The axisymmetric parts of the second, fourth, seventh and the eighth terms vanish.\\

We have obtained the leading terms as series of Legendre polynomials, and we now convert them 
to series involving ${\,}_2Y_{\ell0}$.  We begin with $P_\ell ^{(2)}$.  We use the relation 
between $_s\!Y_{\ell m}$ and $D^\ell_{-sm}(\Theta,\Phi,0)$, the representation matrix for the 
rotation group \cite{goldbergetal} to write $\dis{}_2Y_{\ell0}=Y_{\ell 2}$.   
We next convert the series in terms of Legendre polynomials to series involving ${\,}_2Y_{\ell0}$. 
Then, using 
$P_\ell^{m}
	= \left[\frac{4\pi}{(2\ell+1)}\frac{(\ell+m)!}{(\ell-m)!}\right]^{1/2} Y_{\ell m}$, 
we have 
\beq
P_\ell ^{(2)}(\cos\Theta) = \sum_{\ell'}C_{\ell\ell'}{\ }_2Y_{\ell'0}(\Theta,0), 
\eeq
where
\be
C_{\ell\ell'} = \left[\frac{4\pi(\ell-1)l(\ell+1)(\ell+2)}{(2\ell+1)}\right]^{1/2} \delta_{\ell \ell'}.
\ee
Regarding $P_\ell (\cos\Theta)$ and $P_\ell' (\cos\Theta)$ as elements of $L_2(S^2)$, we find 
\footnote{That is, the coefficients $A_{\ell n}$ and $B_{\ell n}$ are obtained from the 
inner products of $P_n^{(2)}$ with $P_\ell (\cos\Theta)$ and $P_\ell '(\cos\Theta)$, and the relations 
are implied by $L_2$ completeness of $_2Y_{\ell m}$.}  
\bea
P_\ell (\cos\Theta) &=& \sum_{n=2}^\infty A_{\ell n}P_n^{(2)}(\cos\Theta), \\
P_\ell '(\cos\Theta) &=& \sum_{n=2}^\infty B_{\ell n}P_n^{(2)}(\cos\Theta),
\eea
where
\bea
A_{\ell n} & :=& \frac{(2n+1)(n-2)!}{2(n+2)!}\ \left[-\frac{2\ell(\ell-1)}{(2\ell+1)}\delta_{\ell,m} + 4d_{\ell,n}\right], \\
B_{\ell n} &  := &\frac{(2n+1)(n-2)!}{2(n+2)!}\ 2\ell(\ell+1)d_{\ell-1,n},
\eea
with 
\[
d_{\ell, n} = \left\{ \begin{array}{lll} 1, && n -\ell \mbox{ a positive even integer}, \\
					0, &&\mbox{ otherwise}.\end{array}\right. 
\]
The form of these coefficients was found by numerical experiment for $\ell,n<10$ and then checked as rational numbers for larger values.  

The second term in the expression for the leading part of the singular field is proportional 
to the right side of Eq.~(\ref{eq:leadingterm2}); and we now show for each $\ell$ that the 
bracketed expression vanishes as an element of $L_2$.  We have  
\begin{equation}
P_\ell ' + \left(\ell+\frac{1}{2}\right)^2P_\ell = \sum_{\ell'=0}^\infty \left[B_{\ell \ell'}+\left(\ell+\frac{1}{2}\right)^2A_{\ell \ell'}\right]C_{\ell \ell'}\textrm{ }_2Y_{\ell'0}.
\end{equation}
For $\ell'$ even, the sum over $\ell$ in Eq.~(\ref{eq:leadingterm2}) is proportional to 
\begin{equation}
\frac{2(\ell'+2)!}{(2\ell'+1)(\ell'-2)!}\sum_{\ell=0}^\infty  \left[A_{\ell \ell'}+\left(\ell+\frac{1}{2}\right)^2B_{\ell \ell'}\right] 
 = \sum_{\ell=0}^\infty  -\left(\ell+\frac{1}{2}\right)\ell(\ell-1)\delta_{\ell \ell'} 
	+ \sum_{\ell=0,\,\rm even}^{\ell'-2} 4 \left(\ell+\frac{1}{2}\right)^2 
	+ \sum_{\ell=1,\,\rm odd}^{\ell'-1}2\ell(\ell+1) \\
\end{equation}
The second and third sums are given by
\begin{equation}
\sum_{\ell=0,\,\rm even}^{\ell'-2} (2\ell+1)^2   = \sum_{p=0}^{\ell'/2-1}(4p + 1)^2 
 	= \frac16\ell'(4\ell'^2-6\ell'-1),
\end{equation}
\begin{equation}
\sum_{\ell=1,\,\rm odd}^{\ell'-1}2\ell(\ell+1) = \sum_{p=0}^{\ell'/2-1}[2(2p+1)^2 +2(2p+1)]
	= \frac16\ell'(\ell'+2)(2\ell'-1), 
\end{equation}
and the identity
\begin{equation}
\sum_{\ell=0}^\infty  \left[A_{\ell \ell'}+\left(\ell+\frac{1}{2}\right)^2B_{\ell \ell'}\right] = 0, 
\quad \textrm{   for $\ell'$ even},
\end{equation}
follows.  A similar manipulation yields the same identity for $\ell'$ odd.  We conclude that the projection 
of the distribution 
\[
\sum_{\ell=0}^{\ell_{\max}} \left[P_\ell ' + \left(\ell+\frac{1}{2}\right)^2P_\ell \right] 
\]
along $_2Y_{\ell 0}$ vanishes.

Then, of the three terms in Eq.~(\ref{eq:psing-leading}) for $\langle\psi_0^\SING\rangle$, only the term proportional 
to $\sin^2\Theta/\tilde{\rho}^{5/2}$ is nonzero when written as a sum over 
${\,}_2Y_{\ell0}$; and the axisymmetric part of $\psi_0^\SING$ has, to subleading order, the form 
\beq
\langle\psi_0^\SING\rangle_{r_0}(\Theta) = \frac{-{\mathfrak m}(r_0-3M)^{3/2}}{r_0^2(r_0-2M)^{5/2}}\left\langle \frac{1}{\chi^{5/2}} \right\rangle\sum_{\ell=2}^{\infty}\sqrt{\frac{4\pi(\ell+2)!}{(\ell-2)!(2\ell+1)}}{\,}_2Y_{\ell0}(\Theta,0).
\label{eq:psi_sing_final}\eeq


\section{Comparison of analytic and numerical computation of self-force}
\label{ap:sf}
To show how accurately one can recover the leading and the subleading terms in $L$ in the mode sum expression for 
$a^{{\rm s}\ r}$ by numerically matching a power series in $L$ to the values of $a_\ell ^{ret}$, we will present an example where we know A and $B$ analytically: the contribution to the self-force from the part of $h_{11}$ that is axisymmetric 
about a radial line through a particle in circular orbit in a Schwarzschild background. 
The contribution to the self-acceleration from $h_{11}$ is 
\beq
\langle a[h_{11}]\rangle = \frac{(1-2M/r)^2}{8(1-3M/r)}\left[ 2\partial_t - \left(1-\frac{2M}{r} \right)\partial_r - 6\frac{M}{r^2}  \right] \langle h_{11}\rangle
\label{C1}\eeq

To compute the leading and subleading terms analytically requires us to find the leading and subleading terms of the 
radial and time derivatives of $\psi_0^{\rm S}$.  From Eq.~(\ref{finalpsi0sing}), we have 

\begin{eqnarray}
\langle\partial_t \psi_0^{\SING\textrm{-L}}(t=t_0)\rangle & =& \left\langle\frac {6 \mu J^2(r-r_0) e^{-2i\Phi}}{r_0^2f_0\tilde{\rho}^5} + \frac{6\mu E J \sin^3\Theta e^{-i \Phi}}{f_0\tilde{\rho}^5} \right.\nonumber\\
&&  - \frac{15 \mu J E \sin\Theta\cos\Phi}{f_0^2\tilde{\rho}^7}\left. \left(E^2 r_0^2\sin^2\Theta + \frac{J^2}{r_0^2 } (r-r_0)^2 e^{-2i\Phi}  +  2J E \sin\Theta (r-r_0) e^{-i\Phi}\right) \right\rangle\nonumber\\
&=& \left\langle\frac {6 \mu J^2(r-r_0) e^{-2i\Phi}}{r_0^2f_0\tilde{\rho}^5}
- \frac{30 \mu J^2 E^2 (r-r_0)\sin^2\Theta\cos\Phi e^{-i\Phi}}{f_0^2\tilde{\rho}^7} \right\rangle,
\label{eq:psidot}\end{eqnarray}
with only two terms surviving the angle-average over $\Phi$.  
The first term is proportional to $\frac{\delta}{(\delta^2+1-u)^{5/2}}$ where $\delta \propto (r-r_o)$ and $u = \cos\Theta$. Taking $\partial_u\partial_T$ of Eq.~(\ref{eq:dmwd11}), we have

\beq
\lim_{\delta \rightarrow 0+} \frac{\delta}{(\delta^2+1-u)^{5/2}} \doteq \frac{4}{3}\sum_\ell  (\ell+1/2) P_\ell ^\prime (u) = \frac{1}{6} \sum_\ell  P_\ell ^{(2)}(u).
\label{eq:psiprime}\eeq

The second term is proportional to $\frac{\delta \sin^2\Theta}{(\delta^2+1-\cos\Theta)^{7/2}}$. A straightforward 
calculation gives us 

\beq
\lim_{\delta \rightarrow 0^+} \frac{\delta \sin^2\Theta}{(\delta^2+1-\cos\Theta)^{7/2}} \doteq \frac{8}{15} \sum_\ell  (\ell+1/2) P_\ell ^{(2)}(\cos\Theta)
\eeq

Therefore, we get

\bea
\left\langle\partial_t{\psi}_0^{\SING\textrm{-L}}\right\rangle_{t_0,r \rightarrow r_0} & =& \left( \frac{\mu M (r_0-3M)}{4(r_0-2M)^{5/2}r_0^{7/2}} \left\langle \frac{e^{-2i\Phi}}{\chi^2} \right\rangle  + \frac{\mu M(r_0-3M)}{r_0^{7/2}(r_0-2M)^{5/2}} \left( \frac{1}{4}\left\langle \frac{e^{-2i\Phi}}{\chi^2}\right\rangle - \left\langle \frac{e^{-i\Phi}\cos\Phi}{\chi^3}\right\rangle \right) \right) \\ \nonumber
&& \times \sum_\ell  \sqrt{\frac{4\pi(2\ell+1)(\ell+2)!}{(\ell-2)!}} { }_2Y_{\ell,0}(\Theta)
\eea

A similar calculation for the radial derivative of the leading singular field gives us 

\bea
\left\langle\partial_r\psi_0^{\SING\textrm{-L}}\right\rangle_{t_0,r \rightarrow r_0} &=& \left( \frac{\mu(r_0-3M)^2}{2r_0^{5/2}(r_0-2M)^{7/2}}\left\langle \frac{1}{\chi^3} \right\rangle - \frac{\mu M(r_0-3M)}{8r_0^{5/2}(r_0-2M)^{7/2}} \left\langle \frac{e^{-2i\Phi}}{\chi^2} \right\rangle\right) \\ \nonumber
&& \times \sum_\ell  \sqrt{\frac{4\pi(2\ell+1)(\ell+2)!}{(\ell-2)!}} { }_2Y_{\ell,0}(\Theta)
\eea

Here the superscript S-L refers to the singular leading. The dot and prime represent time and radial derivatives, respectively.

To find the t and r derivatives of the subleading terms, we need to assume the following result, directly 
verified only for small $n$.
\beq
\lim_{\delta \rightarrow 0^+}\frac{\delta^{2n}}{(\delta^2+1-\cos\Theta)^{n+3/2}} \doteq 0, \quad n \ge 1.
\label{eq:assume}\eeq

Using Eqs.~(\ref{eq:psidot}) and (\ref{eq:psiprime}), and (\ref{eq:assume}), we have
\bea
\left\langle\partial_t{\psi}_0^{\SING\textrm{-SL}}\right\rangle_{t_0,r \rightarrow r_0} &=& \left( \frac{5\mu M(r_0-3M)^{3/2}}{r_0^4(r_0-2M)^{5/2}}\left\langle \frac{\cos^2\Phi}{\chi^{7/2}} \right\rangle - \frac{3\mu M(r_0-3M)^{3/2}}{r_0^4(r_0-2M)^{5/2}} \left\langle \frac{1}{\chi^{5/2}} \right\rangle \right)\\ \nonumber 
&& \times \sum_\ell  \sqrt{\frac{4\pi(\ell+2)!}{(2\ell+1)(\ell-2)!}} { }_2Y_{\ell,0}(\Theta)
\eea

\bea
\left\langle\partial_r\psi_0^{\SING\textrm{-SL}}\right\rangle_{t_0,r \rightarrow r_0} &=& \left( \frac{15\mu(r_0-3M)^{5/2}}{2^{9/2}r_0^3(r_0-2M)^{7/2}} \left( \left\langle \frac{\sin^2\Phi}{\chi^{7/2}} \right\rangle + \frac{r_0-M}{r_0-3M} \left\langle \frac{\cos^2\Phi}{\chi^{7/2}} \right\rangle \right) + \frac{3\mu M (r_0-3M)^{3/2}}{r_0^3(r_0-2M)^{7/2}} \left\langle\frac{1}{\chi^{5/2}} \right\rangle  \right) \quad \\ \nonumber
&& \times \sum_\ell  \sqrt{\frac{4\pi(\ell+2)!}{(2\ell+1)(\ell-2)!}} { }_2Y_{\ell,0}(\Theta)
\eea

Here the subscript S-SL refers to singular subleading.

From Eqs.~(\ref{eq:hertza1}), (\ref{mpschw}) and (\ref{C1}), we find
\be
 \langle a[h_{11}]\rangle = AL + B + O(L^{-2}), 
\ee
with 
\bea
A &=& \frac{-5c_1M+4(4c_2M+c_5(r_0-3M))}{8r_0^{5/2}(r_0-2M)^{1/2}} \\
B &=& \frac{(r_0-3M)^{1/2}(-20c_3M+2c_4(7M-2r_0)+5c_6(r_0-3M))}{2r_0^3(r_0-2M)^{1/2}}
\eea
where \\
$
c_1 = \left\langle \frac{e^{-2i\Phi}}{\chi^2} \right\rangle,\quad c_2 = \left\langle \frac{\cos^2\Phi}{\chi^3} \right\rangle ,\quad c_3 = \left\langle \frac{\cos^2\Phi}{\chi^{7/2}} \right\rangle ,\quad c_4 = \left\langle \frac{1}{\chi^{5/2}} \right\rangle ,\quad c_5 = \left\langle \frac{1}{\chi^3} \right\rangle ,\quad c_6 =  \left\langle \frac{\sin^2\Phi}{\chi^{7/2}} \right\rangle + \frac{r_0-M}{r_0-3M} \left\langle \frac{\cos^2\Phi}{\chi^{7/2}} \right\rangle
$.\\
In using Eq.~(\ref{eq:hertza1}) to calculate $A$ and $B$, we ignore the term involving $\partial_t\Psi \propto m\Omega\Psi$, because it smaller than the leading term by three powers of $\ell$. 

We numerically calculate $\langle a^{r,\RET}[h_{11}]\rangle$ by matching it to a series in $L$ of the form
\beq
a^{r,\RET}_{h_{11},0} = A L + B + \frac{D}{L^2} + \frac{E}{L^4} + \cdots.
\eeq
Shown below is a table of the fractional error in $A$ and $B$ when found numerically for the self-force's contribution from 
the axisymmetric part of $h_{11}$.

\begin{table}[h]
  \begin{tabular}{|c|c|c|c|c|c|c|}
	\hline
                \multicolumn{1}{|p{0.5cm}|}{\centering $\frac{r_0}{M}$}
              & \multicolumn{1}{|p{2.75cm}|}{\centering $A_\textrm{analytic}$}
              & \multicolumn{1}{|p{2.75cm}|}{\centering $A_\textrm{numerical}$}
              & \multicolumn{1}{|p{2.0cm}|}{\centering $|\Delta A/A|$}
              & \multicolumn{1}{|p{2.75cm}|}{\centering $B_\textrm{analytic}$}
              & \multicolumn{1}{|p{2.75cm}|}{\centering $B_\textrm{numerical}$}
              & \multicolumn{1}{|p{2.0cm}|}{\centering $|\Delta B/B|$}\\
    \hline
	8	&	0.01025269710132717		&		0.010252697099652308		&		$1.634\times10^{-10}$		&	0.005653763844987715		&		0.005653764469505427		&		$1.105\times10^{-7}$	\\
	10	&	0.006104203254814939	&		0.006104203254275933	&		$8.830\times10^{-11}$		&	0.003815279509214152		&		0.003815279708107976		&		$5.213\times10^{-8}$	\\
	15	&	0.002505224308124764	&		0.002505224308053105	&		$2.860\times10^{-11}$		&	0.001847266656645128		&		0.001847266682564611		&		$1.403\times10^{-8}$	\\
	20	&	0.001361796899725571	&		0.001361796899707438	&		$1.332\times10^{-11}$		&	0.001087734862490035		&		0.001087734868899048		&		$5.892\times10^{-9}$	\\
	25	&	0.0008551286914190482	&		0.0008551286914125439	&		$7.606\times10^{-12}$		&	0.0007157902702909011		&		0.0007157902725340001		&		$3.134\times10^{-9}$	\\
	30	&	0.0005866914314647096	&		0.0005866914314628752	&		$3.127\times10^{-12}$		&	0.0005064133828667952		&		0.0005064133835559564		&		$1.361\times10^{-9}$	\\
	35	&	0.0004274390578343503	&		0.0004274390578335089	&		$1.969\times10^{-12}$		&	0.0003770397957432567		&		0.0003770397960593856		&		$8.384\times10^{-10}$	\\
	40	&	0.0003252513409350586	&		0.0003252513409346194	&		$1.350\times10^{-12}$		&	0.0002915634781173744		&		0.0002915634782815453		&		$5.631\times10^{-10}$	\\
	45	&	0.0002557826779337768	&		0.0002557826779335354	&		$9.438\times10^{-13}$		&	0.0002321631294343098		&		0.0002321631295246379		&		$3.891\times10^{-10}$	\\
	50	&	0.0002064159200491534	&		0.0002064159200490082	&		$7.033\times10^{-13}$		&	0.0001892205218241294		&		0.0001892205218781378		&		$2.854\times10^{-10}$	\\
	55	&	0.0001700795176038907	&		0.0001700795176038054	&		$5.015\times10^{-13}$		&	0.0001571744717361175		&		0.0001571744717682680		&		$2.045\times10^{-10}$	\\
	60	&	0.0001425594920264520	&		0.0001425594920263988	&		$3.730\times10^{-13}$		&	0.0001326282836118128		&		0.0001326282836321071		&		$1.530\times10^{-10}$	\\
	70	&	0.0001043336547903106	&		0.0001043336547902868	&		$2.276\times10^{-13}$		&	0.0000980883612046379		&		0.0000980883612137512		&		$9.291\times10^{-11}$	\\
	80	&	0.0000796517646622090	&		0.0000796517646621949	&		$1.770\times10^{-13}$		&	0.0000754722513438541		&		0.0000754722513490826		&		$6.928\times10^{-11}$	\\

    \hline
  \end{tabular}
  \caption{The table compares values of regularization parameters calculated analytically to values obtained numerically by matching the retarded field to a series in $(\ell+1/2)$; 
the quantity that is regularized is $\langle a^{r,\RET}[h_{11}]\rangle$, as described in the text. The first column lists orbital radius in units of Schwarzschild mass; the second and the fifth columns list the analytically computed leading and the subleading regularization parameters A and B; the third and the sixth columns list the numerically obtained values of A and B, and the fourth and the seventh columns list fractional differences between the analytic and numerical values.}
\end{table}
\bibliography{sf2}
\end{document}